\documentclass{aa}    
\usepackage{graphicx,url,twoopt,natbib}
\usepackage[varg]{txfonts}           
\usepackage{hyperref}                                   
\usepackage{caption}							
\usepackage{longtable}							
\usepackage{multicol}							
\hypersetup{
  colorlinks=false,   
  urlcolor=blue,     
  linkcolor=red,     
}

\makeatletter
\renewcommand*\aa@pageof{, page \thepage{} of \pageref*{LastPage}}
\makeatother

\newcommand{\emptynote}[1]{{%
  \let\thempfn\relax
  \footnotetext[0]{#1}
}}

\begin{document}  

\titlerunning{Binary M-dwarf orbital constraints}
\authorrunning{Calissendorff et al.}

\title{Updated orbital monitoring and dynamical masses for nearby M-dwarf binaries}

\author{Per Calissendorff$^{1,2}$ \and 
		Markus Janson$^{2}$ \and Laetitia Rodet$^{3}$ \and Rainer K\"{o}hler$^{4}$ \and
		Micka\"{e}l Bonnefoy$^{5}$ \and Wolfgang Brandner$^{6}$ \and Samantha Brown-Sevilla$^{6}$ \and Ga\"{e}l Chauvin$^{5,7}$ \and Philippe Delorme$^{5}$ \and Silvano Desidera$^{8}$ \and Stephen Durkan$^{2, 9}$ \and Clemence Fontanive$^{10}$ \and Raffaele Gratton$^{8}$ \and Janis Hagelberg$^{11}$ \and Thomas Henning$^{6}$ \and Stefan Hippler$^{6}$ \and Anne-Marie Lagrange$^{12, 5}$ \and Maud Langlois$^{13,14}$ \and Cecilia Lazzoni$^{8}$ \and Anne-Lise Maire$^{6,15}$ \and Sergio Messina$^{16}$ \and Michael Meyer$^{1}$ \and Ole M\"{o}ller-Nilsson$^{11}$ \and Markus Rabus$^{17}$ \and Joshua Schlieder$^{18}$ \and Arthur Vigan$^{14}$ \and Zahed Wahhaj$^{19}$ \and Francois Wildi$^{11}$ \and Alice Zurlo$^{14, 20,21}$
                                }

\institute{
     Department of Astronomy, University of Michigan, Ann Arbor, MI 48109, USA\\
        e-mail:{\bf percal@umich.edu}
\and Department of Astronomy, Stockholm University, 10691, Stockholm, Sweden
\and Cornell Center for Astrophysics and Planetary Science, Department of Astronomy, Cornell University, Ithaca, NY, 14853, USA
\and The CHARA Array of Georgia State University, Mount Wilson Observatory, Mount Wilson, CA 91023, USA
\and University of Grenoble Alpes, CNRS, IPAG, 38000, Grenoble, France
\and Max Planck Institute for Astronomy, Königstuhl 17, 69117, Heidelberg, Germany
\and Unidad Mixta Internacional Franco-Chilena de Astronomía, CNRS/INSU UMI 3386 and Departamento de Astronomía, Universi-
dad de Chile, Casilla 36-D, Santiago, Chile
\and INAF - Osservatorio Astronomico di Padova, Vicolo dell'Osservatorio 5, 35122, Padova		
\and Astrophysics Research Center, Queen's University Belfast, Belfast, Northern Ireland, UK	
\and Center for Space and Habitability, University of Bern, 3012, Bern, Switzerland				
\and Départment d'astronomie de l’Université de Genève, Chemin Pegasi 51, 1290 Versoix, Switzerland 
\and LESIA, Observatoire de Paris, Université PSL, CNRS, Sorbonne Université, Université de Paris, 5 place Jules Janssen, 92195 Meudon, France 
\and CRAL, UMR 5574, CNRS, Université de Lyon, ENS, 9 avenue Charles André, 69561 Saint Genis Laval Cedex, France	
\and Aix Marseille Univ, CNRS, CNES, LAM, Marseille, France	
\and STAR Institute, Université de Liège, Allée du Six Août 19c, B-4000 Liège, Belgium	
\and INAF - Osservatorio Astrofisico di Catania, Via S. Sofia 78, 95123, Catania, Italy	
\and Departamento de Matem\'atica y F\'isica Aplicadas, Facultad de Ingenier\'ia, Universidad Cat\'olica de la Sant\'isima Concepci\'on, Alonso de Rivera 2850, Concepci\'on, Chile	
\and NASA Goddard Space Flight Center, 8800 Greenbelt Road, Greenbelt, MD 20771, USA	
\and European Southern Observatory, Alonso de Cordova 3107, Vitacura, Casilla, 19001, Santiago, Chile	
\and N\'ucleo de Astronom\'ia, Facultad de Ingenier\'ia y Ciencias, Universidad Diego Portales, Av. Ejercito 441, Santiago, Chile	
\and Escuela de Ingenier\'ia Industrial, Facultad de Ingeniería y Ciencias, Universidad Diego Portales, Av. Ejercito 441, Santiago, Chile	
}        

\date{Received 26 November 2021 / Accepted 13 August 2022}

\abstract{
Young M-type binaries are particularly useful for precise isochronal dating by taking advantage of their extended pre-main sequence evolution. Orbital monitoring of these low-mass objects becomes essential in constraining their fundamental properties, as dynamical masses can be extracted from their Keplerian motion. Here, we present the combined efforts of the AstraLux Large Multiplicity Survey, together with a filler sub-programme from the SpHere INfrared Exoplanet (SHINE) project and previously unpublished data from the FastCam lucky imaging camera at the Nordical Optical Telescope (NOT) and the NaCo instrument at the Very Large Telescope (VLT). Building on previous work, we use archival and new astrometric data to constrain orbital parameters for 20 M-type binaries. We identify that eight of the binaries have strong Bayesian probabilities and belong to known young moving groups (YMGs). We provide a first attempt at constraining orbital parameters for 14 of the binaries in our sample, with the remaining six having previously fitted orbits for which we provide additional astrometric data and updated Gaia parallaxes. The substantial orbital information built up here for four of the binaries allows for direct comparison between individual dynamical masses and theoretical masses from stellar evolutionary model isochrones, with an additional three binary systems with tentative individual dynamical mass estimates likely to be improved in the near future. We attained an overall agreement between the dynamical masses and the theoretical masses from the isochrones based on the assumed YMG age of the respective binary pair. The two systems with the best orbital constrains for which we obtained individual dynamical masses, J0728 and J2317, display higher dynamical masses than predicted by evolutionary models. 
}  

\maketitle


\section{Introduction} \label{sec:intro}
The study of the multiplicity of stars is a useful diagnostic for obtaining insight into their formation and dynamical evolution, as it allows for important properties such as binary fraction, semi-major axis distribution, and mass ratios to be constrained \citep[e.g.][]{burgasser_not_2007}. Since low-mass M-dwarf stars form a natural link between the substellar brown dwarfs and the solar-type stars, and the multiplicity frequency tend to decline with lower masses and later spectral types \citep{duchene_stellar_2013, moe_mind_2017, winters_solar_2019}, it becomes even more crucial to discover and characterise low-mass M-dwarf multiples. Hence, a rigorous understanding of the multiplicity characteristics and their evolution within this transitional mass-region that is made up of M dwarfs is vital for constraining formation scenarios of low-mass stars and brown dwarfs. Astrometric monitoring of such binary systems allow for dynamical masses to be derived, which become essential in the efforts of empirical calibrations of fundamental properties such as the mass-luminosity relation and evolutionary models \citep{dupuy_individual_2017, mann_how_2019, rizzuto_dynamical_2020}. This becomes even more important at the lowest stellar masses for which the current theoretical models have been shown to systematically underestimate M-dwarf masses below $M \leq 0.5\,M_\odot$ by $5-50\,\%$ \citep[e.g.][]{hillenbrand_assessment_2004, montet_dynamical_2015, calissendorff_discrepancy_2017, biller_dynamical_2022}. Since M dwarfs evolve slowly and remain in their pre-main sequence phase for $\sim\,100$ Myrs \citep{baraffe_evolutionary_1998}, M-dwarf binaries become valuable benchmark targets for astrophysical calibrations comparing dynamical masses from observational data to isochronal models \citep[e.g.][]{janson_binaries_2017}. As such, groups and associations of stars that can be expected to have originated from the same mutual cluster or region, commonly referred to as young moving groups (YMGs), have seen an increase in interest of late \citep[e.g.][]{torres_young_2008, malo_bayesian_2013}. Thus, M dwarfs residing in YMGs can be isochronally dated, and binaries with estimated dynamical masses can also be used to robustly test the coevality of the YMGs. 

Motivated by these arguments for binary characterisation and low-mass multiplicity studies, the AstraLux Large M-dwarf Multiplicity Survey systematically studied over 1,000 X-ray active M dwarfs with the lucky imaging technique, identifying $\approx 30\,\%$ as multiple systems, many of which are known YMG members \citep{bergfors_lucky_2010, janson_astralux_2012, janson_binaries_2017}. Although most of these binaries have separations which correspond to orbital periods of several decades to hundreds of years, some have periods short enough so that they can already be mapped out after a few years of monitoring.

The SpHere INfrared Exoplanet (SHINE) project utilising the Spectro-Polarimetric High-contrast Exoplanet REsearch \citep[SPHERE;][]{beuzit_sphere_2019} instrument at the Very Large Telescope (VLT) is surveying $~500$ stars with the purpose of directly detecting substellar companions to the stars in order to better understand their formation and early evolution \citep{chauvin_discovery_2017, SHINE_results}. As an auxiliary result from the survey, several low-mass binaries that coincide with the AstraLux survey sample \citep{SHINE_sample} have been observed with high-contrast imaging which provides high quality precision measurements \citep{SHINE_observations}, which are excellent for astrometry and useful for constraining orbital motion.

Here we present the latest results from the combined effort of the AstraLux M-dwarf multiplicity monitoring programme and the SHINE M-dwarf filler programme. We have identified the 20 most prominent systems for fundamental properties to be characterised from the AstraLux Large Multiplicity Survey which have sufficient orbital coverage with which first-hand constraints can be made from orbital fitting routines. Out of these 20 systems, eight have strong indicators to place them in YMGs and thereby have their ages constrained.

The paper is divided up into the following sections that cover the following areas: Section~\ref{sec:obs_data} where we go into detail on how the target sample was collected from the different surveys we combined, and the observations taken along with how data were reduced. In section~\ref{sec:orbs} we explain the orbital fitting procedures and present the main results and discussions in Section~\ref{sec:discussion} where dynamical masses are compared to theoretical isochronal models. Finally we provide a summary and conclusions in Section~\ref{sec:summary}. The collected astrometric data are given in Appendix~\ref{appendixA}, together with the resulting orbital fits for the binaries in Appendix~\ref{appendixB}.

\section{Observations and data reduction} \label{sec:obs_data}
\subsection{Sample selection and observations}
The target list was created from a sample of known binaries and higher hierarchical-systems from the AstraLux M-dwarf multiplicity survey \citep{janson_astralux_2012}, consisting of over 200 M-dwarf multiple systems with separations within $0.08''-6.0''$, and the extended AstraLux sample \citep{janson_noopsorta_2014} of $\approx 60$ multiples with spectral types M5 and later. We selected 20 systems for which we had identified to either have sufficient orbital coverage that dynamical masses could be robustly constrained, or undergone enough monitoring that $\geq 25\,\%$ of the orbit can be mapped out to provide some useful information. We present here new observations from our survey and some previously unpublished observations of the targets.

We compared the space velocities and positions of the targets with those of known YMGs and associations using the BANYAN $\Sigma$-online tool \citep{gagne_banyan_2018}, the LACEwING code \citep{riedel_lacewing_2017} and the GALEX convergence tool \citep{rodriguez_galex_2013}. Unless otherwise specified, parallaxes and proper motions were obtained from the Gaia archive \citep{gaia_collaboration_gaia_2016}, both Gaia Data Release 2 \citep[DR2;][]{gaia_collaboration_gaia_2018} and Gaia Early Data Release 3 \citep[EDR3;][]{gaia_collaboration_gaia_2021}. Spectral types presented in Table~\ref{tab:targets} were derived by \citet{janson_astralux_2012}, using the $(i^{\prime}-z^{\prime})$ photometry obtained from the AstraLux observations and following the methods by \citet{daemgen_discovery_2007}. Some of the systems have additional information from the resolved near-IR medium resolution spectra from SINFONI which \citet{calissendorff_characterising_2020} used to derive near-IR spectral types from the $JHK$ bands and surface-gravity sensitive emission lines. 

As the systems we present in this survey are binaries, we refer to the distance to the systems from us as $d$ in pc, and the separation between the binary components as $s$, either in projected separations of milliarcseconds (mas) or physical as AU. The target binaries all have designated Two Micron All-Sky Survey (2MASS) identifiers, and we abbreviate them by their first four to six digits as Jhhmm(ss). The target systems are presented in Table~\ref{tab:targets}. The YMGs of interest and their adopted ages are shown in Table~\ref{tab:YMGs}, and the target parallaxes and space velocities shown in Table~\ref{tab:YMGprob} which were used to derive membership probabilities for each source. Not all YMGs and associations are included in each of the YMG membership probability tools, and we did not introduce additional YMGs to the existing code.

\begin{table*}[t]
\centering
\caption{Target binary systems}
\begin{tabular}{lcccccc}
\hline \hline
2MASS ID & Alt. Name &  SpT $(\pm0.5)$ & SpT $(\pm0.6)$  & $J$ & $H$ & $K$ \\
 &  &  $(i^{\prime} - z^{\prime})$ & Near IR & mag & mag & mag  \\
\hline
J00085391+2050252 & GJ 3010 & M4.5+M6.0 & & $8.87 \pm 0.03$ & $8.26\pm0.03$ & $8.01\pm0.02$\\
J01112542+1526214 & GJ 3076 & M5+M6 & M3.1+M9.6 & $9.08\pm0.03$ & $8.51\pm0.04$ & $8.21\pm0.03$\\
J02255447+1746467 & LP 410-22 & M4+M5 & & $10.22\pm0.02$ & $9.60\pm0.02$ & $9.33\pm0.02$\\
J02451431-4344102 & LP 993-116 & M4.0+M4.5 & & $8.06\pm0.02$ & $7.53\pm0.04$ & $7.20\pm0.02$ \\
J04373746-0229282 & GJ 3305 & M0+M3 & & $7.30\pm0.02$ & $6.64\pm0.05$ & $6.41\pm0.02$\\
J04595855-0333123 & UCAC4 433-008289 & M4.0+M5.5 & M1.7+M4.5 & $9.76\pm0.02$ & $9.20\pm0.03$ & $8.91\pm0.02$ \\
J05320450-0305291 & V* V1311 Ori & M2.0+M3.5 & & $7.88\pm0.02$ & $7.24\pm0.04$ & $7.01\pm0.02$ \\
J06112997-7213388 & AL 442 & M4.0+M5.0 & M2.9+M5.2 & $9.55\pm0.02$&$8.96\pm0.03$&$8.70\pm0.03$ \\
J06134539-2352077 & HD 43162B & M3.5+M5.0 & & $8.37\pm0.03$&$7.79\pm0.04$&$7.53\pm0.02$ \\
J07285137-3014490 & GJ 2060 & M1.5+M3.5 & M1+M3 & $6.62\pm0.02$&$5.97\pm0.04$&$5.72\pm0.02$\\
J09075823+2154111 & UCAC4 560-047663 & M2.0+M3.5 & & $9.36\pm0.02$&$8.72\pm0.04$&$8.55\pm0.02$ \\
J09164398-2447428 & LP 845-40 & M0.5+M2.5 & & $8.70\pm0.03$ & $8.05\pm0.03$ & $7.83\pm0.02$ \\
J10140807-7636327 & [K2001c] 27 & M4.0+M5.5 & M2.9+M5.2 & $9.75\pm0.02$&$9.16\pm0.03$&$8.87\pm0.02$ \\
J10364483+1521394$^{\dagger}$ & UCAC4 527-051290 & M5.0+M5.0 & M5.8+M4.3 & $9.97\pm0.03$ & $8.97\pm0.03$ & $8.73\pm0.03$ \\
J20163382-0711456 & TYC 5174-242-1 & M0.0+M2.0 & & $8.59\pm0.03$&$7.96\pm0.05$&$7.71\pm0.02$\\
J21372900-0555082 & UCAC4 421-138878 & M3.0+M3.5 & & $8.78\pm0.02$&$8.22\pm0.03$&$7.91\pm0.02$\\
J23172807+1936469 & GJ 4326 & M3.0+M4.5 & & $8.02\pm0.02$&$7.41\pm0.02$&$7.17\pm0.02$ \\
J23261182+1700082 & UCAC4 536-150368 & M4.5+M6.0 & & $9.36\pm0.02$&$8.80\pm0.03$&$8.53\pm0.02$ \\
J23261707+2752034 & UCAC4 590-138502 & M3.0+M3.5 & & $8.46\pm0.02$&$7.87\pm0.02$&$7.64\pm0.02$ \\
J23495365+2427493 & UCAC4 573-135909 & M3.5+M4.5 & M4.1+M5.2 & $9.91\pm0.02$&$9.31\pm0.02$&$9.06\pm0.02$ \\

\hline
\label{tab:targets}
\end{tabular}
{\small \\
$^{\dagger}$ J10364483+1521394 is a resolved triple system. Here we only consider the outer binary pair referred to as the BC components in the literature. The photometry for the BC components are based on SINFONI observations from \cite{calissendorff_characterising_2020}.
}
\end{table*}

\begin{table}[t]
\centering
\caption{Young moving groups}
\begin{tabular}{lccc}
\hline \hline
Group name & Acronym & Age [Myr] & Reference \\ \hline
$\beta$pic & BPMG & $24 \pm 3$ & B15\\
AB Doradus & ABDMG & $149^{+51}_{-19}$ & B15\\
Argus & ARG & $45\pm5$ & Z18 \\
Carina & CAR & $45^{+11}_{-7}$ & B15\\
Carina-Near & CARN & $\sim 200$ & Z06\\
Columba & COL & $42^{+6}_{-4}$ & B15\\
Hyades & HYA & $750\pm100$ & BH15 \\
Octans & OCT & $35\pm5$ & M15 \\
Tucana-Horologium & THA & $45\pm4$ & B15\\
TW Hydrae & TWA & $10 \pm 3$ & B15 \\
Ursa-Majoris & UMA & $\sim 400$ & J15 \\

\hline
\label{tab:YMGs}
\end{tabular}
{\small 

B15 = \citet{bell_self-consistent_2015}; 
BH15 = \citet{brandt_age_2015};
J15 =  \citet{jones_ages_2015};
ML15 = \citet{murphy_new_2015};
Z06 = \citet{zuckerman_carina-near_2006};
Z18 = \citet{zuckerman_nearby_2019}
}
\end{table}

\begin{table*}[t]
\centering
\caption{Young moving group membership probabilities}
\begin{tabular}{lccccccccc}
\hline \hline
Name  & Parallax & pmRA & pmDEC & \multicolumn{2}{c}{BANYAN $\Sigma$} & \multicolumn{2}{c}{LACEwING} & \multicolumn{2}{c}{Convergence} \\
 & [mas] & [mas/yr] & [mas/yr] & YMG & Prob. & YMG & Prob. & YMG & Prob.\\
\hline
J0008$^{\rm a}$ & $55.26 \pm 0.76$ & $-48.64 \pm 1.63$ & $-260.19 \pm 1.54$ & Field & & Field & & Field & \\
J0111$^{\rm b}$ & $58.00 \pm 7.30$ & $192 \pm 8$ & $-130 \pm 8$ & BPMG & 99.7 & BPMG & 84 & BPMG & 79.0 \\
J0225$^{\rm b}$ & $31 \pm 1.9$ & $185 \pm 8$ & $-39 \pm 8$ & ARG & 44 & TWA & 18 & CARN & 92.6\\
J0245$^{\rm c}$ & $87.37 \pm 1.33$ & $24.0 \pm 12.1$ & $-366.1 \pm 9.4$ & Field & & Field & & Field &  \\
J0437 & $36.01 \pm 0.48$ & $54.78 \pm 0.50$ & $-47.31 \pm 0.39$ & BPMG & 98.2 & ARG & 60 & COL & 94.6\\
J0459 & $22.26 \pm 0.63$ & $69.71 \pm 0.55$ & $38.13 \pm 0.43$ & Field & & HYA & 97 & CARN & 38.3 \\
J0532 & $27.22 \pm 0.58$ & $10.10 \pm 0.53$ & $-40.12 \pm 0.39$ & BPMG & 99.2 & BPMG & 36 & ABDMG & 99.1 \\
J0611$^{\rm a}$ & $17.57 \pm 0.41$ & $22.58 \pm 0.90$ & $62.89 \pm 0.73$ & CAR & 97.7 & COL & 49 & CARN & 66.8 \\
J0613 & $59.38 \pm 0.41$ & $-36.87 \pm 0.32$ & $124.76 \pm 0.43$ & ARG & 85.5 & ARG & 100 & CARN & 0.1\\
J0728 & $64.14 \pm 0.47$ & $-112.74 \pm 0.43$ & $-160.97 \pm 0.48$ & ABDMG & 99.6 & ABDMG & 100 & ABDMG & 8.5 \\
J0907 & $27.39 \pm 0.70$ & $-56.98 \pm 0.70$ & $-187.22 \pm 0.50$ & Field & & ABDMG & 25 & THA & 46\\
J0916 & $23.02 \pm 0.24$ & $-194.86 \pm 0.22$ & $77.94 \pm 0.21$ & Field & & Field & & CARN & 38.5\\
J1014$^{\rm d}$ & $14.5 \pm 0.4$ & $-47.2 \pm 1.7$ & $30.6 \pm 3.6$ & CAR & 92.5 & CAR & 91 & CARN & 99.9 \\
J1036 & $49.98 \pm 0.09$ & $110.21 \pm 0.10$ & $-78.94 \pm 0.08$ & Field & & Field & & Field & \\
J2016 & $29.20 \pm 1.20$ & $42.34 \pm 2.00$ & $39.95 \pm 1.65$ & Field & & Field & & CARN & 2.8 \\
J2137$^{\rm e}$ & $62.1 \pm 18.6$ & 19 & 155 & Field & & Field & & Field &  \\
J2317 & $60.77 \pm 0.76$ & $352.05 \pm 0.73$ & $-131.06 \pm 0.59$ & Field & & Field & & THA & 18.2 \\
J232611 & $46.18 \pm 0.45$ & $125.91 \pm 0.37$ & $-43.93 \pm 0.37$ & Field & & Field & & THA & 74.8 \\
J232617 & $38.76 \pm 0.40$ & $-42.38 \pm 0.43$ & $-42.88 \pm 0.32$ & Field & & Field & & Field & \\
J2349 & $21.11 \pm 0.6$ & $121.78 \pm 0.55$ & $-48.08 \pm 0.46$ & Field & & OCT & 16 & TWA & 99.8\\

\hline
\label{tab:YMGprob}
\end{tabular}
{\small\\
Parallax and proper motions were obtained from the Gaia EDR3 catalogue with the exceptions:\\
$^{\rm a} = $ Gaia DR2; 
$^{\rm b} = $\citet{dittmann_trigonometric_2014}; 
$^{\rm c} = $\citet{riedel_solar_2014}; 
$^{\rm d} = $\citet{malo_bayesian_2013}; 
$^{\rm e} = $\citet{lepine_all-sky_2011}\\
}
\end{table*}

\subsubsection{AstraLux}
The AstraLux Large Multiplicity Survey has been ongoing for over a decade, collecting data of numerous visual binaries by applying the lucky imaging technique. The survey primarily employs two principle instruments; AstraLux Norte on the 2.2m telescope in Calar Alto, Spain \citep{hormuth_astralux_2008}, as well as AstraLux Sur at the 3.5m New Technology Telecope (NTT) at La Silla, Chile \citep{hippler_astralux_2009}. The full frame field of view for the respective AstraLux instrument is $\approx\,24'' \times 24''$ for Norte and $\approx\,15.7''\times15.7''$ for Sur, although typical observations utilise subarray redouts in order to minimise readout times. AstraLux observations are mainly carried out in the SDSS $z^{\prime}$-  and $i^{\prime}$-bands, with a preference towards the $z^{\prime}$- band due to its smaller susceptibility to atmospheric refraction compared to the $i^{\prime}$-band \citep{bergfors_lucky_2010}. 

Our observations typically consisted of 10 000 - 20 000 short exposures of just $15-30\,$ms each, adding up to a total of $300\,$s integration time. Both AstraLux Norte and AstraLux Sur data were reduced with the real-time pipeline at the time of the observations, producing a final image from each observation where a subset of $1-20\%$ of the best frames taken were kept. Generally images where $10\%$ of the frames were kept provided a decent trade-off between sensitivity and resolution. Occasionally for closely separated binaries of similar magnitudes the pipeline would centre the frames on the secondary instead of the primary star, leading to a false stellar ghost to appear at the same separation but shifted at a $180^{\circ}$ from the real secondary \citep{bergfors_lucky_2010}.

We performed calibrations for the astrometric measurements with AstraLux by comparing observations of the Orion Trapezium Cluster and M15 to reference observations of the same fields from \citet{mccaughrean_high_1994} and \citet{van_der_marel_italhubble_2002}. The calibrations were performed by measuring the positions of bright stars within the same field that were recognisable and easily identified. We employed between 5-14 reference stars for the calibrations depending on the quality of the point spread function (PSF) and brightness. We assigned the brightest star in the field of view as the main reference, for which we calculated the relative separation and positional angle to for all other reference stars. We then compared the separations and positional angles for our AstraLux measurements to those of the reference observations, taking the average ratio of the separation as the plate scale and the standard deviation as its uncertainty. Correction for True North was performed in similar manner where the average difference in positional angle was used and standard deviation from the average assigned as the uncertainty. The final AstraLux astrometric calibrations calculated here and those obtained from earlier literature are listed in Table.~\ref{tab:calib}. For two epochs, February and April of 2015, we did not have proper reference fields to calibrate the astrometry to with AstraLux Norte, and we assumed a mean pixel scale and correction for True North from the other AstraLux Norte epochs with proper calibration. This only affected two observations of the J1036BC binary and we assumed that the instrument had not changed significantly at this time compared to our other observed epochs. An alteration in plate scale from our smallest to largest calibration values would only change the resulting projected separation for the binary by $\sim 1\,$mas.

\begin{table}[t]
\centering
\caption{Astrometric calibration for AstraLux observations}
\begin{tabular}{lccc}
\hline \hline
 Date & Plate Scale & True North & Reference \\
 & [mas/pxl] &  [deg]   &    \\ 
 \hline
  2008.03 & $23.58 \pm 0.15 $ & $-0.319 \pm 0.18$ & J12 \\
  2008.88 & $23.68 \pm 0.01$ & $0.238 \pm 0.05$ & J14b \\
  2009.13 & $23.55 \pm 0.17$ & $0.224 \pm 0.20$ & This work \\
  2015.17 & $15.23 \pm 0.13$ & $2.87 \pm 0.26$ &  This work \\
  2015.90 & $15.20 \pm 0.12$ & $-2.09 \pm 0.39$ & This work \\
  2015.99 & $15.20 \pm 0.11$ & $-2.41 \pm 0.30$ & This work\\
  2016.38 & $15.27 \pm 0.19$ & $2.64 \pm 0.22$ & This work \\
  2018.39 & $15.26 \pm 0.19$ & $3.43 \pm 0.41$ & This work \\
  2018.63 & $15.13 \pm 0.49$ & $-3.40 \pm 0.39$ & This work\\
\hline
\label{tab:calib}
\end{tabular}\\
{\small
J12 = \citet{janson_astralux_2012}; J14b = \citet{janson_noopsortborbital_2014}
}
\end{table}


\subsubsection{NaCo}
We downloaded the NaCo data from the ESO archive together with their associated calibration files and performed basic reductions using custom scripts with Python. These basic reductions included corrections for bias, dark, flatfield division and combination of multiple frames.

We applied the astrometric corrections from \citet{chauvin_deep_2010} using plate scales $27.01 \pm 0.05$ mas/pxl for observations in the $L'$-band and  $13.25 \pm 0.05$ mas/pxl for shorter wavelengths. We do not correct the observing frames here for True North, but instead add a factor of $\pm 0.20$ deg to the uncertainty for the positional angle of each astrometric data point given by NaCo data, which is in line with the True North corrections obtained by \citet{chauvin_deep_2010}.

\subsubsection{NOT FastCam}
The Lucky Imaging FastCam is an instrument at the Roque de los Muchachos Observatory on La Palma in the Canaries, Spain, designed and capable of obtaining high-resolution images in optical wavelengths from medium-sized ground-based telescopes at the observatory \citep{oscoz_fastcam_2008}. The instrument features a $512 \times 512$ pixels L3CCD from Andor Technology, and a special software package that reduces images in parallel with the data acquisition at the telescope, so that a small fraction of images with minimal atmospheric turbulence can be evaluated in real-time. For our observations, the FastCam instrument was mounted at the 2.56-m Nordic Optical Telescope (NOT), and the observations were carried out in August and November of 2016 using the $I$-band at $820\,$nm.

The astrometric calibrations were made in a similar way to our AstraLux astrometric calibrations, comparing reference fields of the M15 stellar cluster with images taken by the Hubble Space Telescope. We obtained a platescale of $30.6 \pm 0.1$ mas/pixel for the 2016.63 epoch in August, and a platescale of $30.5 \pm 0.1$ mas/pixel for the November epoch of 2016.87. The corresponding corrections for True North were $+3.64 \pm 0.01^\circ$ and $-1.54 \pm 0.1^\circ$ respectively.

\subsubsection{SPHERE}
The SPHERE data were collected as part of a sub-programme for the SHINE survey \citep{chauvin_shine_2017}. The filler programme from which our observations was taken were devoted to astrometric monitoring of tight visual binaries, many of which had been discovered in the AstraLux survey.

The observations were taken with the instrument operating in field-tracking mode without any coronograph. Observations were carried out in the IRDIFS-EXT mode, which enabled for simultaneous observations with the integral field spectrograph \citep[IFS;][]{claudi_sphere_2008, mesa_performance_2015} and the dual-band imaging sub-instrument IRDIS \citep{dohlen_infra-red_2008,vigan_photometric_2010}. The IFS instrument operated in wavelengths between $0.96 - 1.64\,\mu$m in the Y to H bands, while IRDIS observations were predominately performed with the K1 ($\lambda_c = 2.110 \pm 0.102 \,\mu$m) and K2 ($\lambda_c = 2.251 \pm 0.109\,\mu$m) bands, as well as the H2 ($\lambda_c = 1.593 \pm 0.052\,\mu$m) and H3 ($\lambda_c = 1.667 \pm 0.054\,\mu$m) bands.
 
All SPHERE data, both IRDIS and IFS modes, were downloaded and reduced using the SPHERE data centre \citep{pavlov_sphere_2008,delorme_sphere_2017}. The reductions carried out by the automated pipeline included basic corrections for bad pixels, dark current, flat field, as well as corrections for the instrument distortion \citep{maire_first_2016} and rotation. Calibrations of the platescale and for the True North angle were performed in accordance to \citet{maire_sphere_2016}, with typical corrections of $\approx 12.267$ mas/pixel for IRDIS and $\approx 7.46$ mas/pixel for IFS observations, with a True North of $\approx -1.75^\circ$. The specific plate scale and True North corrections handled by the pipeline are stated for each individual observing data point.

From the astrometric measurements described in Section~\ref{sec:astrometry} we also obtained accurate flux-ratios for the components in each system. We summarised the resulting contrast magnitudes in the SPHERE dual-band images in Table~\ref{tab:contrast}. Not all observed epochs had separate dual-band images available and thus missing from the table.

We did not use the IFS data to perform any spectral analysis of the targets here, some of which have superseding spectral information from the SINFONI observations instead \citep{calissendorff_characterising_2020}. Instead, we collapsed the data cubes and performed astrometry on a single frame from the IFS data.

\begin{table}[b]	
\renewcommand{\arraystretch}{1.3}	
\centering
\caption{Contrast in magnitudes for dual band SPHERE observations.}
\begin{tabular}{lccccc}
\hline \hline
Target & Obs. Date & Band & $\Delta$ mag \\
		& yyyy-mm-dd	&		& \\ \hline
J0611	& 2017-02-06	& $K1$	& $0.28 \pm 0.03$ \\
J0611 	& 2017-02-06	& $K2$	& $0.29 \pm 0.01$ \\
J0611	& 2019-03-05	& $K2$	& $0.28 \pm 0.01$ \\
J0728	& 2016-03-27	& $H2$	& $1.07 \pm 0.02$ \\
J0728	& 2016-03-27	& $H3$	& $1.08 \pm 0.01$ \\
J0916	& 2018-01-27	& $K1$	& $0.44 \pm 0.07$ \\
J0916	& 2018-01-27	& $K2$	& $0.44 \pm 0.04$ \\
J0916	& 2018-02-25 	& $K1$ 	& $0.45 \pm 0.07$ \\
J0916	& 2018-02-25	& $K2$	& $0.42 \pm 0.05$ \\
J0916	& 2019-03-06	& $K1$	& $0.27 \pm 0.05$ \\
J0916	& 2019-03-06	& $K2$	& $0.30 \pm 0.04$ \\
J1014	& 2018-05-06	& $H2$	& $0.04 \pm 0.01$ \\
J1014	& 2018-05-06	& $H3$	& $0.06 \pm 0.01$ \\
J1014	& 2019-03-09	& $K1$	& $0.05 \pm 0.01$ \\
J1014	& 2019-03-09	& $K2$	& $0.06 \pm 0.01$ \\
J1036	& 2018-04-17	& $K2$	& $0.02 \pm 0.01$ \\
J2016	& 2015-09-24	& $K1$	& $0.36 \pm 0.01$ \\
J2016	& 2015-09-24	& $K2$	& $0.32 \pm 0.01$ \\
J2317	& 2015-09-25	& $K1$	& $1.22 \pm 0.01$ \\
J2317	& 2015-09-25	& $K2$	& $1.17 \pm 0.01$ \\
\hline
\label{tab:contrast}
\end{tabular}
\end{table}

\subsection{Astrometry}\label{sec:astrometry}
\begin{figure}
	\includegraphics[width=\linewidth]{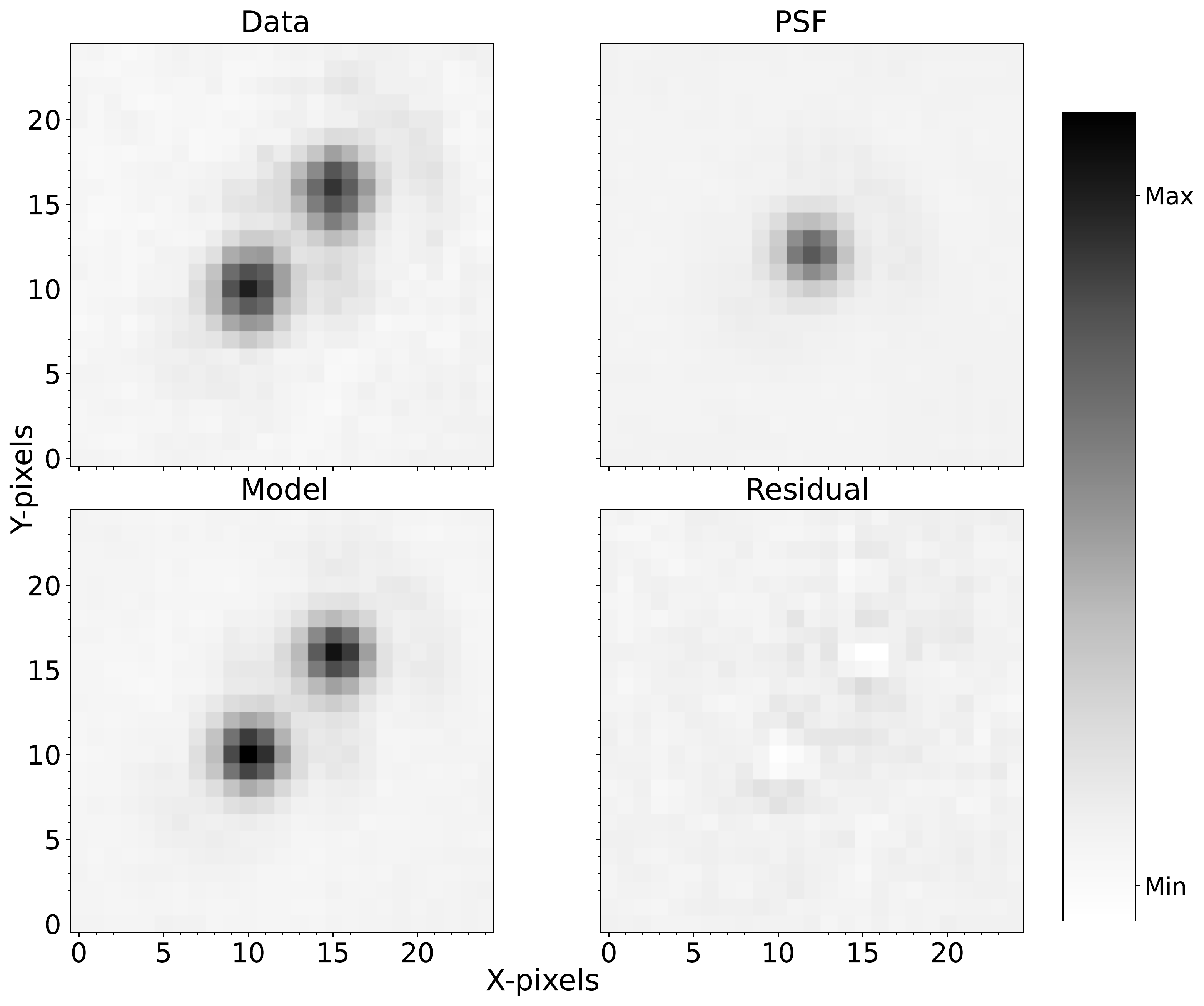}
	\caption{Astrometric extraction example for the J1036BC binary. The upper plots display the observed AstraLux data for the binary pair and a PSF reference. The lower plots show the constructed model using two brightness-scaled and position-shifted PSF models, and the residual after subtracting the model from the observed data. The plots here have been normalised to the peak value of the observed data, are scaled linearly in the plots. The colour-bar markers indicate the peak and bottom values for the observed data. The residual for the AstraLux is typically of the order of $\pm 15\,\%$, whereas with SPHERE the residual is more commonly around $\pm 1\,\%$.}	
	\label{fig:astrometry}
\end{figure}

Astrometric positions were calculated with the same procedure as described in \citet{calissendorff_discrepancy_2017,calissendorff_spectral_2019,calissendorff_characterising_2020}. Concisely, a grid in $x$ and $y$ positions was constructed where we scaled the brightness of a reference PSF, placing two of them on the grid which were sequentially shifted in positions to match the observed data. A residual was then calculated by subtracting the constructed model from the observed data, and the procedure was iterated while scaling the brightnesses and shifting the positions of the model until a minimum residual could be found. The basic workflow of the astrometry extraction procedure is illustrated in Figure~\ref{fig:astrometry} where we used the J1036BC binary and our AstraLux data from April 2015 as an example.

The astrometric extraction was performed in the same manner for all observations and instruments considered here. Generally we would try to obtain a good PSF reference from the same or close to the same epoch as the observation we were extracting astrometry from. In previous AstraLux observing campaigns, designated PSF reference in the form of single stars have been procured \citep{janson_astralux_2012, janson_noopsorta_2014}. However, that was not the case for later epochs. Instead, we identified which binaries or higher hierarchical systems had large relative separations or isolated components, which we then used as PSF references. For AstraLux, FastCam and SPHERE we had access to the primary in the triplet 2MASS J10364483+1521394 system for several epochs, which was close to an ideal PSF reference for our intended purposes given the similar M-dwarf spectral type of the component to the rest of our target sample and that it was available for most out of our observed epochs. The primary component in the system has a projected separation of $\approx 1''$ to the outer binary pair and can be viewed as a single star-proxy in this context. Another benefit from using the primary of J1036 was that whatever aberrations afflicted the observations, altering the PSFs of the binary, would also be seen in the reference PSF so that they could be accounted for. Nevertheless, to increase the statistical certainty of the astrometric measurements we typically used between 3-10 different reference PSFs depending on target quality and instrument applied. We then calculated the mean separation and positional angles, using the standard deviation as the error which we added quadratically together with the instrumental errors (plate scale and True North errors) to the uncertainty. We did not try to mix PSF references from observations taken with different instruments or settings.

Due to the larger field of view and smaller plate scales for the VLT instruments NaCo and SPHERE, we could with ease isolate single components and use them as PSF references. For the AstraLux observations we had the advantage of having a plethora of multiple systems to choose from in the AstraLux Multiplicity Survey. The FastCam observations however did not have quite the same luxury, as the coarser plate scale made it more cumbersome to isolate single components. As such, we mainly used the primaries from the J0111 observations taken in August and the J1036 observations taken in November as PSF references. We also included three additional references for our FastCam astrometry; J0103, J0916 and J1641, which are known tightly bound binaries but appeared as unresolved single sources in the FastCam observations.

Since SINFONI was not calibrated for astrometry we added an extra uncertainty term. We checked the consistency between the SINFONI astrometry presented in \citet{calissendorff_characterising_2020} and our SPHERE astrometry for the binaries which were observed at similar epochs and found no large discrepancy between the two. We therefore included the SINFONI data points into the fitting procedure when they were believed to aid in constraining orbital parameters.


\subsection{Radial velocities}
\begin{figure*}
	\centering
	\includegraphics[width=0.4\textwidth]{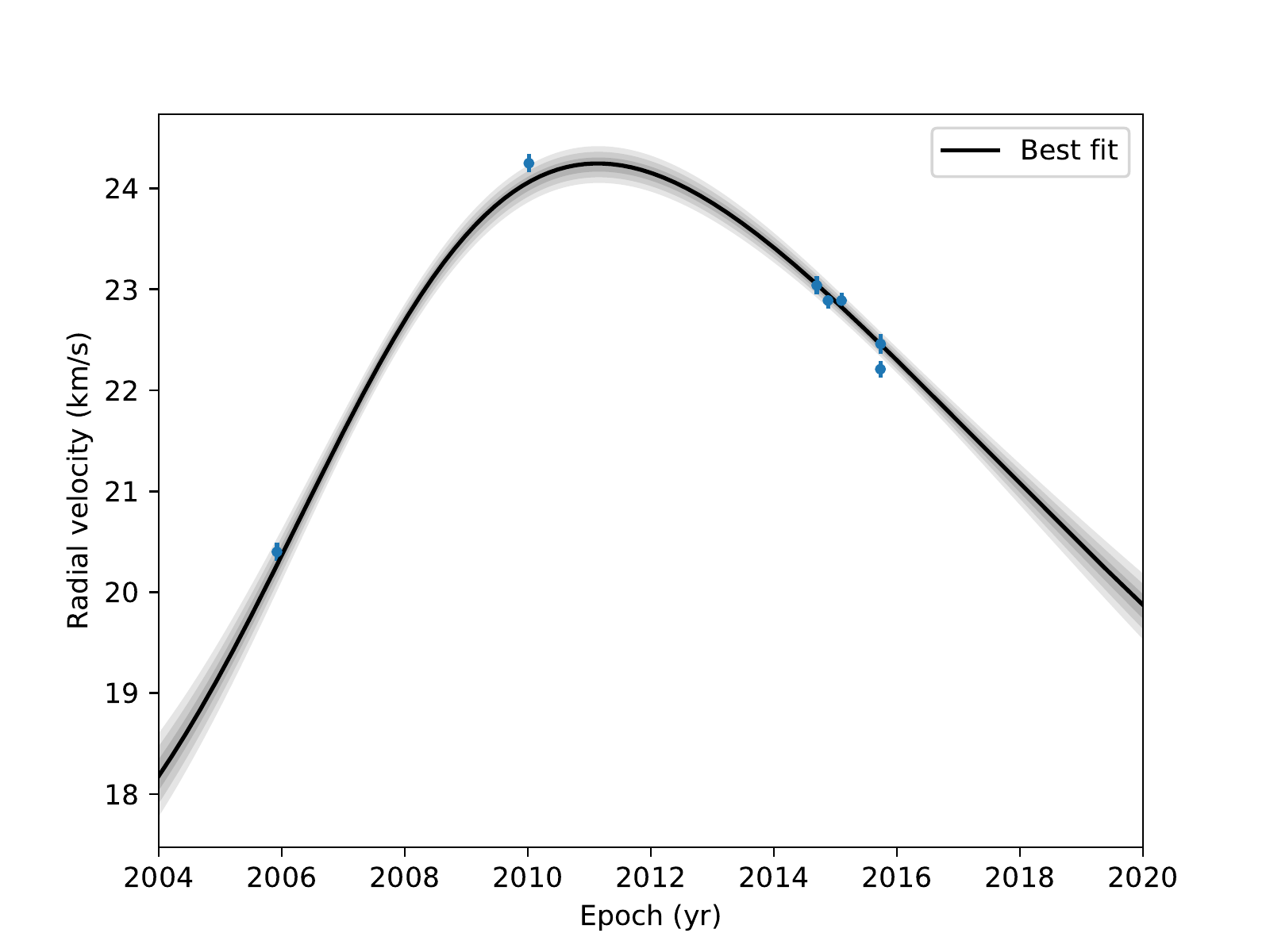}%
	\includegraphics[width=0.4\textwidth]{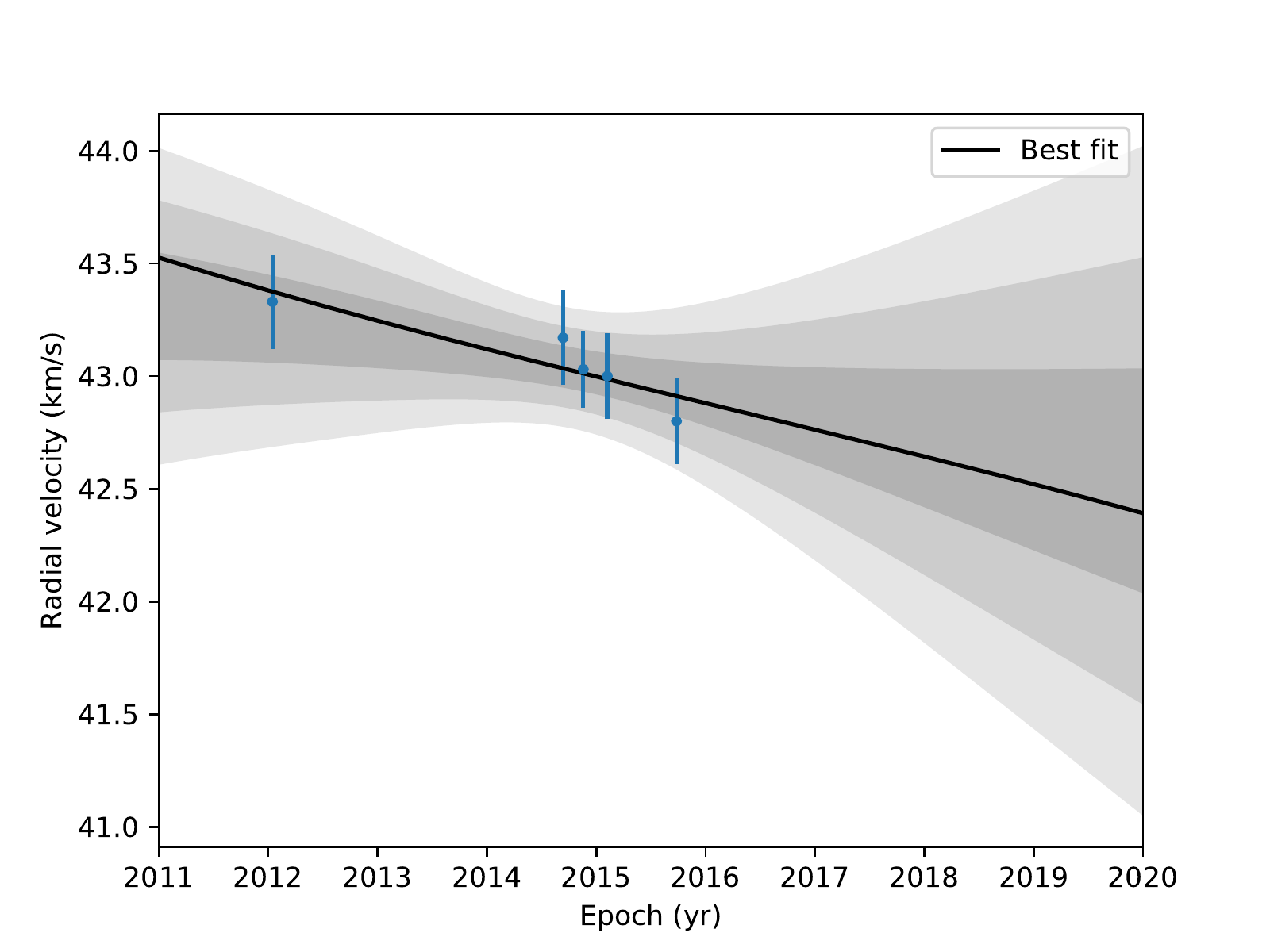}
	\includegraphics[width=0.4\textwidth]{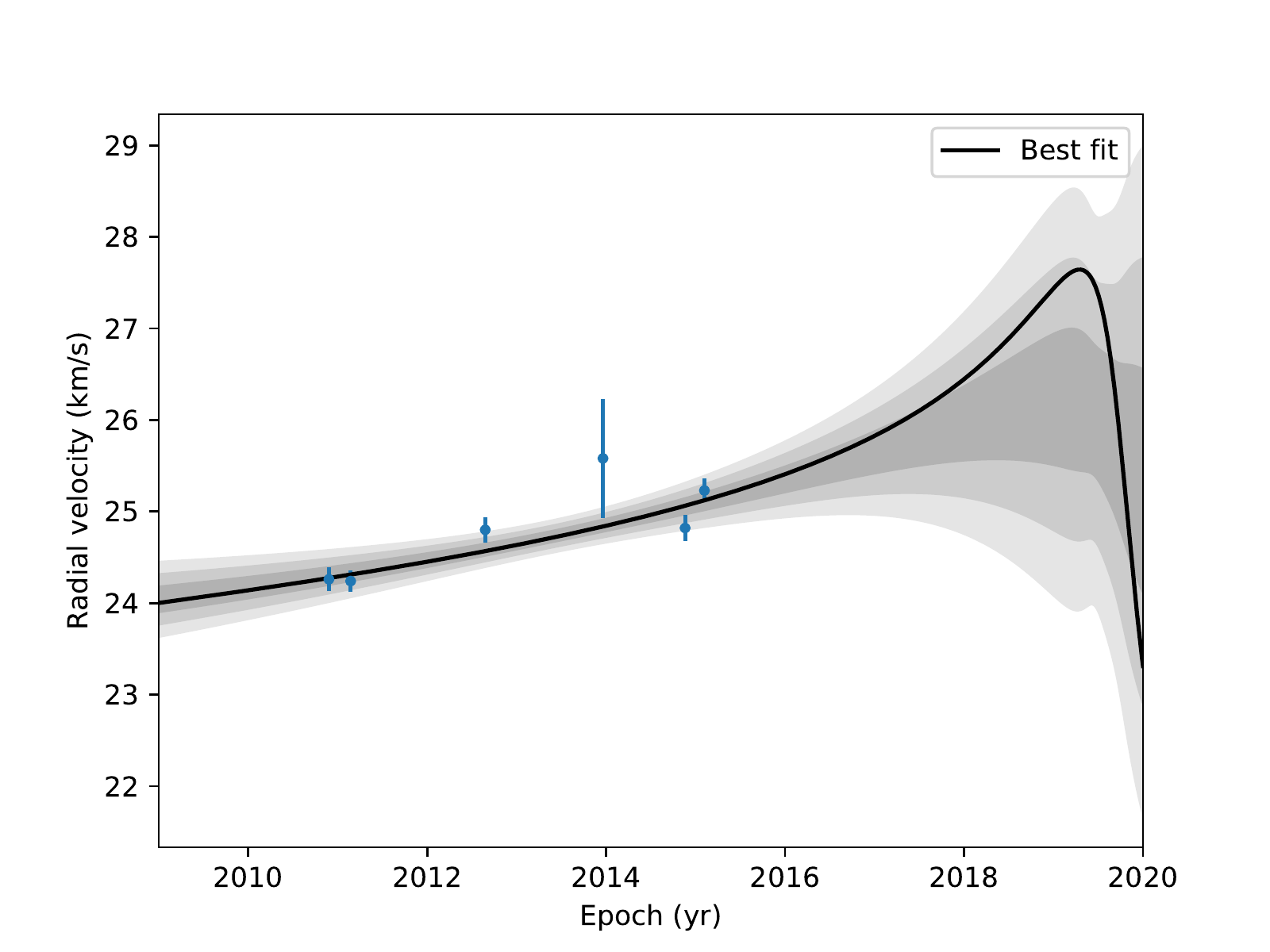}%
	\includegraphics[width=0.4\textwidth]{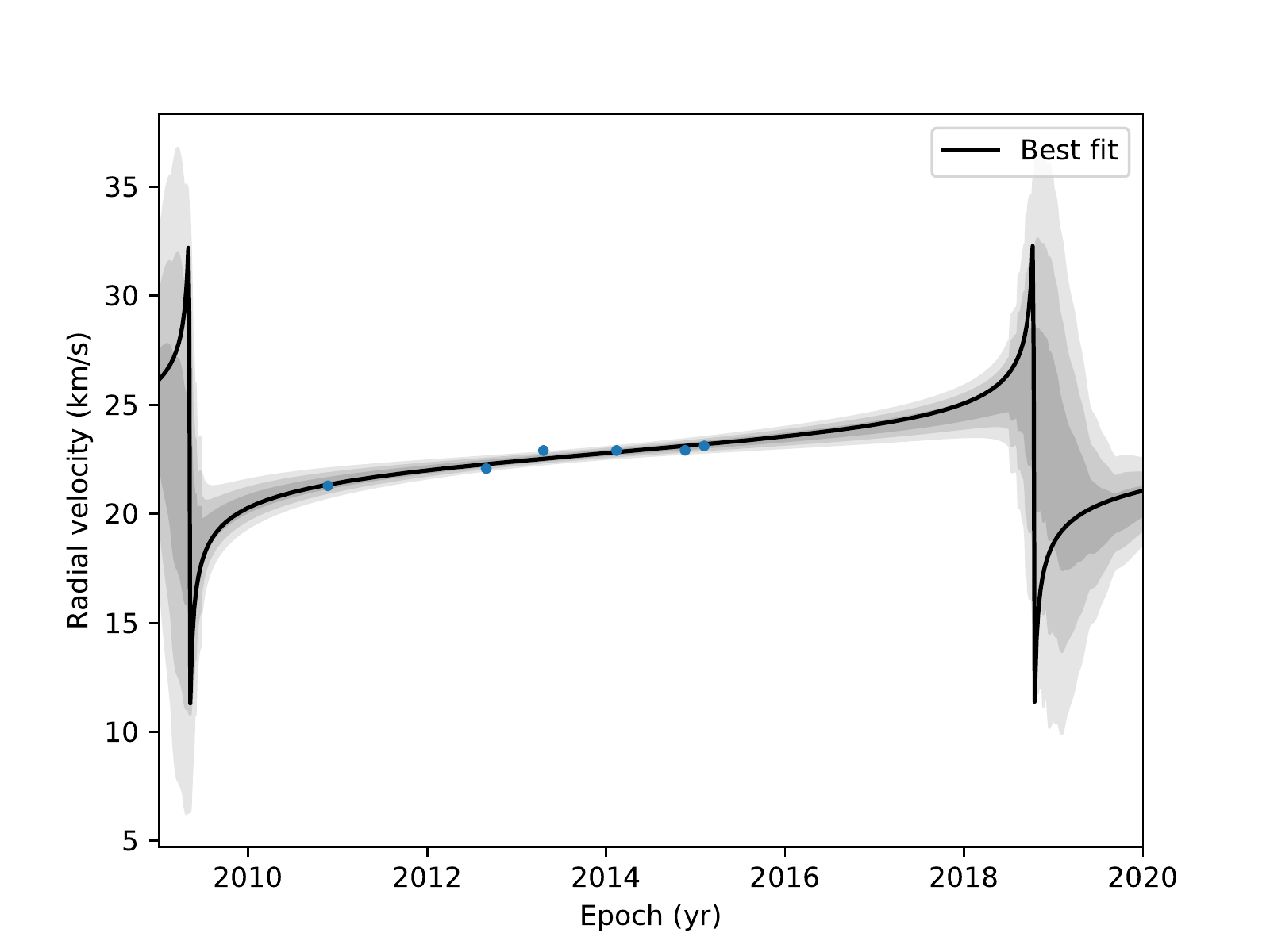}
	\includegraphics[width=0.4\textwidth]{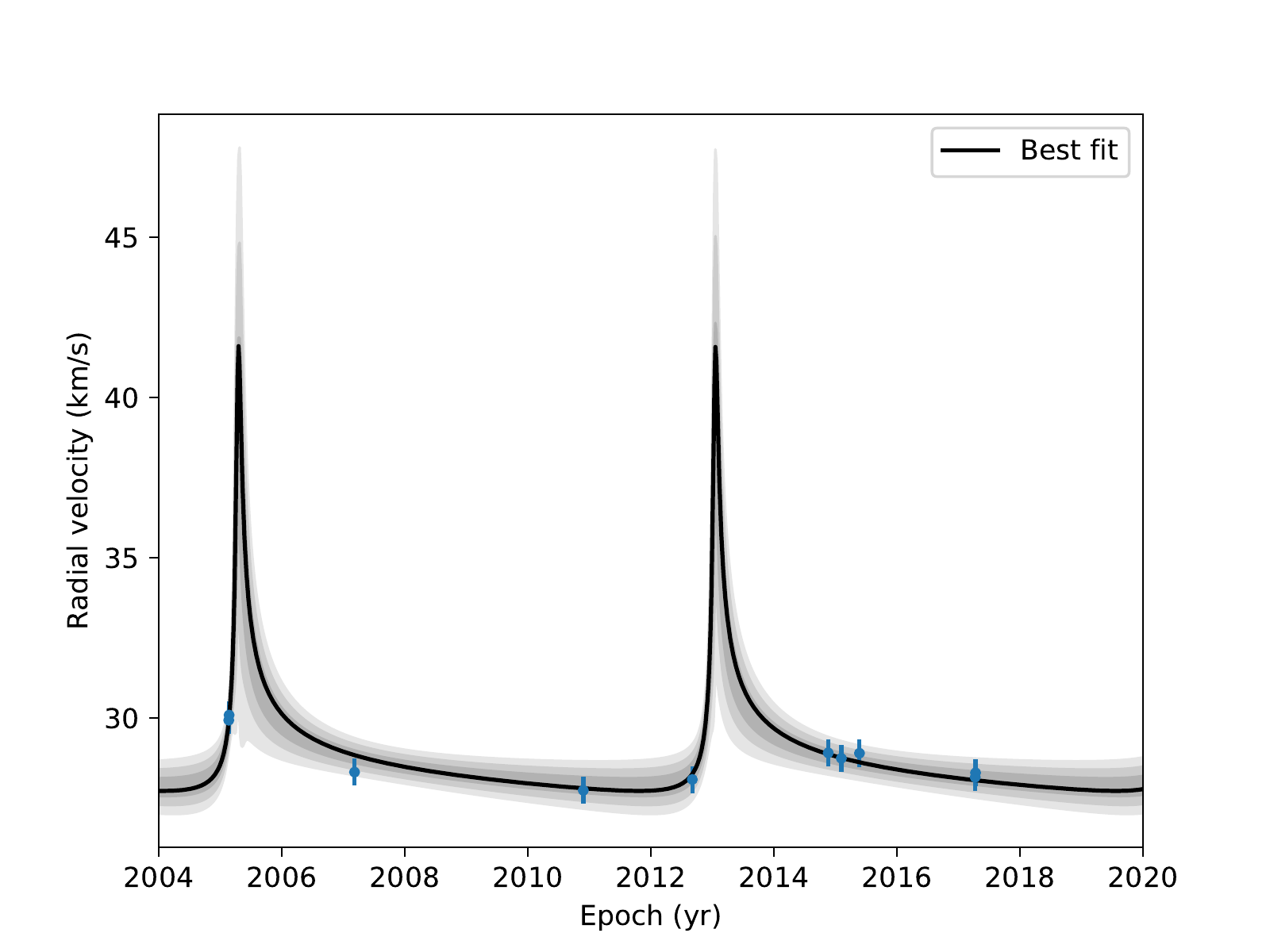}%
	\includegraphics[width=0.4\textwidth]{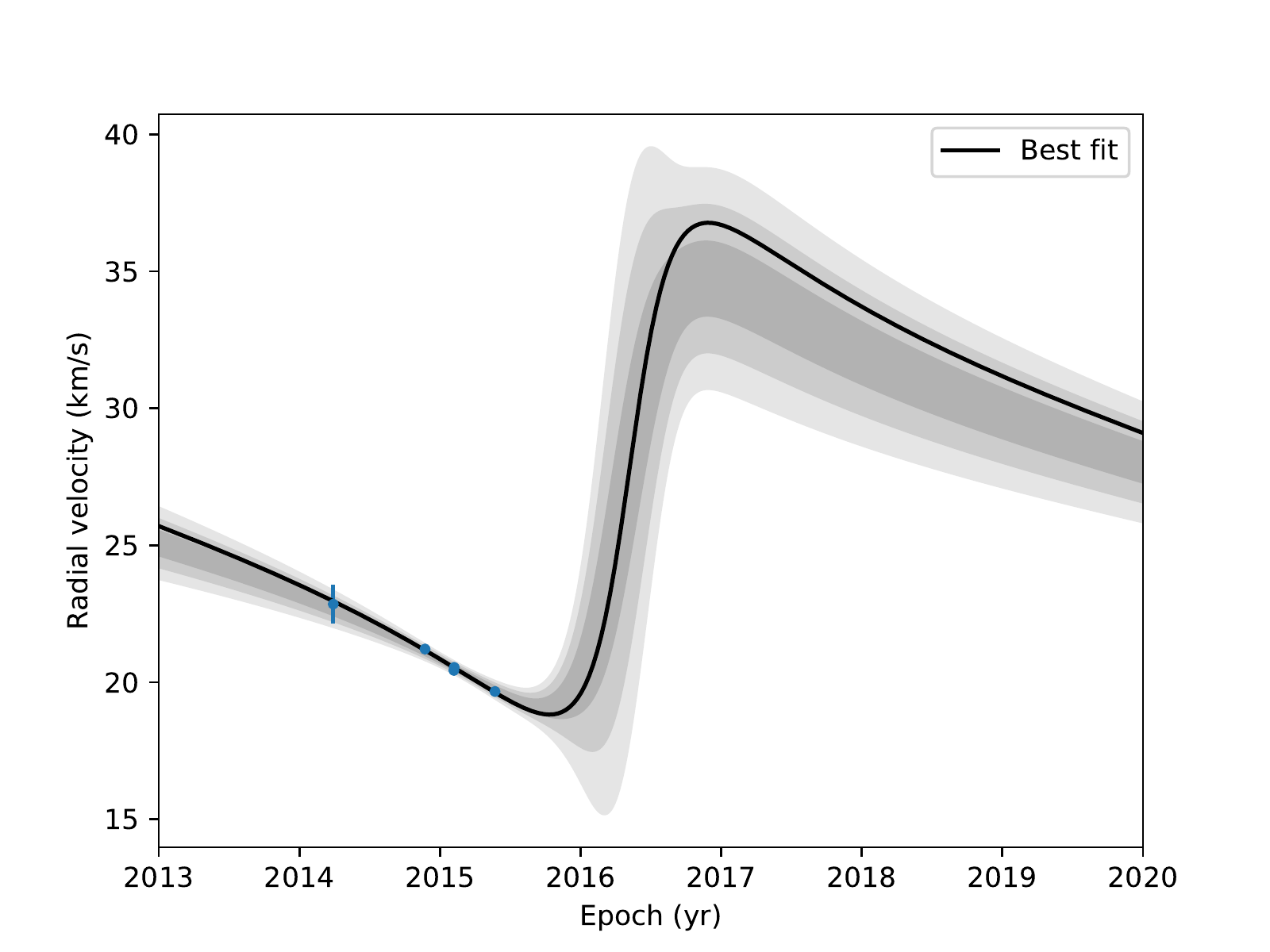}
	\includegraphics[width=0.4\textwidth]{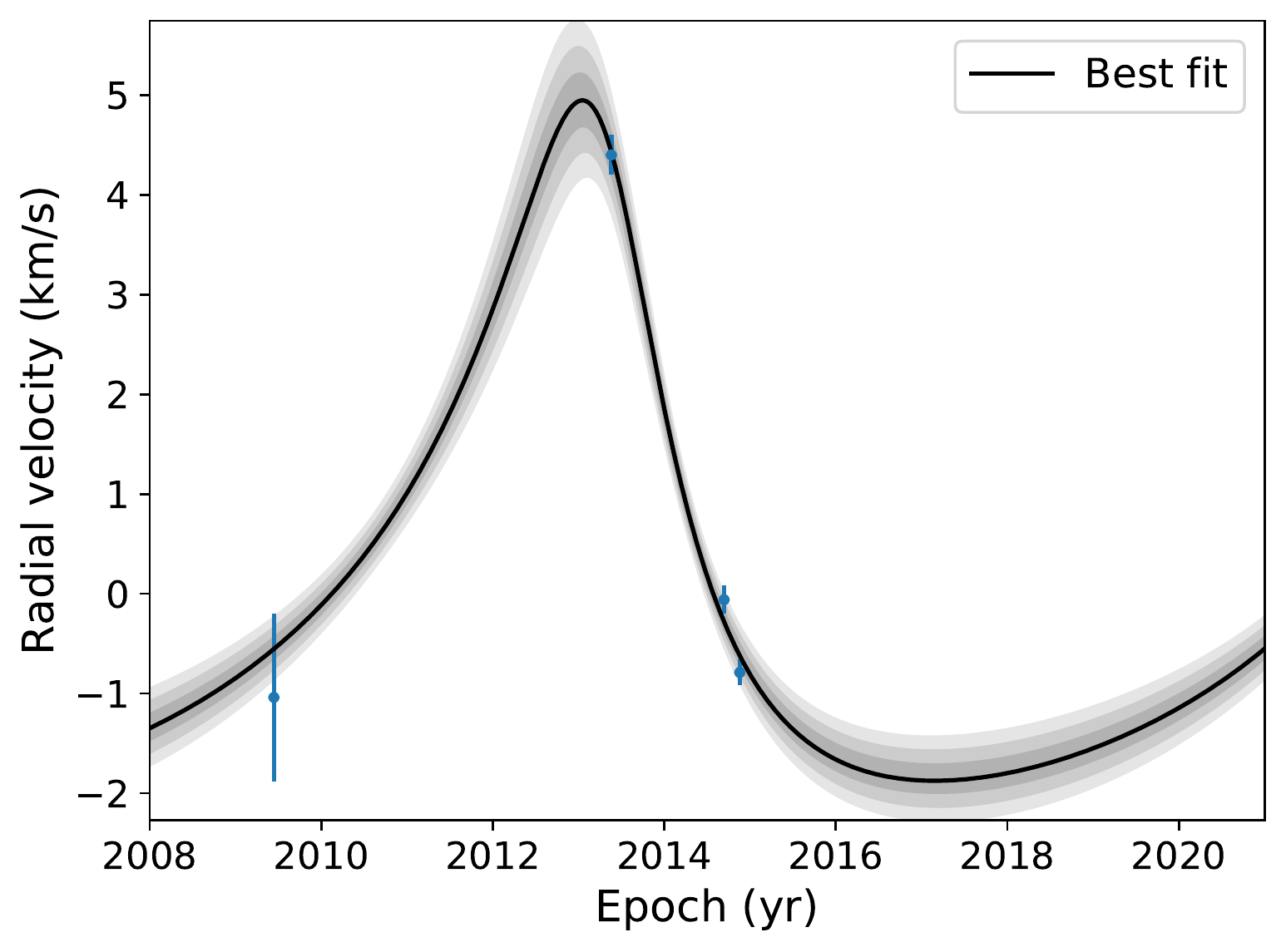}
	\caption{Radial velocity fits from the data gathered in Table~\ref{tab:RV} used to constrain dynamical masses for the binaries.
	The binaries are listed from left to right in the figure as J0437 and J0459 on the first top row; J0532 and J0613 on the second row from the top; J0728 and J0916 on the third row from the top; and J2317 on the bottom row. The shades of grey represent the first, second and third sigma intervals of the values predicted by the fit for each epoch, where sigma is the standard deviation.
	}	
	\label{fig:RV}
\end{figure*}

The orbital motion from the two components in the binaries make them subject to Doppler shifts which can be measured and useful for constraining the orbital motion further. We searched the literature and uncovered unresolved radial velocity (RV) observations for 16 binaries in our sample, which we included into our MCMC fitting to aid the orbital fitting procedure. However, the RV data in the literature is mostly compromised of unresolved measurements in which the two lines from the two components in the binary are blended together. As the strength of these lines are dependent on the spectral template used and fitting method for deriving the RV measurement, which differs from authors and instruments, most RV data were deemed unusable for our the orbital fitting. Hence, only the RV data for seven systems was used in the final orbital fits, listed here in Table~\ref{tab:RV} and shown in their respective fit in Figure~\ref{fig:RV}, where targets with too few RV observations or lack of baseline were omitted. 

For the seven instances where RV data were available and useful, we included two additional parameters to the MCMC code which evaluated the probability density functions  (PDFs) of the offset velocity $v_0$ and RV amplitude $K$. In the adopted formalism which assumed a Keplerian orbit, the radial velocity can be described as

\begin{equation}
v_{\rm rad} =  K \frac{\cos(\theta+\omega) + e\cos(\omega)}{\sqrt{1 - e^2}} +v_0,
\end{equation}
which amplitude for pure SB1 binaries is deduced from the mass fraction of the secondary component $m_{\rm B}/m_{\rm tot}$ as

\begin{equation}
K = \frac{2\pi}{P} \frac{m_{\rm B}}{m_{\rm tot}}a \sin i.
\end{equation}

In principle, the RV data allows for fractional mass and thereby individual masses for the binary components to be derived. However, that is when considering pure SB1 binaries, which is a questionable assumption for the relatively high flux-ratios (and mass-ratios) in our target sample. Therefore, we applied the same method as in \citet{rodet_dynamical_2018}, proposed by \citet{montet_dynamical_2015}, and assumed the sum of two flux-weighted individual RVs to be the considered RV measured. The orbital fit could then fit the RV amplitude as

\begin{eqnarray}
K &=& (1-F)K_{\rm A}-F\,K_{\rm B} \\
  &=& \frac{2\pi}{P} a\sin i\left( (1-F)\frac{m_{\rm B}}{m_{\rm tot}} - F\frac{m_{\rm A}}{m_{\rm tot}}\right)
\end{eqnarray}
with $F = L^{\rm V}_{\rm B}/(L_{\rm A}^{\rm V} + L_{\rm B}^{\rm V})$ being the fractional flux, $L_{\rm A}^{\rm V}$ and $L_{\rm B}^{\rm V}$ being the luminosities in the visible spectrum for each component, and $K_{\rm A}$ and $K_{\rm B}$ the respective RV amplitude. 

The majority of the RV data were obtained from the Fiber-fed Extended Range Optical Spectrograph (FEROS) instrument at the ESO-2.20m telescope \citep{FEROS}, with a wavelength coverage of $\lambda ~3500 - 9200\,$Å and resolving power of $R = 48,000$. We did not perform the RV observations or any reanalysis of the data here, using only the values stated from the given literature cited in Table~\ref{tab:RV}. We estimated the flux ratios from the magnitude difference in the $i^{\prime}$-band from the AstraLux observations, as the $i^{\prime}$ span the wavelength $\lambda \sim 6700 - 8400\,$Å, thereby encompassing the most similar wavelength range as the FEROS RV data. For the J2016 system we did not posses any flux-ratio in the $i^{\prime}$-band and applied the flux-ratio in the $I$-band from the FastCam/NOT observations instead, which has a reference wavelength of $\lambda_{\rm ref} \sim 8200\,$Å. The flux ratios used in our calculations are shown in Table~\ref{tab:RVflux}. The reported uncertainties of the flux-ratios are likely underestimated due to the different wavelength coverage by the photometric bands and that of FEROS.

This approach of weighing RV signals by the flux-ratio proved to have some limitations, where some estimated flux-ratios resulted in fractional-masses with higher dynamical mass for the secondary component B compared to the primary A component. We kept the results we obtained from our calculations, but highlight the caveat of the method not being fully reliable for our target sample, mainly serving as a first-order method. In order to disentangle the two lines for more precise estimates require more refined methods, for example tracing back individual RVs \citep[e.g.][]{czekala_disentangling_2017}.

\begin{table}[t]
\centering
\caption{Radial velocity data}
\begin{tabular}{lccc}
\hline \hline
Target & MJD & Instrument & RV [km$/$s]  \\ \hline

J0437 & 53707.288 & FEROS & $20.40 \pm 0.09$  \\
	  &	55203.146 & FEROS & $24.25 \pm 0.09$  \\
	  &	56912.331 & FEROS & $23.04 \pm 0.09$  \\
	  &	56979.236 & FEROS & $22.89 \pm 0.08$  \\
	  &	57059.095 & FEROS & $22.89 \pm 0.08$  \\
	  &	57290.319 & FEROS & $22.21 \pm 0.08$  \\
	  &	57291.257 & FEROS & $22.46 \pm 0.10$  \\[.3em]
	  
J0459 & 55942.000 & MIKE  &	$43.33 \pm 0.21$  \\
	  & 56912.344 & FEROS & $43.17 \pm 0.21$  \\
	  & 56980.125 & FEROS & $43.03 \pm 0.17$  \\
	  & 57060.127 & FEROS & $43.00 \pm 0.19$  \\
	  & 57291.272 & FEROS & $42.80 \pm 0.19$  \\[.3em]
	  
J0532 & 55526.282 & FEROS & $24.26 \pm 0.13$  \\	
	  & 55615.041 & FEROS &	$24.24 \pm 0.12$  \\
	  & 56164.407 & FEROS & $24.80 \pm 0.14$  \\
	  & 56645.000 & DuPont& $25.58 \pm 0.65$  \\
	  & 56980.258 & FEROS & $24.82 \pm 0.14$  \\
	  & 57059.134 & FEROS & $25.23 \pm 0.13$  \\[.3em]

J0613 & 55522.312 & FEROS &	$21.28 \pm 0.21	$  \\
	  & 56168.403 & FEROS & $22.07 \pm 0.25	$  \\
	  & 56402.000 & CRIRES &$22.90 \pm 0.20 $  \\
	  & 56700.142 & FEROS & $22.90 \pm 0.19	$  \\
	  & 56980.335 & FEROS & $22.91 \pm 0.23	$  \\
	  & 57058.209 & FEROS & $23.11 \pm 0.23 $  \\[.3em]

J0728 & 53421.159 & FEROS &	$29.93 \pm 0.10$  \\
	  & 53423.153 & FEROS &	$30.09 \pm 0.10$  \\
	  & 54168.043 & FEROS &	$28.31 \pm 0.10$  \\
	  & 55526.355 & FEROS &	$27.74 \pm 0.11$  \\
	  & 56173.407 & FEROS &	$28.08 \pm 0.10$  \\
	  & 56980.349 & FEROS &	$28.91 \pm 0.12$  \\
	  & 57058.295 & FEROS &	$28.74 \pm 0.12$  \\
	  & 57166.001 & FEROS &	$28.90 \pm 0.13$  \\
	  & 57853.031 & FEROS &	$28.15 \pm 0.08$  \\
	  & 57855.144 & FEROS &	$28.29 \pm 0.10$  \\[.3em]

J0916 & 56746.000 & DuPont&	$22.85 \pm 0.70$  \\
	  & 56984.343 & FEROS & $21.21 \pm 0.12$  \\
	  & 57059.297 & FEROS & $20.43 \pm 0.15$  \\
	  & 57060.209 & FEROS & $20.54 \pm 0.15$  \\
	  & 57166.015 & FEROS & $19.66 \pm 0.16$  \\[.3em]

J2317 & 54995.000 & DuPont &$-1.04 \pm 0.84$  \\
	  & 56432.000 & ESPaDOnS&$4.40 \pm 0.20$  \\
	  & 56912.207 & FEROS &	$-0.06 \pm 0.14$  \\
	  & 56979.091 & FEROS &	$-0.79 \pm 0.13$  \\

\hline 
\label{tab:RV}
\end{tabular}
\\
{\small 
FEROS ($3500-9200\,$Å) data from \citet{durkan_radial_2018}; DuPpont ($3700-7000\,$Å), ESPaDOnS ($3700-10500\,$Å) and MIKE ($4900-10000\,$ Å) from \citet{schneider_acronym_2019}; CRIRES ($15306-15688\,$Å) from \citet{malo_banyan_2014}.
}
\end{table}


\begin{table}[t]
\centering
\caption{Flux ratios for radial velocities}
\begin{tabular}{lc}
\hline \hline
FEROS  & $3500-9200\,$Å\\
AstraLux $i^{\prime}$ & $6689-8389\,$Å \\
\hline \hline
Target & $i^{\prime}$-band flux ratio  \\ \hline
J0437 & $0.03 \pm 0.01$ \\	
J0459 & $0.19 \pm 0.01$ \\		
J0532 & $0.24 \pm 0.02$ \\		
J0613 & $0.20 \pm 0.01$ \\		
J0728 & $0.21 \pm 0.01$ \\		
J0916 & $0.30 \pm 0.04$ \\		
J2317 & $0.17 \pm 0.02$ \\	

\hline 
\label{tab:RVflux}
\end{tabular}
\\
\end{table}

\section{Orbital fitting} \label{sec:orbs}
For the orbital fitting procedure we fit the relative orbit of the fainter component in the binary, typically denoted as B here, with respect to the brighter A component. We assumed Keplerian orbits projected on the plane of the sky, so that in the chosen formalism the astrometric position of the companion B could be written as:

\begin{eqnarray}
x &=& \Delta {\rm Dec} = r (\cos(\omega+\theta)\cos\Omega-\sin(\omega+\theta)\cos i \sin \Omega)\\
y &=& \Delta {\rm Ra} = r (\cos(\omega+\theta)\sin \Omega + \sin(\omega+\theta)\cos i \cos \Omega)
\end{eqnarray}
with $\omega$ being the argument of the periastron, $\theta$ the true anomaly, $\Omega$ the longitude of the ascending node and $i$ the inclination. Here, $r = a(1 - e^2)/(1 + e\cos\theta)$ is the radius with $a$ being the semi-major axis and $e$ the eccentricity. The orbital fits were then performed using the observed astrometries to derive the most likely seven orbital parameters; $a$, $e$, $\omega$, $\Omega$, $i$, period $P$ and time of periastron $t_p$.
 
In order to derive and constrain orbital parameters for our target binaries we applied two complementary approaches and codes. Initially, we applied a grid-search of the seven orbital parameters, the same as described in \citet{calissendorff_discrepancy_2017} and \citep{kohler_orbits_2008,kohler_orbits_2012,kohler_orbits_2013,kohler_orbits_2016}. The procedure determined Thiele-Innes elements for points in a grid by solving a linear fit to the astrometric data utilising singular value decomposition. The grid-search was then repeated until a minimum was found and refined for a smaller grid step size. The best-fitted parameters were then determined by comparing the reduced $\chi^2$ from the resulting orbit obtained from the fitted parameters and the relative astrometry for the binary components as
$$
\chi^2_\nu = \frac{\chi^2}{2 N_{\rm obs} - 7}
$$
with
$$
\chi^2 = \sum_i \left( 
\left(\frac{s_{\rm obs,\,i} - s_{\rm mod,\,i}}{\sigma_{s,{\rm i}}}\right)^2 +
\left(\frac{{\rm PA}_{\rm obs,\,i} - {\rm PA}_{\rm mod,\,i}}{\sigma_{\rm PA,\,i}}\right)^2
\right),
$$
where 7 is the number of orbital parameters fitted (9 parameters for systems where RV measurements were applied), $s$ and PA are the separation and positional angles respectively, and $\sigma$ their uncertainty. The algorithm is based on a Levenberg-Marquardt $\chi^2$ minimisation \citep{press_numerical_1992}, and relies heavily on the starting values which may bias certain orbital parameters if given insufficient orbital coverage or poor initial conditions. We stress that a low $\chi^2_\nu$-value is not necessarily a good indicator for a good orbital fit by itself, rather that the measured astrometric data points are well fitted to the calculated orbit. As such, a high $\chi^2_\nu$ value is not inevitably a bad orbital fit, but could indicate for the uncertainty in the measured astrometric data points to be underestimated, and therefore also the uncertainty in the derived orbital parameters and resulting dynamical mass estimate. In order to address the underestimation of the uncertainty we scaled the astrometric errors for orbits in the grid by $\sqrt{\chi^2_\nu}$ and refitted the orbits while ensuring that the $\chi^2_\nu$ was equal to 1.

For the second approach for constraining the orbital parameters we employed an Monte-Carlo Markov Chain (MCMC) Bayesian analysis technique \citep{ford_quantifying_2005,ford_improving_2006}, the same as used in \citet{rodet_dynamical_2018}. From the MCMC code we obtained the PDFs for the parameters. A sample of 500,000 orbits were randomly picked following the convergence criterion of the applied Gelman-Rubin statistics in the fitting. The sample was assumed as representative of the PDFs of the orbital elements given the initial priors, which were chosen to be uniform in $p = (\ln a, \ln P, e, \cos i, \Omega+\omega, \omega-\Omega, t_p)$. In this way, any orbital solution with the couples $(\omega, \Omega)$ and $(\omega+\pi, \Omega+\pi)$ yield the same astrometries, and thus the algorithm fits $\Omega+\omega$ and $\omega-\Omega$ to avoid this degeneracy. Nevertheless, due to how the $(\Omega,\,\omega)$ pair is defined, some degeneracy remains and the MCMC occasionally found two families of solutions for some orbits, which is interpreted as a $\pm 180\,^{\circ}$ ambiguity by the routine. The actual uncertainty is more centred upon the probability peak for each family of solutions. Therefore, for systems which lack RV and are subjected to this degeneracy we cut $90^{\circ}$ around the most probable peak and computed the error around that single interval, allowing for a well-defined uncertainty when the distribution is clearly peaked. The introduction of RV breaks the degeneracy of the ($\Omega, \omega$) couple, so that unique values for these variables could be derived for the systems which had sufficient RV data. 

The observations with associated astrometric measurements gathered in this work are listed at the end of the paper in the Appendix~\ref{appendixA}, where $s$ is the separation between the binaries in mas, PA the positional angle in degrees. In the table we also listed the deviation between the orbital fit from the grid-method and the observations as $|\Delta s|/\sigma_s$ and $|\Delta {\rm PA}|/\sigma_{\rm PA}$, which were calculated as $\sqrt{({\rm obs} - {\rm fit})^2}/ \sigma_{\rm obs}$. The $\chi^2$ is then related as the sum of the squares of the deviations.

The resulting orbits from the grid-search orbital fitting procedure are shown in Figures~\ref{fig:orb_J0008} - \ref{fig:orb_J2349} together with their associated best-fit parameters, and the $68\,\%$ confidence interval around the probability peak for the MCMC. The astrometric measurements are included in the figures as black dots, with grey ellipses representing their associated uncertainty at the 1-$\sigma$ level before scaling, and blue lines connect their expected positions from the fit. Most epochs are labelled with their date of observations, but some plots feature less explicitly spelled out dates to avoid cluttering the figure. The semi-major axis $a$ and total system mass $M_{\rm s}$ are listed in both AU and solar masses, as well as in units of mas and mas$^2$/yr$^3$ in order to give the values without distance measurements and uncertainties incorporated. The $(\omega, \Omega)$ couple which have confidence intervals defined with a $\pm 180^{\circ}$ degeneracy for systems that lack RV measurements are marked with an asterisk $(^*)$. The results from the MCMC with associated probability peaks and orbits are given in Appendix~\ref{appendixB2} for the previously unpublished constraints of the J0613, J0916 and J232617 systems.

We noted that the minimised $\chi_\nu^2$-value was not sufficient alone to objectively disclose the robustness of a given orbital fit. Instead, we adopted a custom-made grading system loosely based on the orbital fit grading criteria from \citet{worley_fourth_1983} and \citet{hartkopf_2001_2001} in order to quantitatively assess how well-constrained each orbital fit was. A direct comparison to the grading criteria from \citet{worley_fourth_1983} could not be made, as for example it does not account for the additional RV data we possessed for some of our systems, nor did we weight the observations or literature epochs when estimating our grades. For each orbit we calculated a value from a linear combination of the largest gap for the positional angle and phase\footnote{The phase coverage is calculated from the time of periastron and the period of the orbit, and used to better represent orbits with high inclination where the positional angle does not change much.} coverage for the observed epochs of the orbit, divided by the number of orbital revolutions and observed epochs. The values for all systems were sequentially scaled between 1 and 5 from lowest (best, J0008) to highest (worst, J0611) value, effectively creating five equally sized bins. The details of the calculations are outlined in Appendix~\ref{appendixC}, with final grades presented in Table~\ref{tab:grades}. Because the grid-search method included more astrometric data points and to be consistent with the method used for the main results,  we employed this strategy for the orbits obtained from the grid-search approach only, and adopted the following grading scale: Reliable (1), the orbit is well-constrained; good (2), only minor changes to the orbital elements are expected; tentative (3), no major changes are expected for the orbit; preliminary (4), substantial revisions for the orbit are likely; indetermined (5), the orbital elements are not reliable or necessarily approximately correct.


One of the important factors allowing for constrained dynamical masses in our survey is due to to the updated parallaxes from the Gaia mission, as the distance to the system is typically the main contribution to the uncertainty. With that in mind, it is worth noting that the current Gaia data releases are not yet optimised for handling binaries as they are not photometrically resolved, and further improvements are expected to follow in future releases. The impact of the Gaia parallax measurements is more thoroughly explained for the individual systems in Appendix~\ref{appendixD}.

\section{Results and discussion}\label{sec:discussion}
We were able to derive individual dynamical masses for the binary components for seven systems in our target sample. Furthermore, we were able to procure luminosities from the resolved observations and age estimates from their adopted YMG membership for the respective systems, which  together with their dynamical masses we compared to pre- main sequence (PMS) evolutionary models, probing their accuracy.

\begin{figure*}
	\sidecaption
	\includegraphics[width=12cm]{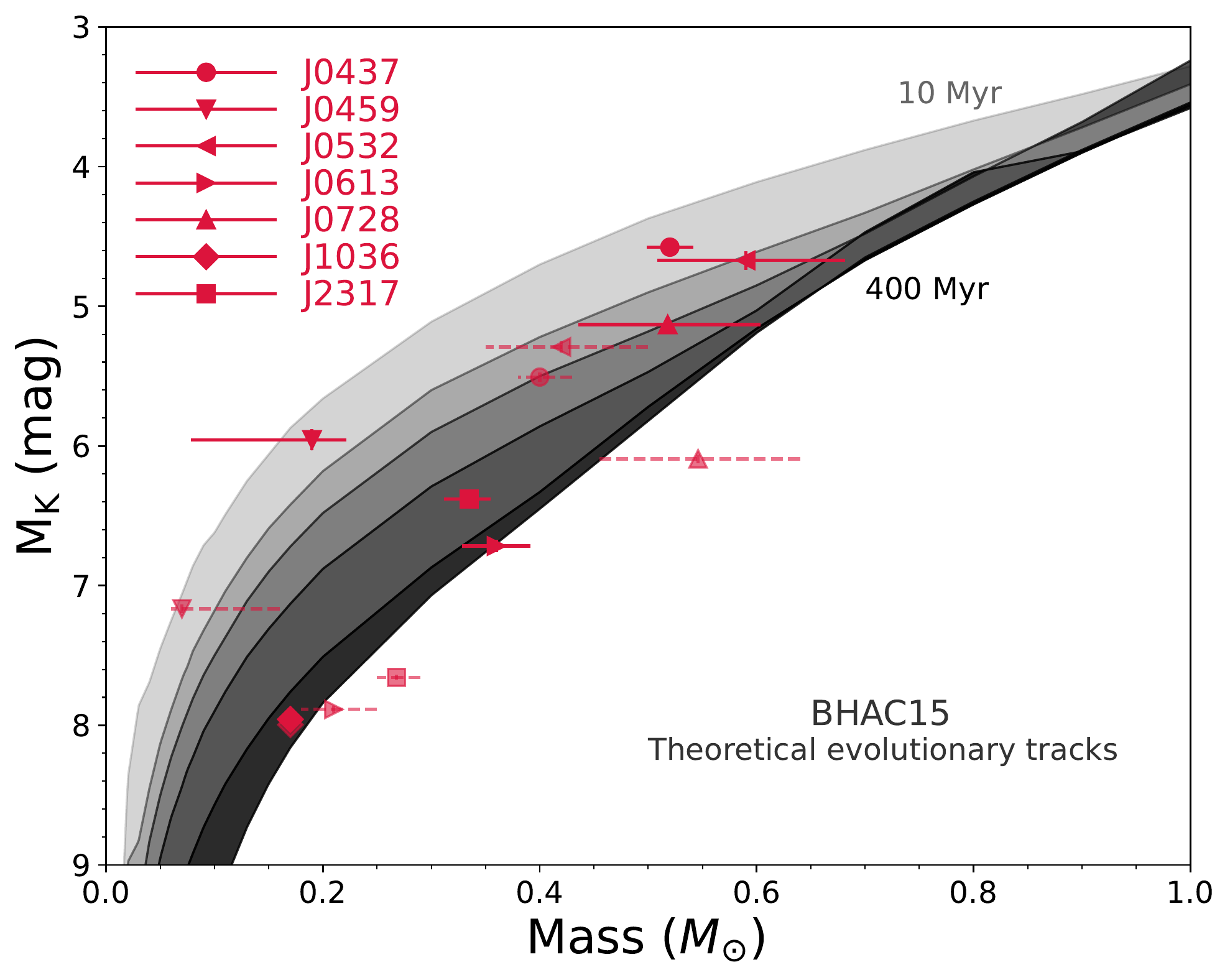}
	\caption{Mass-magnitude diagram of the individual components for the targets we were able to resolve and derive dynamical masses for. The filled areas display isochrones with ages ranging from 10, 20, 30, 50, 120 to 400 Myrs from the BHAC15 models. Each binary pair is represented by a different set of symbols, with the secondary component being slightly faded out and having dashed lines for the uncertainty. For J1036 the masses and brightnesses are equal and the two components are overplotted on top of eachother. Magnitudes here are shown as absolute magnitudes, which are also listed in Table~\ref{tab:mass} together the distances and corresponding theoretical masses for the approximate age-range of the associated YMG.}
	\label{fig:isochrones}
\end{figure*}

Although many evolutionary models exist in the literature, here we adopted the evolutionary models from \citet[hereafter BHAC15]{baraffe_new_2015}, as they are well-suited for lower mass stars and younger ages, reflecting our target sample. The dynamical mass estimates for the binaries with individual masses are plotted against stellar isochrones from the BHAC15 models in Figure~\ref{fig:isochrones} in a mass-luminosity diagram. The individual mass data-points with distances to the respective system, corresponding absolute magnitudes, their approximate associate age, and estimated theoretical mass from the BHAC15 models are listed in Table~\ref{tab:mass}. The age-ranges listed in the Table~\ref{tab:mass} show the ages of the isochrones used to calculate the theoretical mass, and not the given age-range of the respective YMG shown in Table~\ref{tab:targets}. The absolute magnitudes were calculated from the unresolved 2MASS $K$-band magnitudes of the systems, together with the $K$-band flux-ratios from our SPHERE observations or from previous SINFONI observations \citep{calissendorff_characterising_2020}. We therefore prescribe $K^{\prime}$ name convention for the absolute magnitude in the fourth column of Table~\ref{tab:mass} in in order to highlight this difference.

\begin{table*}[b]	
\renewcommand{\arraystretch}{1.3}	
\centering
\caption{Dynamical masses for individual binary components}
\begin{tabular}{lccccc}
\hline \hline
Target & Dynamical. Mass & Distance   &  App. Mag. & Age & Theoretical Mass \\
		& $[M_\odot]$ & [pc] & $[K^{\prime}]$ & [Myrs] & $[M_\odot]$ \\ \hline
J0437A	& $0.52 \pm 0.02$ 		 & $27.77 \pm 0.37$ & $6.79 \pm 0.04$ & 20 - 30 & 0.60 - 0.69 \\
J0437B	& $0.40^{+0.03}_{-0.02}$ & $27.77 \pm 0.37$ & $7.72 \pm 0.03$ & 20 - 30 & 0.32 - 0.41 \\
J0459A	& $0.19^{+0.03}_{-0.11}$ & $44.93 \pm 1.26$ & $9.22 \pm 0.06$ & $>500$ & 0.47 - 0.49\\
J0459B	& $0.07^{+0.09}_{-0.01}$ & $44.93 \pm 1.26$ & $10.43 \pm 0.03$ & $>500$ & 0.28 - 0.29\\
J0532A	& $0.59^{+0.09}_{-0.08}$ & $36.74 \pm 0.78$ & $7.50 \pm 0.05$ & 20 - 30 & 0.56 - 0.66 \\
J0532B	& $0.42^{+0.08}_{-0.07}$ & $36.74 \pm 0.78$ & $8.12 \pm 0.05$ & 20 - 30 & 0.37 - 0.48\\
J0613A	& $0.36^{+0.29}_{-0.09}$ & $16.84 \pm 0.12$ & $7.85 \pm 0.03$ & 30 - 50  & 0.17 - 0.23 \\
J0613B	& $0.21^{+0.12}_{-0.04}$ & $16.84 \pm 0.12$ & $9.02 \pm 0.03$ & 30 - 50 & 0.07 - 0.10 \\
J0728A	& $0.52 \pm 0.08$ & $15.59 \pm 0.11$ & $6.09 \pm 0.01$ & 120 - 200 & 0.60 - 0.61\\
J0728B	& $0.55 \pm 0.09$ & $15.59 \pm 0.11$ & $7.06 \pm 0.03$ & 120 - 200 & 0.43 - 0.46\\
J1036B	& $0.17 \pm 0.01$ & $20.01 \pm 0.03$ & $9.46 \pm 0.01$ & 300 - 500 & 0.19 \\
J1036C	& $0.17 \pm 0.01$ & $20.01 \pm 0.03$ & $9.50 \pm 0.01$ & 300 - 500 & 0.18 - 0.19 \\
J2317A	& $0.34 \pm 0.02$ & $16.46 \pm 0.21$ & $7.46 \pm 0.03$ & $>500$ & 0.41 \\
J2317B	& $0.27 \pm 0.02$ & $16.46 \pm 0.21$ & $8.74 \pm 0.02$ & $>500$ & 0.22 - 0.23\\
\hline
\label{tab:mass}
\end{tabular}\\
{\small
The age ranges listed were used to calculate theoretical masses in the last column predicted by the models for the given absolute magnitude in 2MASS $K$-band. The magnitudes listed in the fourth column are derived from the unresolved 2MASS $K$-band magnitudes of the system and the flux ratios in the SPHERE or SINFONI $K$-bands, and the name convention of $K^{\prime}$ is applied to mark this difference. For J0437 we used the flux ratio from the Keck/NIRC2 $K-$band observations from \citet{montet_dynamical_2015}.
}
\end{table*}

Overall we found a good consistency between the dynamical mass estimates and the theoretical mass from the models in Figure~\ref{fig:isochrones}. Most systems have dynamical masses which correspond well with the ages from their respective YMG according to the isochrone tracks. We found two outliers from the prediction of the isochrones in the mass-magnitude diagram, J0459 and J0532. The space velocities for J0459 does not suggest it to belong to any known YMG or association, albeit the binary components are placed amongst the younger isochrones at around $\approx 40$ Myrs in the mass-magnitude diagram in Figure~\ref{fig:isochrones}. Nevertheless, the orbit for the system is not constrained to such a degree that stringent dynamical masses could be procured, and it is likely that our estimate is low. A higher mass would bring the system more in line with an older age. For J0532 we found a tentative orbital fit, displaying some degeneracy in the period-separation space which caused large uncertainties on the dynamical mass. Even with the large formal error bars we found a higher dynamical mass compared to what was expected by the model isochrones for the given young age. This discrepancy is likely attributed to the uncertain method of using the flux-ratio for the RV signals to derive the mass-ratio, but could also potentially be explained by the the individual components being unresolved binaries themselves, causing the observed source to appear underluminous for its mass. Nevertheless, such unseen companions would have to be in close-in orbits of less than one AU to avoid detection from our SPHERE observations.

The RV data for the J0916 system aided in constraining the orbit, and was consistent with a Keplerian orbit. However, the flux-weighted approximation for inferring mass-fraction did not seem to apply for this particular system, where we obtained a mass-fraction $\gg0.5$ for the fainter companion. As such, we omitted the system from the mass-magnitude altogether. The flux-weighted approach also seemed dubious in other instances as well, including J0437, and to some extent J0728 where the mass-fraction suggested a slightly higher companion mass than primary, but well within the $68\,\%$ confidence interval and consistent with the results from \citet{rodet_dynamical_2018} as well as with the isochrones, confirming the inferred age-range for the system belonging to the AB Doradus moving group of 120-200 Myrs. We did not find the same mass-discrepancy as \citet{rodet_dynamical_2018}, however, our method of testing dynamical against theoretical mass only consisted of the $K-$band and the age range $120-200$ Myrs for a single theoretical model. For a younger age of 50 Myrs we obtain the same discrepancy of $\approx 15\,\%$ missing mass, with the models underpredicting the total mass of the system compared to the dynamical mass. \citet{rodet_dynamical_2018} explored the possibility for an unseen companion to explain the missing mass, which could remain hidden and stable if it were closer in than 0.1 AU from one of the other components. From our flux-weighted RV analysis of the system we obtain a slightly higher mass-fraction for the secondary component than the primary, which is also something that would be expected from a higher order hierarchical system such as a triplet. Resolved spectroscopic data could potentially discover such a companion.

No RV data were available for the J1036 binary that could aid the orbital fit. Regardless, since the system is a known triplet we could take advantage of the orbit of the outer binary pair around their common centre of mass along their path on their common orbit around the primary A component in the system. This was previously done in \citet{calissendorff_discrepancy_2017} and we adopt their results of the outer binary being of equal mass. Given the new parallax measurements for the system provided by the Gaia EDR3, the uncertainty in the distance to the system was reduced. In turn, our dynamical mass estimate for the system is the most robust in our sample. The new dynamical mass from the improved orbit and distance is also well aligned with the prediction from the theoretical model isochrones, thus also abolishing the former discrepancy reported by \citet{calissendorff_discrepancy_2017}.

The recent efforts by the Gaia mission made it possible to provide updated dynamical masses and absolute magnitudes for some of these systems which previous distance measurements remained somewhat uncertain. In principle, Gaia astrometry could also help to further constrain the orbits of binaries from exploiting the instantaneous acceleration in proper motions, for example by comparing to Hipparcos measurements \citep[e.g.][]{calissendorff_improving_2018, brandt_hipparcos-gaia_2018, brandt_precise_2019}. However, only one system in our sample, J0728, exists in both the Gaia and Hipparcos catalogues, and the baseline of 24.5 years between Hipparcos and Gaia far exceeds the orbital period of the system of $\,\approx 7.8$ years, causing some degeneracy when including proper motion acceleration into the orbital fit. Nonetheless, future data releases from Gaia may provide more advantageous information for binary systems with tentative orbital constraints that are unresolved by the space telescope itself. At the same time, one of the troubles for Gaia is to properly identify the photo-centre for unresolved binaries \citep{lindegren_gaia_2018}, which the ground-based astrometry could remedy to some extent.

Some systems exhibit a discrepancy in mass when compared to the evolutionary models. A portion of this may be attributed to undetected close-in companions, and given our sample of 20 Mdwarf binaries with the expected multiplicity frequency of $\approx 25\,\%$ and companion rate $\geq 30\,\%$ \citep{winters_solar_2019}, it is likely that some of these systems contain yet unknown companions. Furthermore, systems which contain more mass, and thereby have faster orbits, are of notable interest for orbital monitoring programmes as the orbits can be mapped out in relatively shorter period of time, and the sample could be afflicted by a selection bias due to this. Observations with for example interferometry that can achieve smaller angular resolutions could place more stringent constraints on the parameter space for a potential visual companions, while spectroscopic monitoring could rule out spectroscopic companions at very low separations.

\section{Summary and conclusions}\label{sec:summary}
We considered 20 systems of astrometric M-dwarf binaries for which we present over 75 previously unpublished astrometric data points from AstraLux Norte/CAHA, AstraLux Sur/NTT, FastCam/NOT, NaCo/VLT and SPHERE/VLT. The new astrometric data allowed us to constrain Keplerian orbits and derive dynamical masses for the binaries. We constructed our own relative scale for grading the orbital fits in an attempt to obtain a more objective assessment of the results, with improving grades based on the positional angle and phase coverage as well as revolutions made and number of epochs observed. Provided our modest sample of 20 binaries used for our grading criteria, is likely to be skewed and not directly comparable to the orbital grading system demonstrated on the $\gtrsim 900$ binaries by \citet{worley_fourth_1983}. The grades provided some quantitative measurement of the orbital fit as a whole, but did not always reflect the uncertainties of the individual orbital parameters or dynamical mass estimate. For example, the orbit for J0459 was labelled as grade 4 (preliminary) but showed low errors for the dynamical mass. Thus, the grades can be interpreted as how well the orbital fit can be trusted, and whether we expect small or large improvements to be made for the orbit, not how stringent the uncertainties are. We summarised the information on the orbital period, semi-major axis, total dynamical and theoretical masses from both orbital fitting methods, along with grades for each orbit in Appendix~\ref{appendixE}.

This was the first time orbital constraints were attempted and reported for 14 of our targeted systems. We found good orbital constraints for three systems that had not previously been published, J0613, J0916 and J232617. The PDFs for the orbital parameters obtained from the MCMC fitting procedure for three of these systems together with their associated orbit predictions and mass-distributions are presented in Appendix~\ref{appendixB2}.

Six systems, J0008, J0437, J0532, J0728, J1036 and J2317 all had previous orbital parameters, and our additional data mainly confirmed the earlier results, with a slight improvement for J0437 and J0728 with updated Gaia parallaxes. J1036 and J2317 had previously more uncertain dynamical mass estimates due to the lack of good distance measurements, which we redressed here in this work. The improved dynamical mass estimates for J1036 and J2317 now stand among the most robust in our sample. The remaining 11 systems still require further monitoring to provide reliable dynamical masses, and our new data pave the way for future orbital constraints for these systems.

Out of the 20 binaries in our sample, the four systems J0008, J0728, J1036 and J2317 received reliable orbital constraints. We determined tentative orbits for six systems; J0111, J0245, J0907, J0916, J1014 and J232611, which are expected to yield more robust orbital parameter constraints in the near future if additional astrometric measurements can be procured. Six systems, J0459, J0532, J2016, J2137, J232611 and J2349, only had their orbits constrained to a preliminary level, which may help to indicate an approximate period to some degree, while most of the orbital parameters are yet too uncertain to place robust dynamical masses. For the J0225 and J0611 systems the orbits were completely undetermined.

We searched the literature for RV data for the target sample presented here, finding RV measurements that were consistent with Keplerian orbits for seven of our binaries. Since the binaries were not of pure SB1 single lined spectroscopic binaries we assumed a flux-weighted method to infer individual dynamical masses. The method proved unreliable for the J0916 binary, and dubious at most for J0437. We therefore argue that only their total dynamical masses are reliable, and that a different approach is necessary to disentangle the individual lines and masses.

When we compared the derived individual dynamical masses with theoretical masses from the BHAC15 model isochrones we found an overall good consistency between the empirical and theoretical masses. The largest discrepancy was found for J0459, for which the isochrones predict ages between 10-50 Myrs, but Bayesian membership probabilities suggest the system more likely to be belonging to the field and no known young association or group, thus expected to be much older. We could explain this discrepancy from the degeneracy in the orbital fit, and it is possible that the period is overestimated while the semi-major axis underestimated. The near-IR spectral type for the primary being an earlier M$1.7\pm0.6$ does also suggest the system to be older and more massive than what our derived dynamical-mass advocates. However, spectral analysis by \citet{calissendorff_characterising_2020} shows a discrepancy between the near-IR bands, where the $J-$band suggest a lower surface-gravity, and thereby younger age, compared to the other bands. Our grid-search method suggested a greater mass and confidence interval than the MCMC orbital fit for this system, and we are likely to see further improvements to the orbit with just a single more astrometric data point in the future, which may also aid to constrain the age of the system through isochronal dating.

Out of our sample of 20 low-mass binaries, eight have strong indicators of being young and members of YMGs and associations (excluding J1036 which has uncertain Ursa-Majoris affiliation and it is questionable whether the age estimate of $\sim 400$ Myrs can be considered young in this context). These systems will continue to prove to be important calibrators for evolutionary models, and as orbital monitoring continues, better estimates for important formation diagnostics such as semi-major axis and eccentricity distributions will be acquired. With the increasing sample size of young binaries with stringent orbital parameter constraints, we will soon achieve a set of empirical isochrones, which can then be utilised to evaluate more precise ages of nearby young moving groups.


\begin{acknowledgements} 
The authors thank the anonymous referee for the comments which helped improve the paper. This project received funding by the Swedish Royal Academy (KVA). M.J. gratefully acknowledges funding from the Knut and Alice Wallenberg foundation. S.D. gratefully acknowledges support from the Northern Ireland Department of Education and Learning. This work has made use of data from the European Space Agency (ESA) mission {\it Gaia} (\url{https://www.cosmos.esa.int/gaia}), processed by the {\it Gaia}Data Processing and Analysis Consortium (DPAC,\url{https://www.cosmos.esa.int/web/gaia/dpac/consortium}). Funding for the DPAC has been provided by national institutions, in particular the institutions participating in the {\it Gaia} Multilateral Agreement.  This work has made use of the SPHERE Data Centre, jointly operated by OSUG/IPAG (Grenoble), PYTHEAS/LAM/CeSAM (Marseille), OCA/Lagrange (Nice), Observatoire de Paris/LESIA (Paris), and Observatoire de Lyon/CRAL, 

\end{acknowledgements}

\bibliographystyle{aa} 
\bibliography{references-binary_orbits.bib}      

\begin{appendix}

\clearpage
\onecolumn
\section{Astrometric data}\label{appendixA}

\begin{center}
\begin{longtable}{lccccccccc}
\caption{Observations and astrometry.}\label{tab:astrometry}\\
\hline \hline
\multicolumn{1}{l}{Target} & \multicolumn{1}{c}{Epoch} & \multicolumn{1}{c}{$s$} & \multicolumn{1}{c}{PA} & \multicolumn{1}{c}{\underline{$|\Delta s|$}} & \multicolumn{1}{c}{\underline{$|\Delta {\rm PA}|$}} & \multicolumn{1}{c}{Instrument} & \multicolumn{1}{c}{Filter} & \multicolumn{1}{c}{Contrast} & \multicolumn{1}{c}{Reference} \\
\multicolumn{1}{l}{2MASS} & \multicolumn{1}{c}{} & \multicolumn{1}{c}{[mas]} & \multicolumn{1}{c}{[deg]} & \multicolumn{1}{c}{$\sigma_s$} & \multicolumn{1}{c}{$\sigma_{\rm PA}$} & \multicolumn{1}{c}{} & \multicolumn{1}{c}{} & \multicolumn{1}{c}{[$\Delta$mag]} & \multicolumn{1}{c}{} \\ \hline 
\endfirsthead

\multicolumn{10}{c}%
{{\tablename\ \thetable{}. continued.}} \\
\hline \hline \multicolumn{1}{l}{Target} & \multicolumn{1}{c}{Epoch} & \multicolumn{1}{c}{$s$} & \multicolumn{1}{c}{PA} & \multicolumn{1}{c}{\underline{$|\Delta s|$}} & \multicolumn{1}{c}{\underline{$|\Delta {\rm PA}|$}} & \multicolumn{1}{c}{Instrument} & \multicolumn{1}{c}{Filter} & \multicolumn{1}{c}{Contrast} & \multicolumn{1}{c}{Reference} \\
\multicolumn{1}{l}{2MASS} & \multicolumn{1}{c}{} & \multicolumn{1}{c}{[mas]} & \multicolumn{1}{c}{[deg]} & \multicolumn{1}{c}{$\sigma_s$} & \multicolumn{1}{c}{$\sigma_{\rm PA}$} & \multicolumn{1}{c}{} & \multicolumn{1}{c}{} & \multicolumn{1}{c}{[$\Delta$mag]} & \multicolumn{1}{c}{} \\ \hline 
\endhead

\multicolumn{1}{l}{{...}} & \multicolumn{1}{c}{{...}} & \multicolumn{1}{c}{{...}} & \multicolumn{1}{c}{{...}} & \multicolumn{1}{c}{{...}} & \multicolumn{1}{c}{{...}} & \multicolumn{1}{c}{{...}} & \multicolumn{1}{c}{{...}} & \multicolumn{1}{c}{{...}} & \multicolumn{1}{c}{{...}} \\ \hline
\endfoot

\hline \hline
\endlastfoot

J00085391+2050252  & 2001.60 & $111 \pm 5$ & $169.9 \pm 0.5$ & 0.58 & 0.62 & CFHT & $H$  & 0.46 & B04 \\
	   & 2003.94 & $138 \pm 4$ & $37.0 \pm 1.1$  & 0.84 & 0.93 & NaCo       & NB$_{1.64}$ & $0.28 \pm 0.04$ & This work \\
	   & 2007.56 & $110 \pm 1$ & $167.1 \pm 2.5$ & 1.40 & 0.23 & NaCo       & $H$      &  $0.81 \pm 0.08$& This work \\
	   & 2012.02 & $133 \pm 5$ & $271.9 \pm 1.7$ & 0.80 & 0.11 & AstraLux  & $z^{\prime}$ & $1.20 \pm 0.07$ & J14a \\
	   & 2012.02 &			   &				 &		&	   & AstraLux  & $i^{\prime}$ & $1.59\pm0.10$ & J14a\\
	   & 2014.61 & $147 \pm 2$ & $97.5 \pm 1.0$  & 0.24 & 5.03 & AstraLux  & $z^{\prime}$ &  & J14b \\
	   & 2016.63 & $130 \pm 3$ & $351.3 \pm 1.5$ & 0.72 & 2.02 & FastCam        & $I$			& $0.19 \pm 0.09$ & This work \\
	   & 2016.87 & $122 \pm 4$ & $325.1 \pm 2.6$ & 1.59 & 3.10 & FastCam        & $I$			& $0.17 \pm 0.24$ &  This work \\
\hline
\end{longtable}
\begin{minipage}{\textwidth}
{\small
K01 = \citep{kohler_multiplicity_2001}; B04 = \citet{beuzit_new_2004}; D07 = \citet{daemgen_discovery_2007}; K07 =  \citet{kasper_novel_2007} ;  B10 = \citet{bergfors_lucky_2010}; D12 = \citet{delorme_high-resolution_2012}; J12 = \citet{janson_astralux_2012}; J14a = \citet{janson_noopsorta_2014}; J14b = \citet{janson_noopsortborbital_2014}; M15 = \citet{montet_dynamical_2015}; R18 = \citet{rodet_dynamical_2018}; T19 = \citet{tokovinin_speckle_2019}; C20 = \citet{calissendorff_characterising_2020}; T20 = \citet{tokovinin_speckle_2020}; T21 = \citet{tokovinin_speckle_2021}; T22 = \citet{tokovinin_family_2022} \\[.2em]
$^{\dagger}$ Photometry subjected to lucky imaging ghosts and therefore contrast magnitudes overestimated.\\
$^{\ddagger}$ Not included in the MCMC fitting.
}
\end{minipage}

\end{center}


\section{Orbital fits}\label{appendixB}
\clearpage
\twocolumn

\begin{figure*}
	\begin{minipage}[b]{0.49\textwidth}
	    \centering  
		\renewcommand{\arraystretch}{1.3}
		\begin{tabular}{lr@{}lr@{}l}
		\noalign{\vskip1pt\hrule\vskip1pt}
		Orbital element         & \multicolumn{2}{c}{Grid}  & \multicolumn{2}{c}{MCMC}  \\
		\noalign{\vskip1pt\hrule\vskip1pt}
$t_p$ (JD)							& 2018.91 & $\,_{-0.01}^{+0.04}$	
									& 2018.95  & $ \,_{-  0.08}^{+  0.09}$ \\
Period $P$ (years)              	& $   5.92$ & $\,_{-  0.01}^{+  0.01}$ 
									& $  5.94$ & $\,_{-  0.01}^{+  0.02}$ \\
$a$ (mas)                      		& $    143.8$ & $\,_{-  0.8}^{+  1.6}$
									& $  144.4$ & $\,_{-  2.5}^{+  2.5}$\\
$a$ (AU)                        	& $   2.60$ & $\,^{+0.05}_{-0.04}$
									& $   2.61$ & $\,_{-  0.01}^{+  0.01}$\\
$e$                            		& $  0.10$ & $\,^{+0.01}_{-0.01}$
									& $ 0.08$ & $\,_{-  0.01}^{+  0.01}$\\
$\omega$ ($^\circ$)     			& $    -119$ & $\,^{+   1}_{  -4}$
									& $ -115.7$ & $\,_{-  8.4}^{+ 7.9}\,^*$\\
$\Omega$ ($^\circ$)      		    & $   84.2$ & $\,^{+0.2}_{-1.6}$
									& $  88.0$ & $\,_{-  6.5}^{+  6.7}\,^*$\\
$i$ ($^\circ$)                 		& $  145.3$ & $\,^{+0.7}_{-1.6}$
 									& $  144.3$ & $\,_{-  1.7}^{+  1.8}$\\
$M_S$ ($\rm 10^{-5} mas^3/year^2$)  & $   8.49$ & $\,_{-  0.2}^{+  0.1}$
									& $   8.54$ & $\,_{-  0.43}^{+  0.39}$\\
$M_S$ ($M_\odot$)             	    & $   0.50$ & $\,_{-  0.02}^{+  0.02}$
									& $  0.50$ & $\,_{-  0.03}^{+  0.04}$\\

\hline
Grade								& \multicolumn{2}{c}{1} & \multicolumn{2}{c}{Reliable}\\
		\noalign{\vskip1pt\hrule\vskip1pt}
		\end{tabular}
	\end{minipage}%
	\hfill
	\begin{minipage}{0.49\textwidth}
		\centering
		\includegraphics[width=\columnwidth]{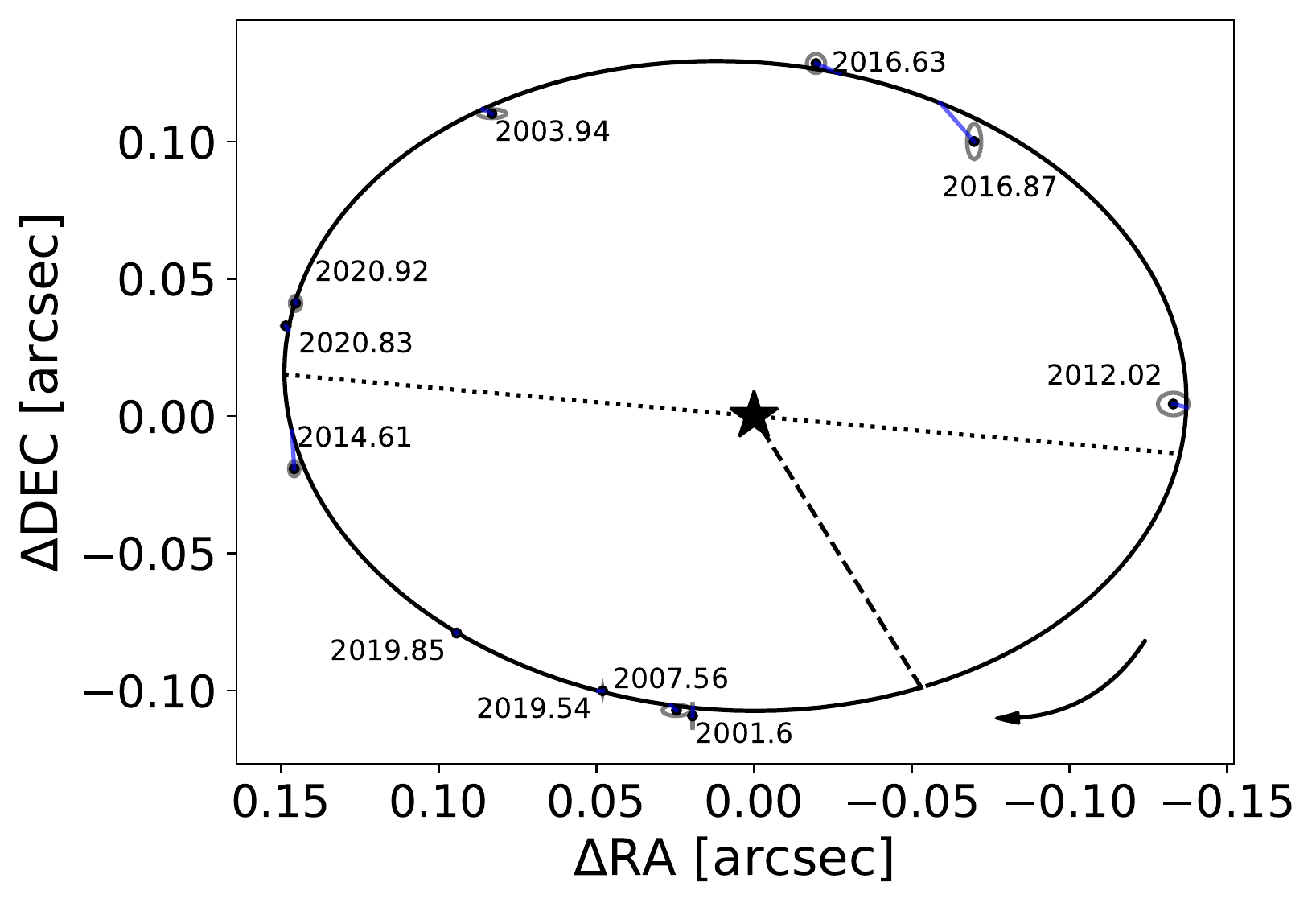}	
	\end{minipage}
\caption{Orbital parameters and fit for J0008. The right panel show the best fit from the grid approach, with black dots representing the astrometric data and the grey ellipses their uncertainty. The periastron is represented with the dashed line, with the dotted line the line of nodes. The blue lines connect the observations with their expected fitted values. The orbital solution from the MCMC are given in Appendix~\ref{appendixA}. The asterisk $(^*)$ next to $\omega$ and $\Omega$ values for the MCMC indicate the pair to be defined with a $\pm 180^{\circ}$ degeneracy.}
\label{fig:orb_J0008}
\end{figure*}

\begin{figure*}
	\begin{minipage}[b]{0.49\textwidth}
	    \centering  
		\renewcommand{\arraystretch}{1.3}
\begin{tabular}{lr@{}lr@{}l}
		\noalign{\vskip1pt\hrule\vskip1pt}
		Orbital element         & \multicolumn{2}{c}{Grid}  & \multicolumn{2}{c}{MCMC}  \\
		\noalign{\vskip1pt\hrule\vskip1pt}
$t_p$ (JD)							& 2031.75 & $\,^{+0.12}_{ -0.11}$
									&  2045.4 & $\,_{-2.3}^{+3.45}$\\
Period $P$ (years)              	& $     41$ & $\,^{+  20}_{ -10}$ 
									& $  56$ & $\,_{-15}^{+1}$\\
$a$ (mas)                      		& $   450$ & $\,^{+20}_{-10}$
									& $  437$ & $\,_{-6}^{+15}$\\
$a$ (AU)                        	& $    7.8$ & $\,^{+1.0}_{-1.0}$
									& $    7.5$ & $\,_{-1.1}^{+1.2}$\\
$e$                            		 & $    0.3$ & $\,^{+0.3}_{-0.3}$
									& $ 0.06$ & $\,_{-0.04}^{+0.15}$\\
$\omega$ ($^\circ$)     			& $    -84$ & $\,^{+  25}_{  -4}$
									& $  -85$ & $\,_{-14}^{+33}\,^*$\\
$\Omega$ ($^\circ$)      			 & $    127$ & $\,^{+   3}_{  -3}$
									& $ 127.86$ & $\,_{-5.4}^{+5.4}\,^*$\\
$i$ ($^\circ$)                 		& $     62$ & $\,^{+  10}_{  -8}$
 									& $  51.05$ & $\,_{-1.43}^{+6.41}$\\
$M_S$ ($\rm 10^{-5} mas^3/year^2$)  & $   5.53$ & $\,_{-  1.8}^{+  0.8}$
									& $   3.02$ & $\,_{-0.30}^{+1.34}$\\
$M_S$ ($M_\odot$)             	    & $   0.27$ & $\,_{-  0.10}^{+  0.11}$
									& $    0.15$ & $\,_{-0.05}^{+0.11}$\\

\hline
Grade								& \multicolumn{2}{c}{3} & \multicolumn{2}{c}{Tentative}\\
		\noalign{\vskip1pt\hrule\vskip1pt}
		\end{tabular}
	\end{minipage}%
	\hfill
	\begin{minipage}{0.49\textwidth}
		\centering
		\includegraphics[width=\columnwidth]{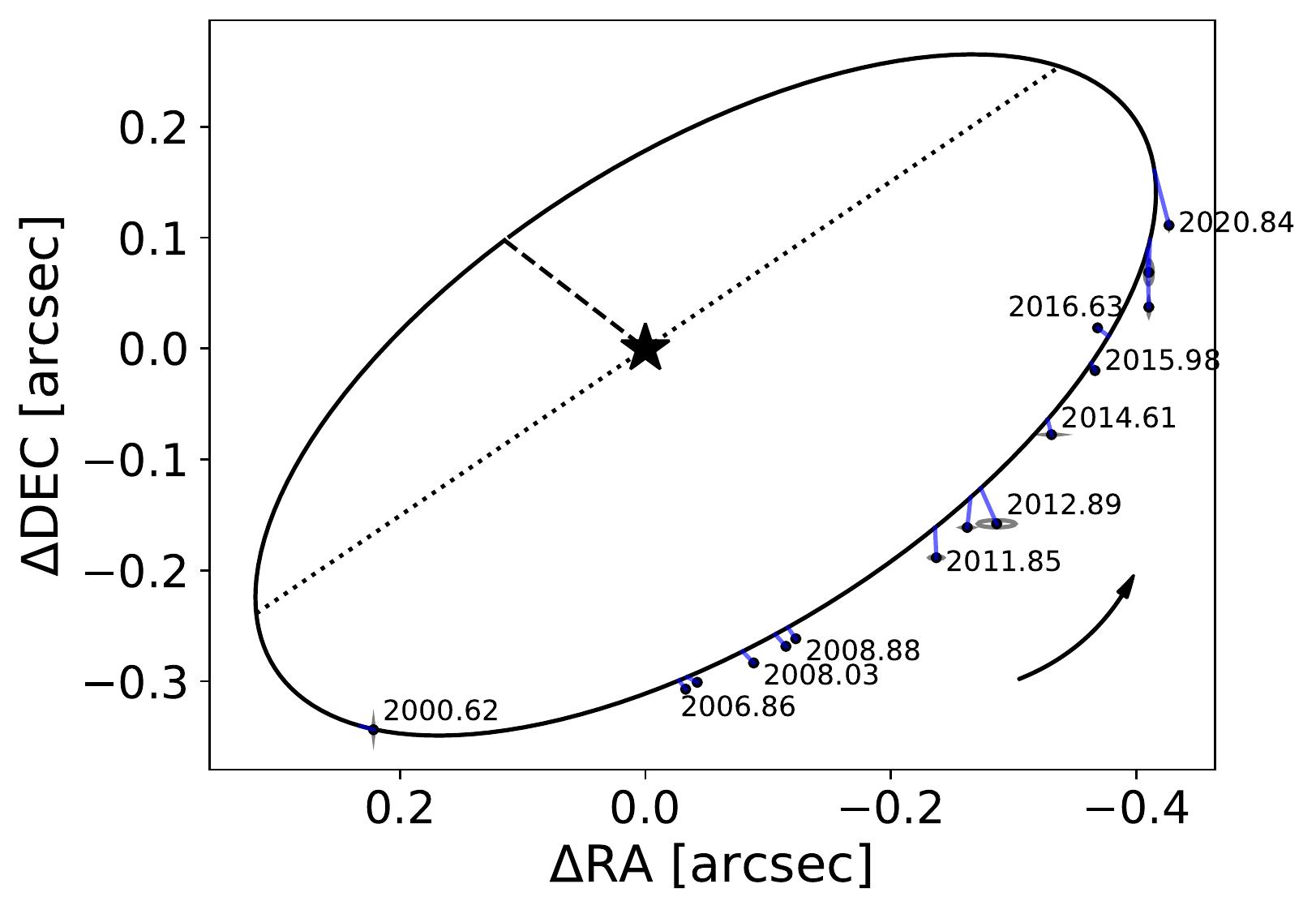}	
	\end{minipage}
\caption{Orbital parameters and fit for J0111.}
\label{fig:orb_J0111}
\end{figure*}
\begin{figure*}
	\begin{minipage}[b]{0.49\textwidth}
	    \centering  
		\renewcommand{\arraystretch}{1.3}
		\begin{tabular}{lr@{}lr@{}l}
		\noalign{\vskip1pt\hrule\vskip1pt}
		Orbital element         & \multicolumn{2}{c}{Grid}  & \multicolumn{2}{c}{MCMC}  \\
		\noalign{\vskip1pt\hrule\vskip1pt}
$t_p$ (JD)							& 2031.16 & $\,^{+0.03}_{ -0.01}$
									& 2010.83 & $\,_{-0.19}^{+0.71}$ \\
Period $P$ (years)              	& $     20$ & $\,^{+  10}_{  -3}$
									& $  49$ & $\,_{-14.5}^{+43.6}$\\
$a$ (mas)                      		& $  148$ & $\,^{+2}_{-5}$
									& $ 215$ & $\,_{-41}^{+91}$\\
$a$ (AU)                        	 & $    4.8$ & $\,^{+0.3}_{-0.3}$
									& $    6.94$ & $\,_{-1.75}^{+3.36}$\\
$e$                            		& $   0.54$ & $\,^{+0.01}_{-0.06}$
									& $  0.82$ & $\,_{-0.08}^{+0.06}$\\
$\omega$ ($^\circ$)     			& $  -123.7$ & $\,^{+2.6}_{-0.7}$
									& $  -86$ & $\,_{-19}^{+22}\,^*$\\
$\Omega$ ($^\circ$)      			 & $  132.5$ & $\,^{+8.4}_{-0.3}$
									& $ -102$ & $\,_{-22}^{+19}\,^*$\\
$i$ ($^\circ$)                 		& $     48$ & $\,^{+   5}_{  -2}$
 									& $  37$ & $\,_{-14}^{+9}$\\
$M_S$ ($\rm 10^{-5} mas^3/year^2$)  & $   0.81$ & $\,_{-0.28}^{+0.12}$
									& $   0.37$ & $\,_{-0.07}^{+0.09}$\\
$M_S$ ($M_\odot$)             	    & $   0.27$ & $\,_{-  0.06}^{+  0.05}$
									& $    0.12$ & $\,_{-0.03}^{+0.04}$\\

\hline
Grade								& \multicolumn{2}{c}{5} & \multicolumn{2}{c}{Indetermined}\\
		\noalign{\vskip1pt\hrule\vskip1pt}
		\end{tabular}
	\end{minipage}%
	\hfill
	\begin{minipage}{0.49\textwidth}
		\centering
		\includegraphics[width=\columnwidth]{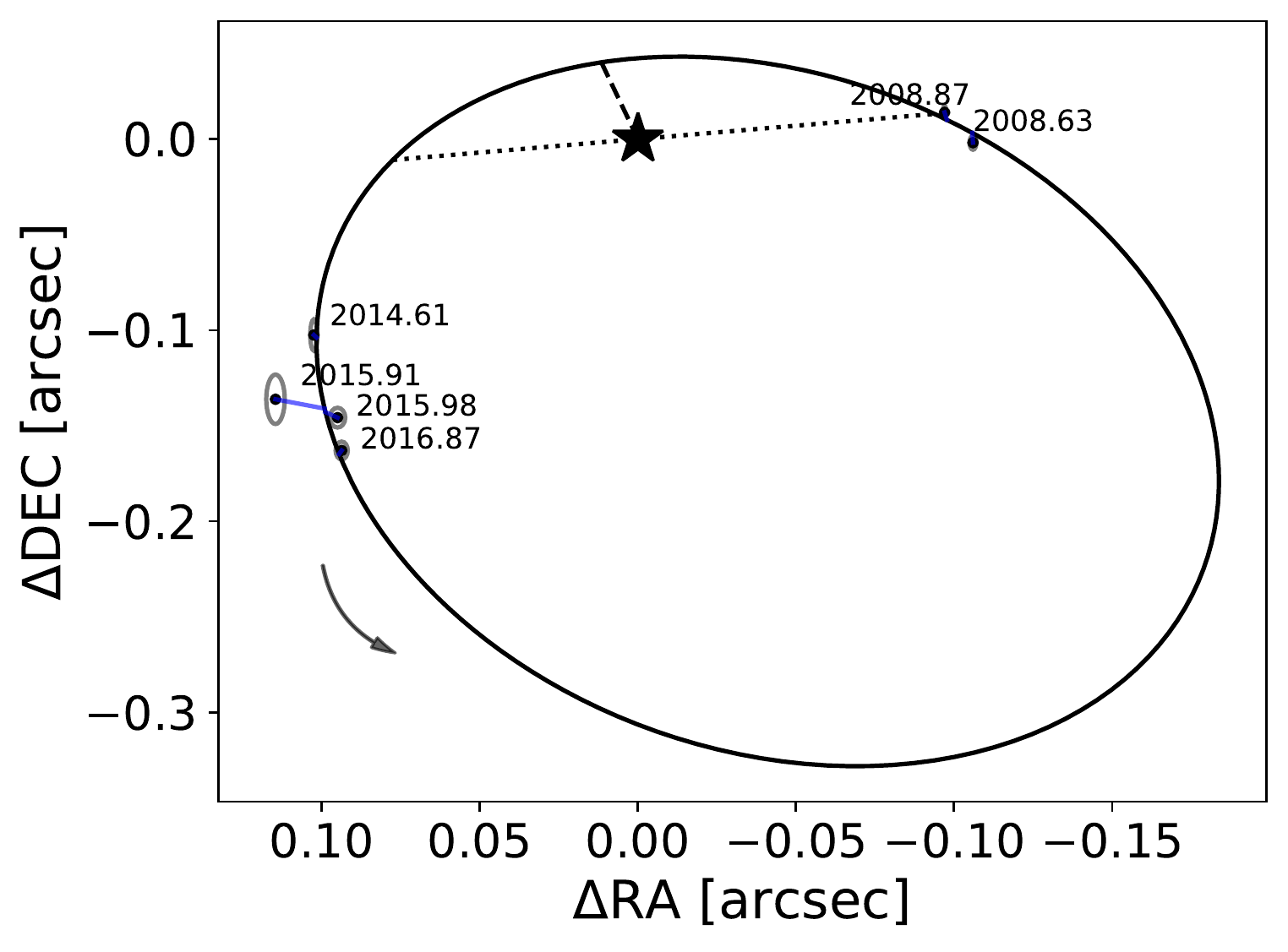}	
	\end{minipage}
\caption{Orbital parameters and fit for J0225.}
\label{fig:orb_J0225}
\end{figure*}
\begin{figure*}
	\begin{minipage}[b]{0.49\textwidth}
	    \centering  
		\renewcommand{\arraystretch}{1.3}
		\begin{tabular}{lr@{}lr@{}l}
		\noalign{\vskip1pt\hrule\vskip1pt}
		Orbital element         & \multicolumn{2}{c}{Grid}  & \multicolumn{2}{c}{MCMC}  \\
		\noalign{\vskip1pt\hrule\vskip1pt}
$t_p$ (JD)							& 2010.02 & $\,^{+ 0.01}_{  -0.01}$
									& 2011.96 & $\,_{-0.27}^{+0.19}$ \\
Period $P$ (years)              	 & $     30$ & $\,^{+   2}_{  -2}$
									& $  69$ & $\,_{-19}^{+43}$\\
$a$ (mas)                      		& $ 670$ & $\,^{+11}_{-11}$
									& $ 1242$ & $\,_{-29}^{+45}$\\
$a$ (AU)                        	& $    7.7$ & $\,^{+0.2}_{-0.2}$
									& $    14.22$ & $\,_{-0.55}^{+0.73}$\\
$e$                            		 & $  0.260$ & $\,^{+0.009}_{-0.011}$
									& $  0.64$ & $\,_{-0.12}^{+0.09}$\\
$\omega$ ($^\circ$)     			 & $    -49$ & $\,^{+   2}_{ -17}$
									& $  -18.2$ & $\,_{-6.5}^{+6.1}\,^*$\\
$\Omega$ ($^\circ$)      			& $  159.8$ & $\,^{+2.1}_{-1.0}$
									& $ 158.4$ & $\,_{-2.2}^{+5.2}\,^*$\\
$i$ ($^\circ$)                 		& $  115.5$ & $\,^{+1.0}_{-0.8}$
 									& $  108.7$ & $\,_{-1.0}^{+1.4}$\\
$M_S$ ($\rm 10^{-5} mas^3/year^2$)  & $  34$ & $\,_{-2}^{+2}$
									& $   37$ & $\,_{-2}^{+2}$\\
$M_S$ ($M_\odot$)             	    & $  0.51$ & $\,_{-0.03}^{+0.03}$
									& $    0.55$ & $\,_{-0.03}^{+0.04}$\\

\hline
Grade								& \multicolumn{2}{c}{3} & \multicolumn{2}{c}{Tentative} \\
		\noalign{\vskip1pt\hrule\vskip1pt}
		\end{tabular}
	\end{minipage}%
	\hfill
	\begin{minipage}{0.49\textwidth}
		\centering
		\includegraphics[width=\columnwidth]{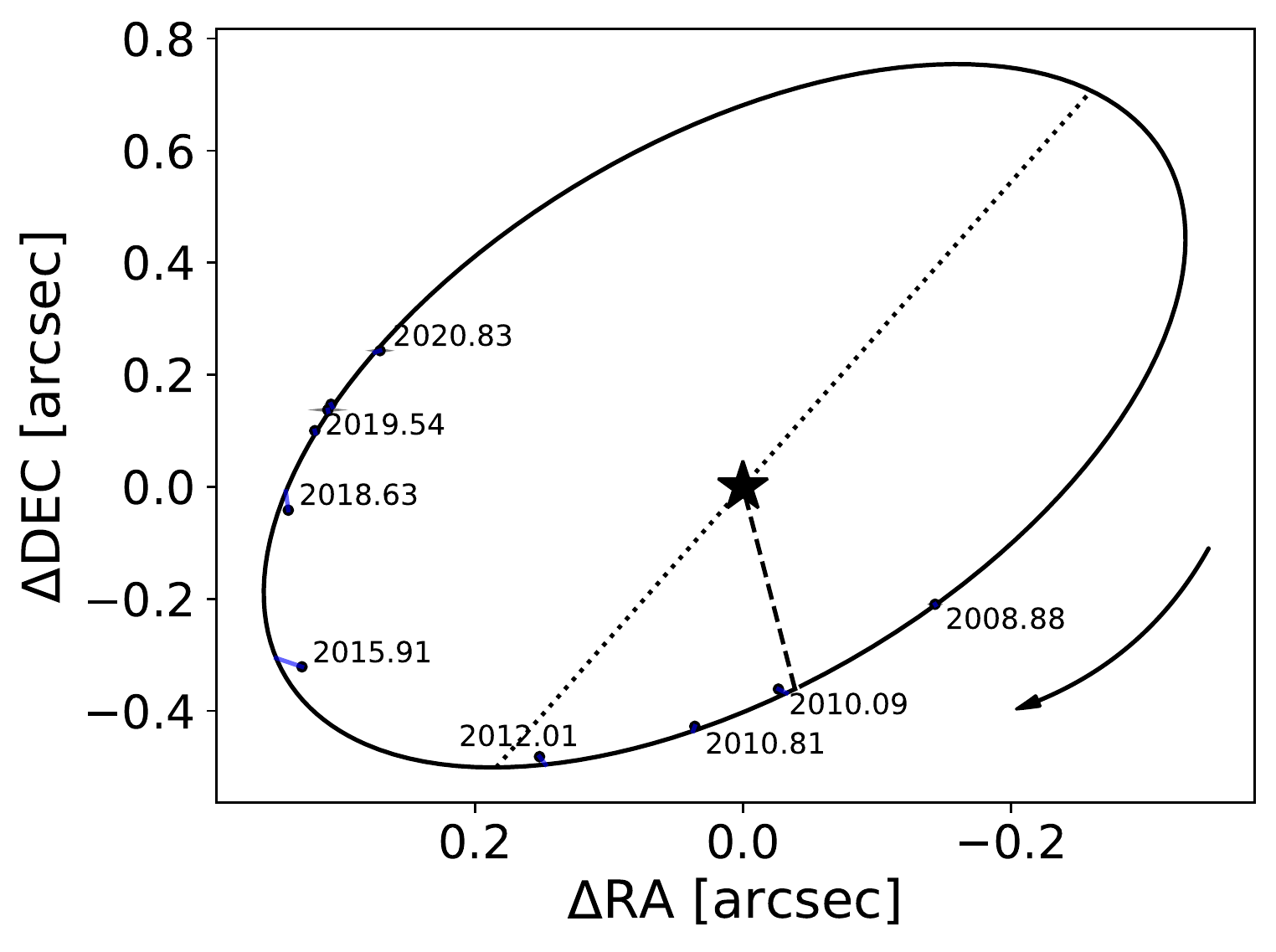}	
	\end{minipage}
\caption{Orbital parameters and fit for J0245.}
\label{fig:orb_J0245}
\end{figure*}
\begin{figure*}
	\begin{minipage}[b]{0.49\textwidth}
	    \centering  
		\renewcommand{\arraystretch}{1.3}
		\begin{tabular}{lr@{}lr@{}l}
		\noalign{\vskip1pt\hrule\vskip1pt}
		Orbital element         & \multicolumn{2}{c}{Grid}  & \multicolumn{2}{c}{MCMC}  \\
		\noalign{\vskip1pt\hrule\vskip1pt}
$t_p$ (JD)							& 2032.44 & $\,^{+ 0.01}_{  -0.54}$
									& 2036.67& $\,_{-0.14}^{+0.14}$ \\
Period $P$ (years)              	 & $     26$ & $\,^{+   6}_{  -1}$
									& $  29.39$ & $\,_{-0.44}^{+0.53}$\\
$a$ (mas)                      		 & $  308$ & $\,^{+2}_{-6}$
									& $ 335$ & $\,_{-3}^{+4}$\\
$a$ (AU)                        	& $    8.6$ & $\,^{+0.1}_{-0.2}$
									& $    9.30$ & $\,_{-0.21}^{+0.24}$\\
$e$                            		 & $   0.16$ & $\,^{+0.04}_{-0.03}$
									& $  0.19$ & $\,_{-0.01}^{+0.01}$\\
$\omega$ ($^\circ$)     			 & $ -86.90$ & $\,^{+1.68}_{-0.08}$
									& $  -66.16$ & $\,_{-2.31}^{+2.54}$\\
$\Omega$ ($^\circ$)      			& $   18.9$ & $\,^{+0.1}_{-0.3}$
									& $ 18.88$ & $\,_{-0.1}^{+0.1}$\\
$i$ ($^\circ$)                 		 & $   93.2$ & $\,^{+0.2}_{-0.2}$
 									& $  92.17$ & $\,_{-0.12}^{+0.1}$\\
$v_0$ (km/s)			      &  &  & $   20.03$ & $\,_{-0.09}^{+0.10}$\\	
$M_S$ ($\rm 10^{-5} mas^3/year^2$)  & $  4.28$ & $\,_{-0.07}^{+0.10}$
									& $   4.34$ & $\,_{-0.07}^{+0.06}$\\
$M_S$ ($M_\odot$)             	    & $  0.92$ & $\,_{-0.06}^{+0.04}$
									& $    0.93$ & $\,_{-0.04}^{+0.04}$\\
$M_{\rm{A}}$ ($M_\odot$)      &  &  & $    0.52$ & $\,_{-0.02}^{+0.02}$\\
$M_{\rm{B}}$ ($M_\odot$)      &  &  & $    0.40$ & $\,_{-0.02}^{+0.03}$\\

\hline
Grade								& \multicolumn{2}{c}{2} & \multicolumn{2}{c}{Good}\\
		\noalign{\vskip1pt\hrule\vskip1pt}
		\end{tabular}
	\end{minipage}%
	\hfill
	\begin{minipage}{0.49\textwidth}
		\centering
		\includegraphics[width=\columnwidth]{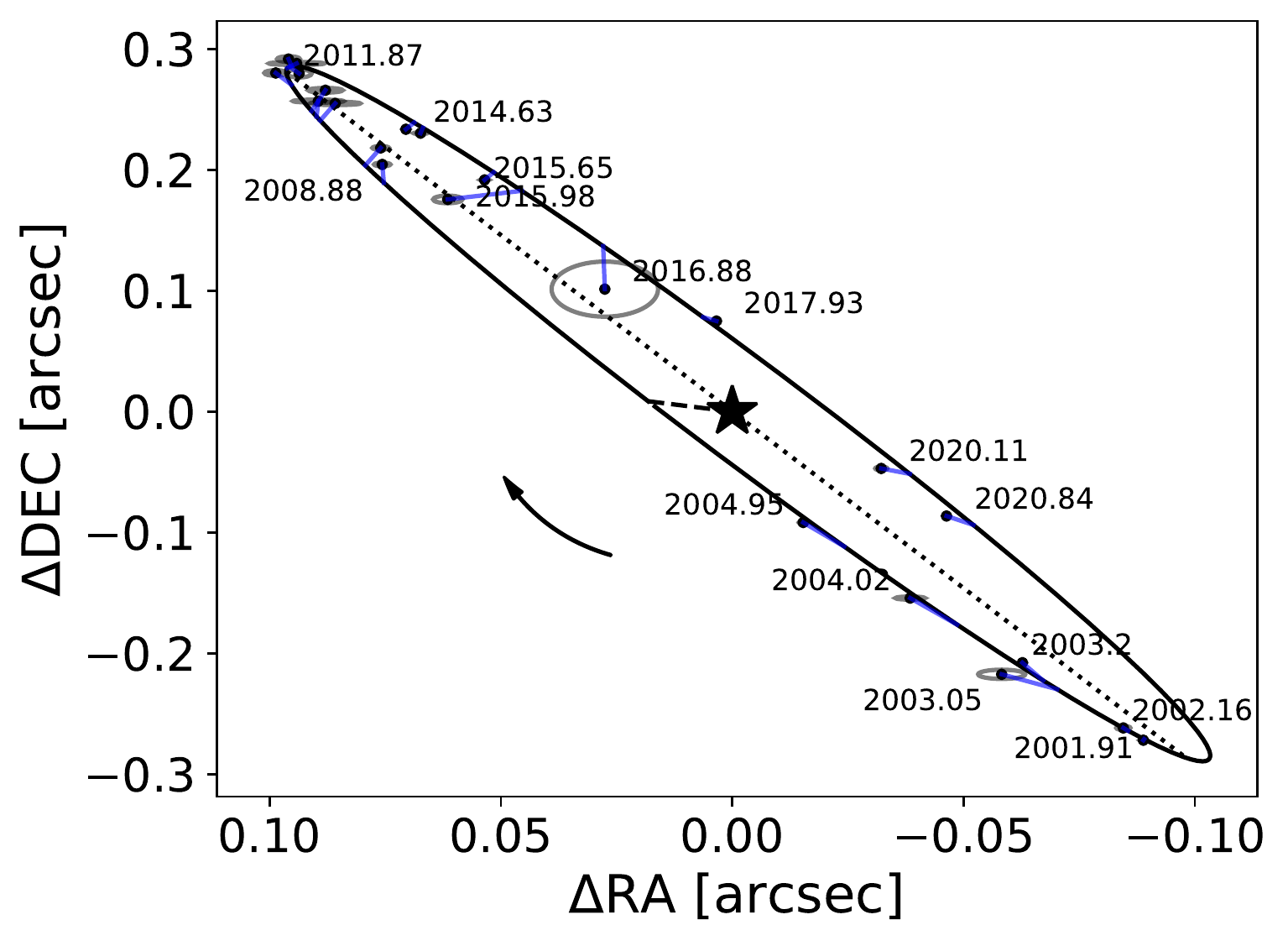}	
	\end{minipage}
\caption{Orbital parameters and fit for J0437. Not all epochs are explicitly labelled to avoid cluttering.}
\label{fig:orb_J0437}
\end{figure*}
\begin{figure*}
	\begin{minipage}[b]{0.49\textwidth}
	    \centering  
		\renewcommand{\arraystretch}{1.3}
		\begin{tabular}{lr@{}lr@{}l}
		\noalign{\vskip1pt\hrule\vskip1pt}
		Orbital element         & \multicolumn{2}{c}{Grid}  & \multicolumn{2}{c}{MCMC}  \\
		\noalign{\vskip1pt\hrule\vskip1pt}
$t_p$ (JD)							& 2026.56 & $\,^{+0.84}_{-8.64}$
									& 2032.2 & $\,_{-6.0}^{+5.7}$ \\
Period $P$ (years)              	 & $     28$ & $\,^{+   6}_{ -16}$
									& $  32$ & $\,_{-5}^{+1}$\\
$a$ (mas)                      		& $  134$ & $\,^{+9}_{-6}$
									& $ 142$ & $\,_{-12}^{+2}$\\
$a$ (AU)                        	& $    6.0$ & $\,^{+0.4}_{-0.3}$
									& $   6.38$ & $\,_{-0.72}^{+0.27}$\\
$e$                            		& $   0.10$ & $\,^{+0.08}_{-0.09}$
									& $  0.01$ & $\,_{-0.01}^{+0.09}$\\
$\omega$ ($^\circ$)     			& $    116$ & $\,^{+  65}_{ -39}$
									& $   -93$ & $\,_{-84}^{+55}$\\
$\Omega$ ($^\circ$)      			& $     31$ & $\,^{+  62}_{ -49}$
									& $ -27$& $\,_{-145}^{+22}$\\
$i$ ($^\circ$)                 		& $     17$ & $\,^{+   8}_{ -50}$
 									& $  18$ & $\,_{-9}^{+5.8}$\\
$v_0$ (km/s)			      &  &  & $   43.02$ & $\,_{-0.66}^{+0.82}$\\	
$M_S$ ($\rm 10^{-5} mas^3/year^2$)  & $  0.31$ & $\,_{-0.05}^{+1.15}$
									& $   0.29$ & $\,_{-0.02}^{+0.04}$\\
$M_S$ ($M_\odot$)             	    & $  0.28$ & $\,^{+0.05}_{-0.03}$
									& $    0.26$ & $\,_{-0.04}^{+0.06}$\\
$M_{\rm{A}}$ ($M_\odot$)      &  &  & $    0.19$ & $\,_{-0.11}^{+0.03}$\\
$M_{\rm{B}}$ ($M_\odot$)      &  &  & $    0.07$ & $\,_{-0.01}^{+0.09}$\\

\hline
Grade								& \multicolumn{2}{c}{4} & \multicolumn{2}{c}{Preliminary}\\
		\noalign{\vskip1pt\hrule\vskip1pt}
		\end{tabular}
	\end{minipage}%
	\hfill
	\begin{minipage}{0.49\textwidth}
		\centering
		\includegraphics[width=\columnwidth]{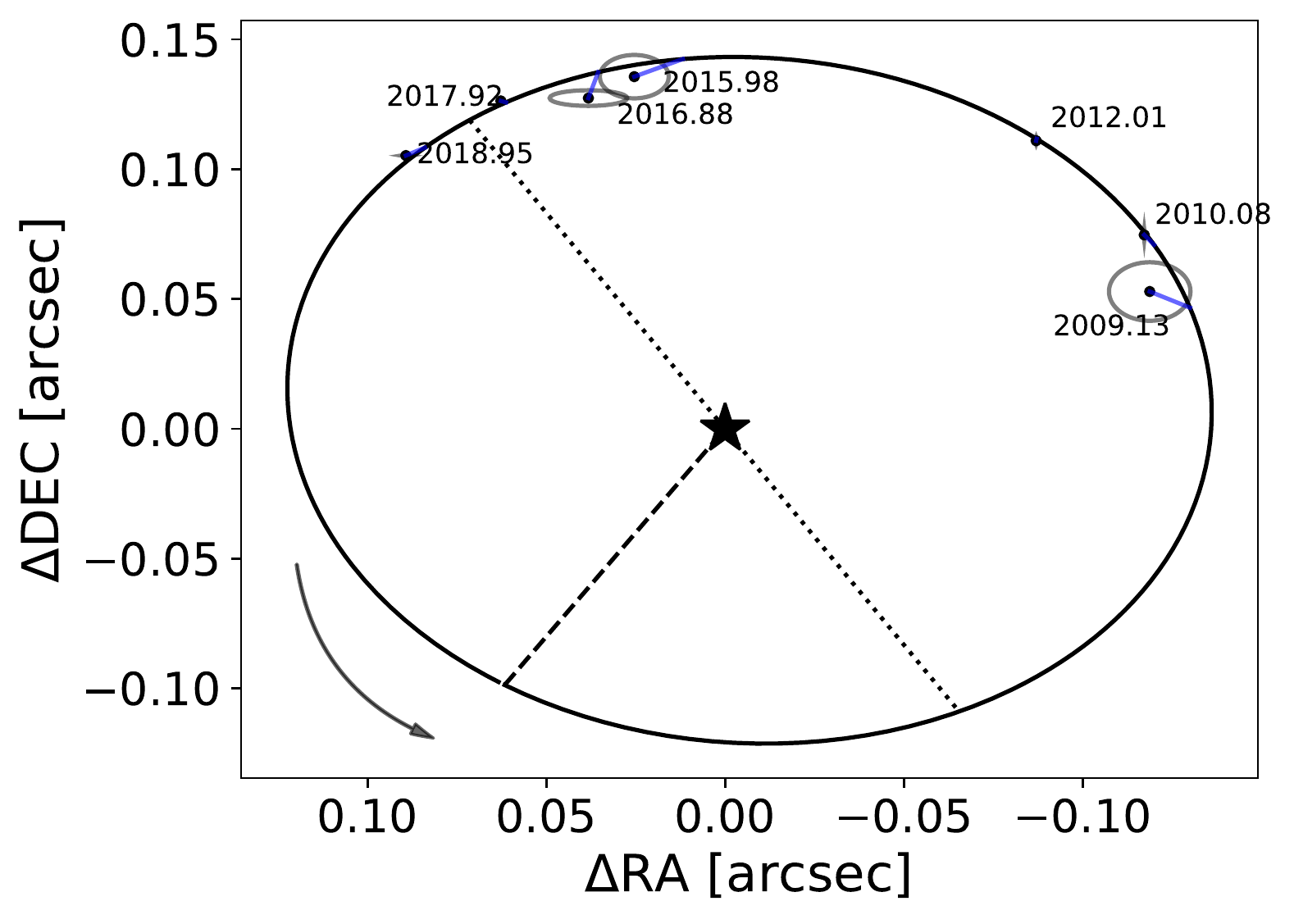}	
	\end{minipage}
\caption{Orbital parameters and fit for J0459.}
\label{fig:orb_J0459}
\end{figure*}
\begin{figure*}
	\begin{minipage}[b]{0.49\textwidth}
	    \centering  
		\renewcommand{\arraystretch}{1.3}
		\begin{tabular}{lr@{}lr@{}l}
		\noalign{\vskip1pt\hrule\vskip1pt}
		Orbital element         & \multicolumn{2}{c}{Grid}  & \multicolumn{2}{c}{MCMC}  \\
		\noalign{\vskip1pt\hrule\vskip1pt}
$t_p$ (JD)							& 2019.53& $\,^{+0.01}_{ -0.08}$
									& 2019.67 & $\,_{-0.43}^{+0.22}$ \\
Period $P$ (years)              	& $     87$ & $\,^{+  12}_{  -4}$
									& $   26$ & $\,_{-7}^{+15}$\\
$a$ (mas)                      		 & $  537$ & $\,^{+6}_{-6}$
									& $ 379$ & $\,_{-47}^{+64}$\\
$a$ (AU)                        	 & $   19.7$ & $\,^{+0.5}_{-0.5}$
									& $   13.92$ & $\,_{-2.02}^{+2.65}$\\
$e$                            		 & $  0.80$ & $\,^{+0.01}_{-0.01}$
									& $  0.81$ & $\,_{-0.03}^{+0.05}$\\
$\omega$ ($^\circ$)     			 & $   48.7$ & $\,^{+0.4}_{-0.4}$
									& $   73$ & $\,_{-10}^{+7}$\\
$\Omega$ ($^\circ$)      			 & $  133.7$ & $\,^{+0.7}_{-0.8}$
									& $ 147.4$& $\,_{-6.2}^{+5.1}$\\
$i$ ($^\circ$)                 		 & $   64.5$ & $\,^{+0.4}_{-0.4}$
 									& $  68.7$ & $\,_{-2.0}^{+3.6}$\\
$v_0$ (km/s)			      &  &  & $   23.78$ & $\,_{-0.41}^{+0.28}$\\
$M_S$ ($\rm 10^{-5} mas^3/year^2$)  & $ 2.03$ & $\,_{-0.07}^{+0.07}$
									& $   3.8$ & $\,_{-1.0}^{+0.2}$\\
$M_S$ ($M_\odot$)             	    & $  1.01$ & $\,_{-0.07}^{+0.07}$
									& $    1.87$ & $\,_{-0.49}^{+0.10}$\\
$M_{\rm{A}}$ ($M_\odot$)      &  &  & $    0.98$ & $\,_{-0.23}^{+0.18}$\\
$M_{\rm{B}}$ ($M_\odot$)      &  &  & $   0.65$ & $\,_{-0.11}^{+0.12}$\\

\hline
Grade								& \multicolumn{2}{c}{4} & \multicolumn{2}{c}{Preliminary}\\
		\noalign{\vskip1pt\hrule\vskip1pt}
		\end{tabular}
	\end{minipage}%
	\hfill
	\begin{minipage}{0.49\textwidth}
		\centering
		\includegraphics[width=\columnwidth]{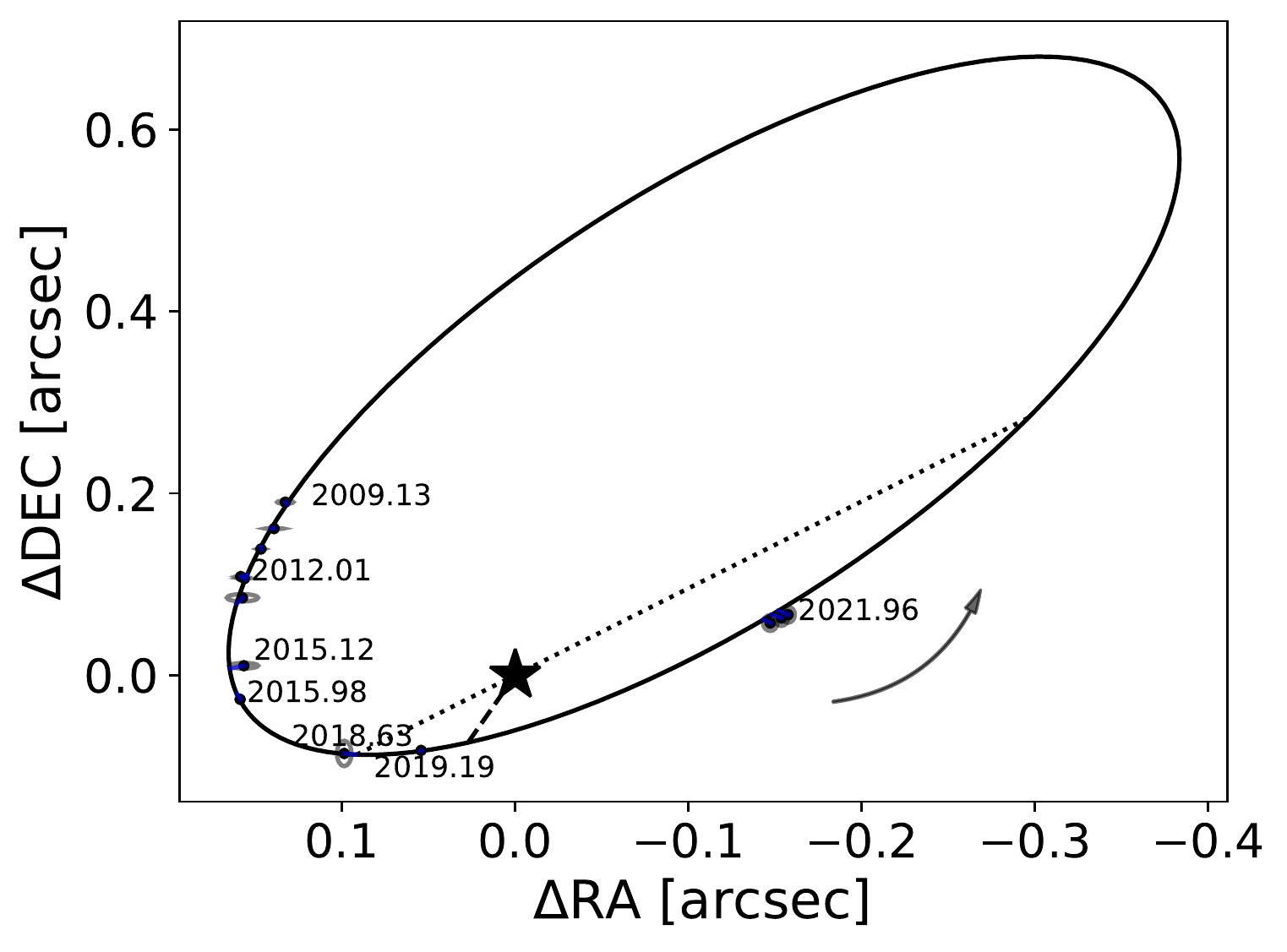}	
	\end{minipage}
\caption{Orbital parameters and fit for J0532.}
\label{fig:orb_J0532}
\end{figure*}
\begin{figure*}
	\begin{minipage}[b]{0.49\textwidth}
	    \centering  
		\renewcommand{\arraystretch}{1.3}
		\begin{tabular}{lr@{}lr@{}l}
		\noalign{\vskip1pt\hrule\vskip1pt}
		Orbital element         & \multicolumn{2}{c}{Grid}  & \multicolumn{2}{c}{MCMC}  \\
		\noalign{\vskip1pt\hrule\vskip1pt}
$t_p$ (JD)							& 2007.86 & $\,^{+ 1.46}_{ -3.40}$
									& 2006.55 & $\,_{-0.69}^{+0.95}$ \\
Period $P$ (years)              	 & $    121$ & $\,^{+3304}_{ -99}$
									& $   57$ & $\,_{-21}^{+41}$\\
$a$ (mas)                      		 & $    900$ & $\,^{+>1000}_{-700}$
									& $ 340$ & $\,_{-87}^{+155}$\\
$a$ (AU)                        	 & $     54$ & $\,^{+5635}_{ -40}$
									& $  19$ & $\,_{-5}^{+8}$\\
$e$                            		 & $    0.9$ & $\,^{+0.1}_{-0.4}$
									& $ 0.73$ & $\,_{-0.16}^{+0.17}$\\
$\omega$ ($^\circ$)     			& $    117$ & $\,^{+2226}_{ -22}$
									& $   118$ & $\,_{-13}^{+34}\,^*$\\
$\Omega$ ($^\circ$)      			& $    155$ & $\,^{+  19}_{ -11}$
									& $ -162.7$& $\,_{-8.3}^{+10.1}\,^*$\\
$i$ ($^\circ$)                 		& $    102$ & $\,^{+  51}_{ -12}$
 									& $  112$ & $\,_{-7}^{+10}$\\
$M_S$ ($\rm 10^{-5} mas^3/year^2$)  & $ 5.81$ & $\,_{-4.37}^{+>10}$
									& $   0.60$ & $\,_{-0.21}^{+0.32}$\\
$M_S$ ($M_\odot$)             	    & $  11$ & $\,_{-8}^{+11}$
									& $   1.10$ & $\,_{-0.39}^{+0.60}$\\

\hline
Grade								& \multicolumn{2}{c}{5} & \multicolumn{2}{c}{Indetermined}\\
		\noalign{\vskip1pt\hrule\vskip1pt}
		\end{tabular}
	\end{minipage}%
	\hfill
	\begin{minipage}{0.49\textwidth}
		\centering
		\includegraphics[width=\columnwidth]{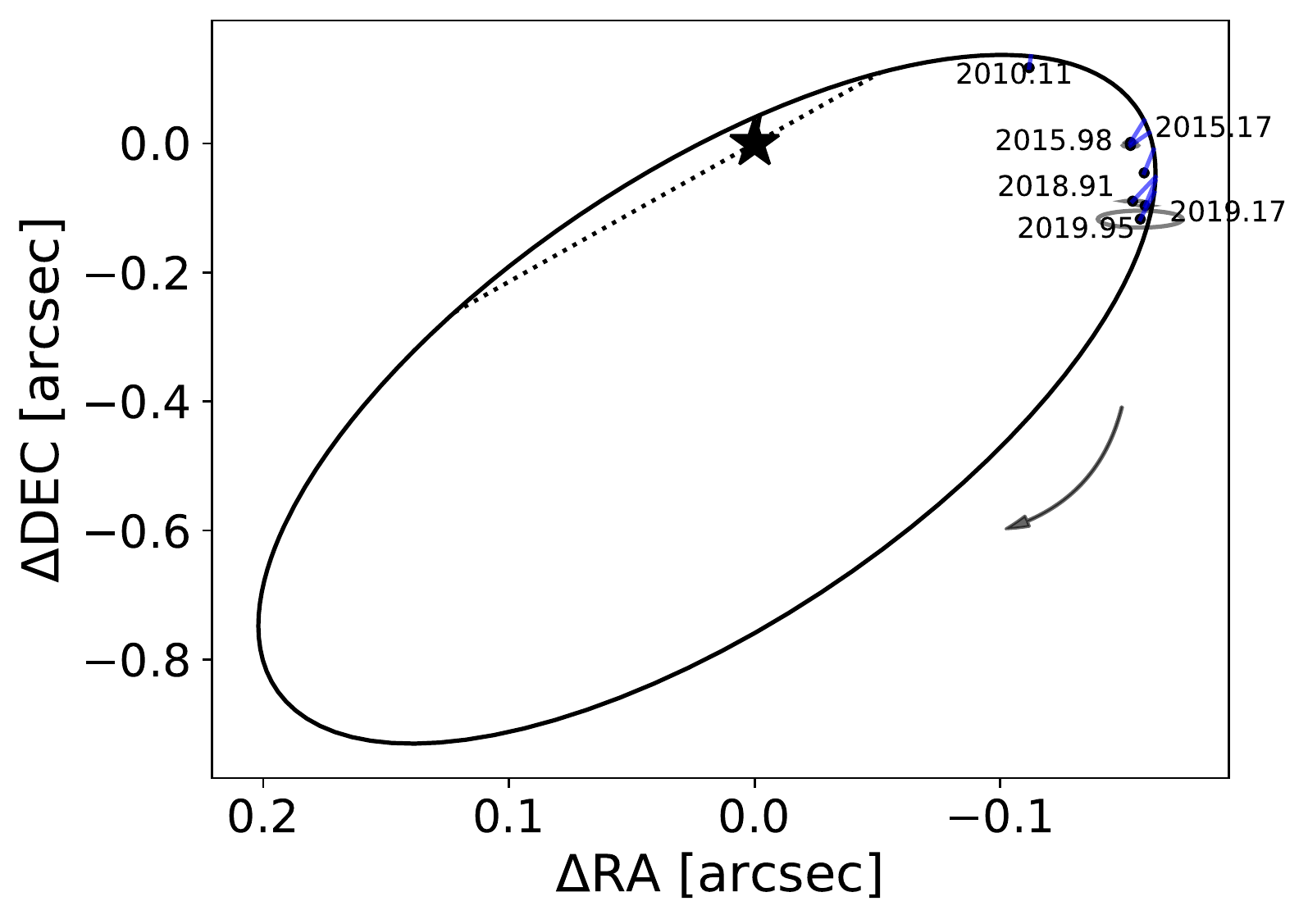}	
	\end{minipage}
\caption{Orbital parameters and fit for J0611.}
\label{fig:orb_J0611}
\end{figure*}
\begin{figure*}
	\begin{minipage}[b]{0.49\textwidth}
	    \centering  
		\renewcommand{\arraystretch}{1.3}
		\begin{tabular}{lr@{}lr@{}l}
		\noalign{\vskip1pt\hrule\vskip1pt}
		Orbital element         & \multicolumn{2}{c}{Grid}  & \multicolumn{2}{c}{MCMC}  \\
		\noalign{\vskip1pt\hrule\vskip1pt}
$t_p$ (JD)							& 2020.47 & $\,^{+0.01}_{ -0.01}$
									& 2020.34 & $\,_{-0.40}^{+0.33}$ \\
Period $P$ (years)              	& $   13.2$ & $\,^{+0.2}_{-0.4}$
									& $   11.56$ & $\,_{-0.73}^{+0.90}$\\
$a$ (mas)                      		& $  275$ & $\,^{+3}_{-2}$
									& $ 232$ & $\,_{-8}^{+51}$\\
$a$ (AU)                        	& $   4.62$ & $\,^{+0.06}_{-0.04}$
									& $   3.91$ & $\,_{-0.11}^{+0.83}$\\
$e$                            		& $  0.65$ & $\,^{+0.01}_{-0.01}$
									& $ 0.77$ & $\,_{-0.11}^{+0.10}$\\
$\omega$ ($^\circ$)     			 & $  -67.2$ & $\,^{+0.6}_{-1.1}$
									& $    -68.0$ & $\,_{-7.8}^{+9.1}$\\
$\Omega$ ($^\circ$)      			& $   25.1$ & $\,^{+0.9}_{-0.8}$
									& $ -13.2$& $\,_{-4.9}^{+6.6}$\\
$i$ ($^\circ$)                 		& $  116.7$ & $\,^{+0.6}_{-0.7}$
 									& $  122.2$ & $\,_{-6.1}^{+5.2}$\\
$v_0$ (km/s)			      &  &  & $  22.72$ & $\,_{-0.22}^{+0.15}$\\	
$M_S$ ($\rm 10^{-5} mas^3/year^2$)  & $ 11.86$ & $\,_{-0.2}^{+0.3}$
									& $   8.5$ & $\,_{-2.9}^{+7.6}$\\
$M_S$ ($M_\odot$)             	    & $  0.57$ & $\,_{-0.02}^{+0.02}$
									& $    0.42$ & $\,_{-0.15}^{+0.38}$\\
$M_{\rm{A}}$ ($M_\odot$)      &  &  & $    0.23$ & $\,_{-0.09}^{+0.29}$\\
$M_{\rm{B}}$ ($M_\odot$)      &  &  & $    0.18$ & $\,_{-0.04}^{+0.12}$\\

\hline
Grade								& \multicolumn{2}{c}{2} & \multicolumn{2}{c}{Good}\\
		\noalign{\vskip1pt\hrule\vskip1pt}
		\end{tabular}
	\end{minipage}%
	\hfill
	\begin{minipage}{0.49\textwidth}
		\centering
		\includegraphics[width=\columnwidth]{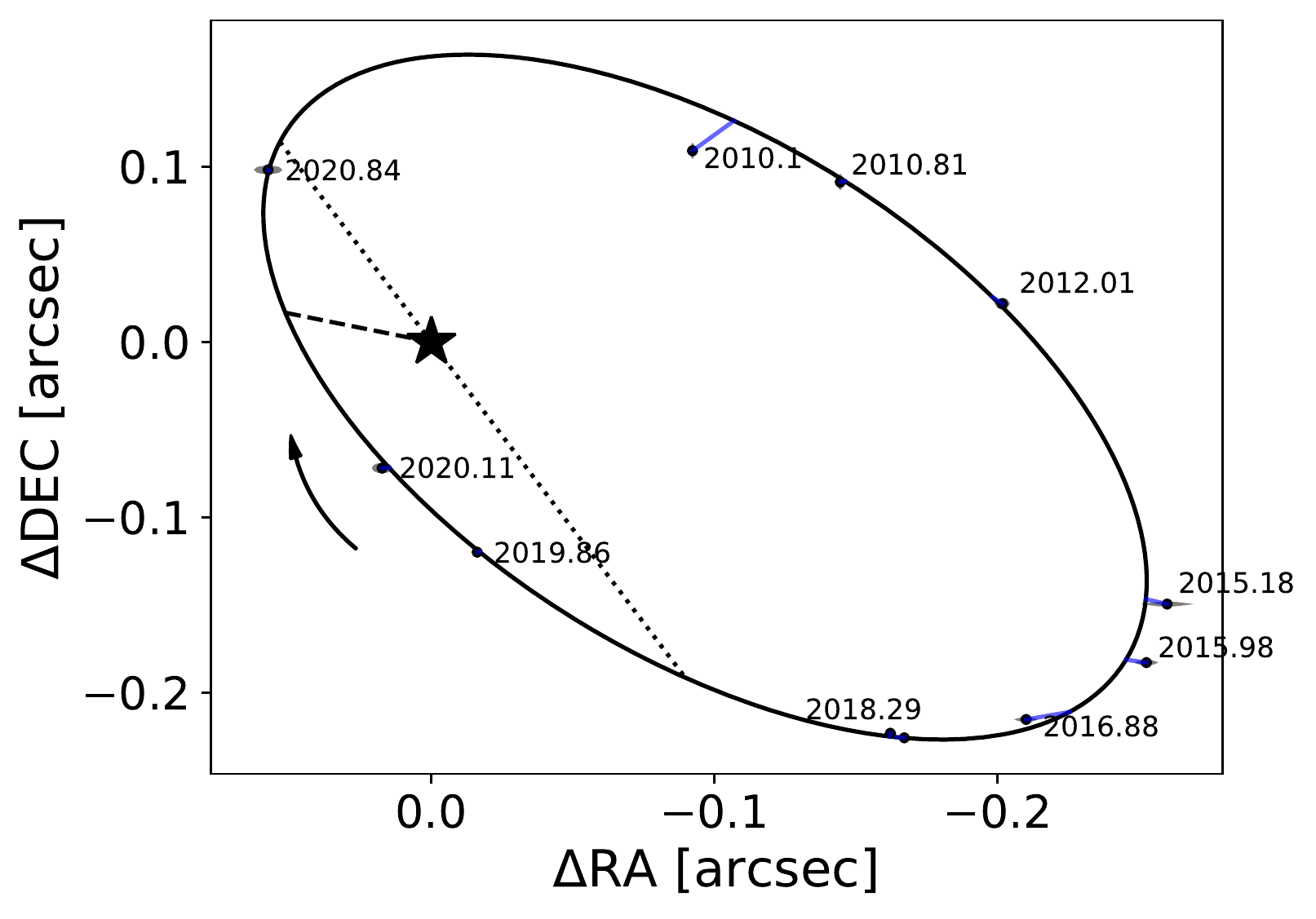}	
	\end{minipage}
\caption{Orbital parameters and fit for J0613.}
\label{fig:orb_J0613}
\end{figure*}
\begin{figure*}
	\begin{minipage}[b]{0.49\textwidth}
	    \centering  
		\renewcommand{\arraystretch}{1.3}
		\begin{tabular}{lr@{}lr@{}l}
		\noalign{\vskip1pt\hrule\vskip1pt}
		Orbital element         & \multicolumn{2}{c}{Grid}  & \multicolumn{2}{c}{MCMC}  \\
		\noalign{\vskip1pt\hrule\vskip1pt}
$t_p$ (JD)							& 2020.83 & $\,^{+0.01}_{-0.01}$
									& 2020.81 & $\,_{-0.02}^{+0.02}$ \\
Period $P$ (years)              	 & $   7.79$ & $\,^{+0.03}_{-0.03}$
									& $   7.76$ & $\,_{-0.02}^{+0.02}$\\
$a$ (mas)                      		 & $  258$ & $\,^{+2}_{-2}$
									& $ 255.9$ & $\,_{-1.3}^{+1.3}$\\
$a$ (AU)                        	& $   4.03$ & $\,^{+0.04}_{-0.04}$
									& $  3.99$ & $\,_{-0.01}^{+0.01}$\\
$e$                            		& $  0.903$ & $\,^{+0.001}_{-0.001}$
									& $ 0.895$ & $\,_{-0.004}^{+0.003}$\\
$\omega$ ($^\circ$)     			 & $    -27$ & $\,^{+   8}_{  -8}$
									& $    -19.5$ & $\,_{-2.9}^{+2.9}$\\
$\Omega$ ($^\circ$)      			& $     16$ & $\,^{+   8}_{  -6}$
									& $ 8.1$& $\,_{-2.5}^{+2.5}$\\
$i$ ($^\circ$)                 		 & $   32.3$ & $\,^{+1.3}_{-1.0}$
 									& $ 34.8$ & $\,_{-1.5}^{+1.3}$\\
$v_0$ (km/s)			      &  &  & $  28.8$ & $\,_{-0.16}^{+0.16}$\\
$M_S$ ($\rm 10^{-5} mas^3/year^2$)  & $ 28.41$ & $\,_{-0.2}^{+0.3}$
									& $   27.87$ & $\,_{-0.41}^{+0.45}$\\
$M_S$ ($M_\odot$)             	    & $  1.08$ & $\,_{-0.03}^{+0.03}$
									& $    1.06$ & $\,_{-0.03}^{+0.03}$\\
$M_{\rm{A}}$ ($M_\odot$)      &  &  & $    0.52$ & $\,_{-0.10}^{+0.08}$\\
$M_{\rm{B}}$ ($M_\odot$)      &  &  & $    0.55$ & $\,_{-0.09}^{+0.09}$\\

\hline
Grade								& \multicolumn{2}{c}{1} & \multicolumn{2}{c}{Reliable}\\
		\noalign{\vskip1pt\hrule\vskip1pt}
		\end{tabular}
	\end{minipage}%
	\hfill
	\begin{minipage}{0.49\textwidth}
		\centering
		\includegraphics[width=\columnwidth]{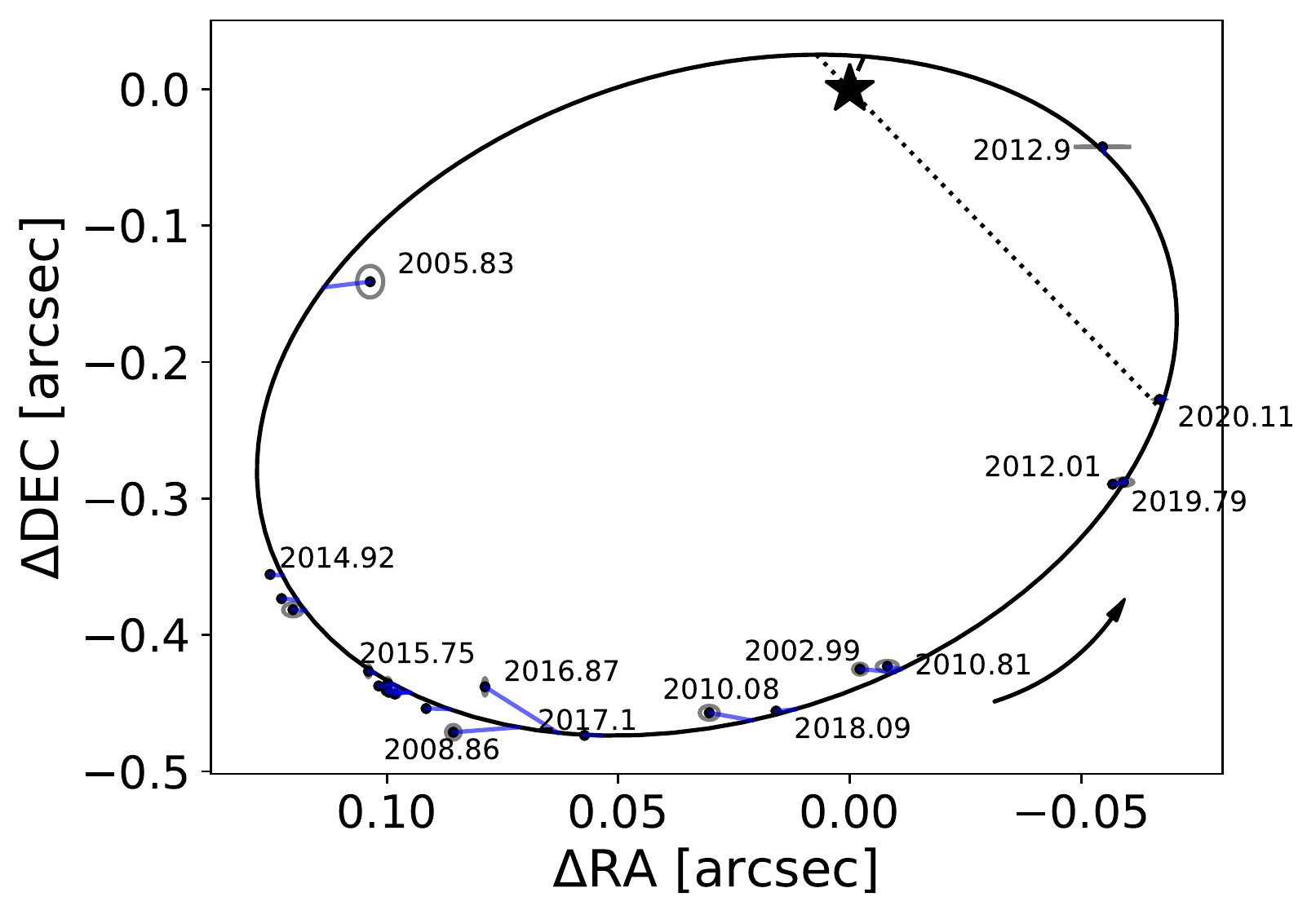}	
	\end{minipage}
\caption{Orbital parameters and fit for J0728.}
\label{fig:orb_J0728}
\end{figure*}
\begin{figure*}
	\begin{minipage}[b]{0.49\textwidth}
	    \centering  
		\renewcommand{\arraystretch}{1.3}
		\begin{tabular}{lr@{}lr@{}l}
		\noalign{\vskip1pt\hrule\vskip1pt}
		Orbital element         & \multicolumn{2}{c}{Grid}  & \multicolumn{2}{c}{MCMC}  \\
		\noalign{\vskip1pt\hrule\vskip1pt}
$t_p$ (JD)							& 2024.51 & $\,^{+0.17}_{   -0.01}$
									& 2023.57 & $\,_{-0.89}^{+0.78}$ \\
Period $P$ (years)              	& $   10.2$ & $\,^{+2.0}_{-0.8}$
									& $  10.8$ & $\,_{-0.6}^{+0.8}$\\
$a$ (mas)                      		 & $ 309$ & $\,^{+3}_{-16}$
									& $ 127$ & $\,_{-19}^{+34}$\\
$a$ (AU)                        	& $   11.3$ & $\,^{+0.3}_{-0.7}$
									& $  4.64$ & $\,_{-0.81}^{+1.36}$\\
$e$                            		 & $  0.900$ & $\,^{+0.013}_{-0.005}$
									& $ 0.68$ & $\,_{-0.11}^{+0.11}$\\
$\omega$ ($^\circ$)     			 & $ -88.93$ & $\,^{+0.12}_{-0.10}$
									& $    -100.7$ & $\,_{-9.0}^{+8.6}\,^*$\\
$\Omega$ ($^\circ$)      			& $  106.6$ & $\,^{+0.2}_{-0.9}$
									& $  102$& $\,_{-12}^{+9}\,^*$\\
$i$ ($^\circ$)                 		 & $   79.5$ & $\,^{+0.3}_{-0.2}$
 									& $ 60$ & $\,_{-11}^{+9}$\\
$M_S$ ($\rm 10^{-5} mas^3/year^2$)  & $ 28.57$ & $\,_{-0.8}^{+1.1}$
									& $   1.6$ & $\,_{-0.6}^{+1.1}$\\
$M_S$ ($M_\odot$)             	    & $  14$ & $\,_{-1}^{+1}$
									& $    0.78$ & $\,_{-0.35}^{+0.60}$\\

\hline
Grade								& \multicolumn{2}{c}{3} & \multicolumn{2}{c}{Tentative}\\									
		\noalign{\vskip1pt\hrule\vskip1pt}
		\end{tabular}
	\end{minipage}%
	\hfill
	\begin{minipage}{0.49\textwidth}
		\centering
		\includegraphics[width=\columnwidth]{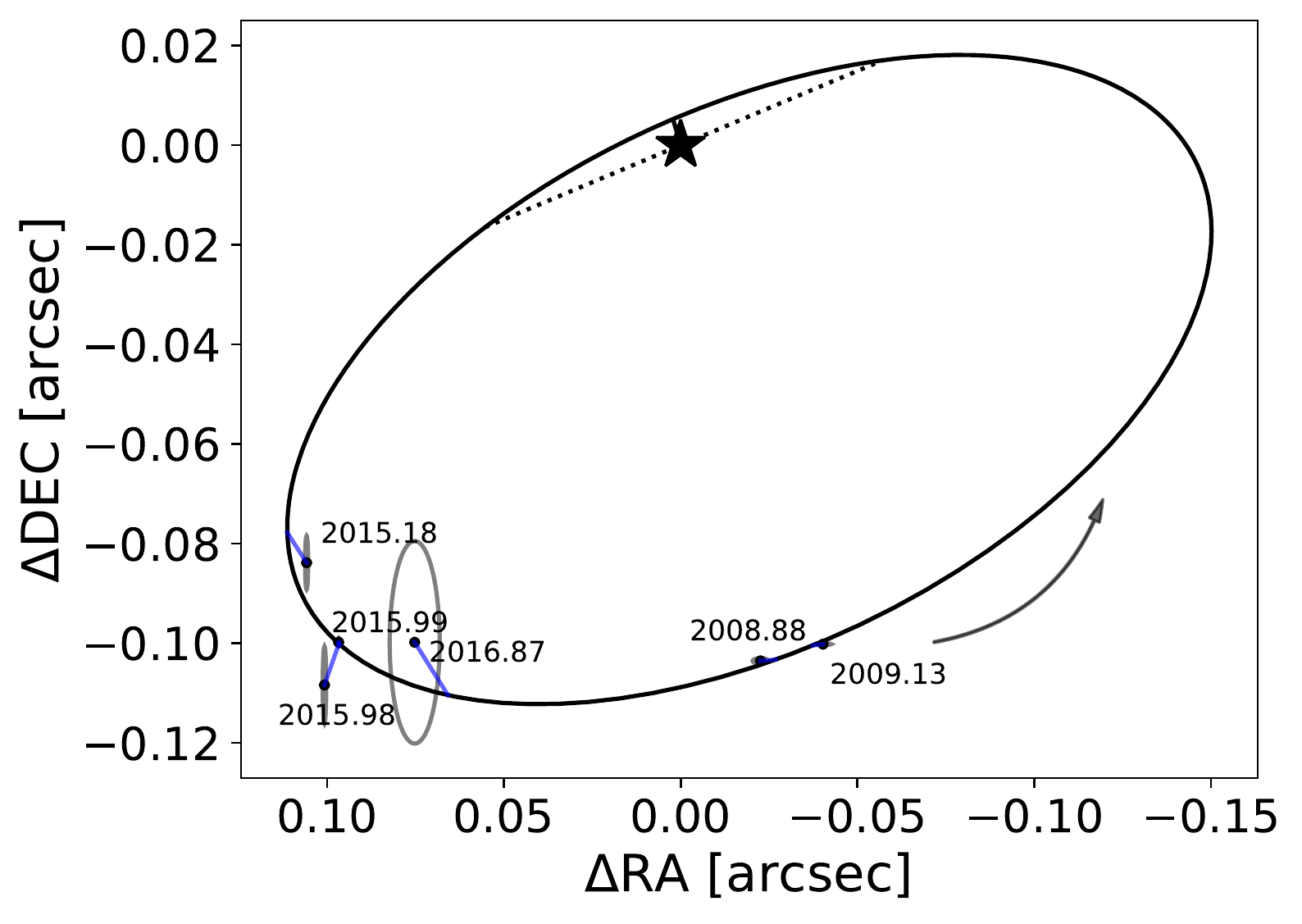}	
	\end{minipage}
\caption{Orbital parameters and fit for J0907.}
\label{fig:orb_J0907}
\end{figure*}
\begin{figure*}
	\begin{minipage}[b]{0.49\textwidth}
	    \centering  
		\renewcommand{\arraystretch}{1.3}
		\begin{tabular}{lr@{}lr@{}l}
		\noalign{\vskip1pt\hrule\vskip1pt}
		Orbital element         & \multicolumn{2}{c}{Grid}  & \multicolumn{2}{c}{MCMC}  \\
		\noalign{\vskip1pt\hrule\vskip1pt}
$t_p$ (JD)							& 2025.02& $\,^{+ 0.02}_{ -0.21}$
									& 2025.10 & $\,_{-0.13}^{+0.11}$ \\
Period $P$ (years)              	& $    8.6$ & $\,^{+0.1}_{-0.2}$
									& $   8.76$ & $\,_{-0.08}^{+0.09}$\\
$a$ (mas)                      		& $  98$ & $\,^{+3}_{-2}$
									& $ 98.3$ & $\,_{-2.8}^{+0.3}$\\
$a$ (AU)                        	 & $   4.24$ & $\,^{+0.12}_{-0.11}$
									& $  4.27$ & $\,_{-0.17}^{+0.06}$\\
$e$                            		& $   0.62$ & $\,^{+0.02}_{-0.02}$
									& $ 0.629$ & $\,_{-0.013}^{+0.015}$\\
$\omega$ ($^\circ$)     			 & $   90.4$ & $\,^{+0.8}_{-0.2}$
									& $    90.6$ & $\,_{-0.6}^{+0.6}$\\
$\Omega$ ($^\circ$)      			& $   16.6$ & $\,^{+0.4}_{-3.0}$
									& $ 18.1$& $\,_{-2.8}^{+2.3}$\\
$i$ ($^\circ$)                 		 & $  116.4$ & $\,^{+1.1}_{-0.5}$
 									& $  116.3$ & $\,_{-1.2}^{+1.1}$\\
$v_0$ (km/s)			      &  &  & $   27.21$ & $\,_{-0.70}^{+0.49}$\\	
$M_S$ ($\rm 10^{-5} mas^3/year^2$)  & $ 1.25$ & $\,_{-0.01}^{+0.01}$
									& $   1.22$ & $\,_{-0.09}^{+0.14}$\\
$M_S$ ($M_\odot$)             	    & $  1.02$ & $\,_{-0.07}^{+0.08}$
									& $    1.01$ & $\,_{-0.09}^{+0.12}$\\
$M_{\rm{A}}$ ($M_\odot$)      &  &  & $    0.21$ & $\,_{-0.07}^{+0.09}$\\
$M_{\rm{B}}$ ($M_\odot$)      &  &  & $   0.80$ & $\,_{-0.09}^{+0.09}$\\

\hline
Grade								& \multicolumn{2}{c}{3} & \multicolumn{2}{c}{Tentative}\\
		\noalign{\vskip1pt\hrule\vskip1pt}
		\end{tabular}
	\end{minipage}%
	\hfill
	\begin{minipage}{0.49\textwidth}
		\centering
		\includegraphics[width=\columnwidth]{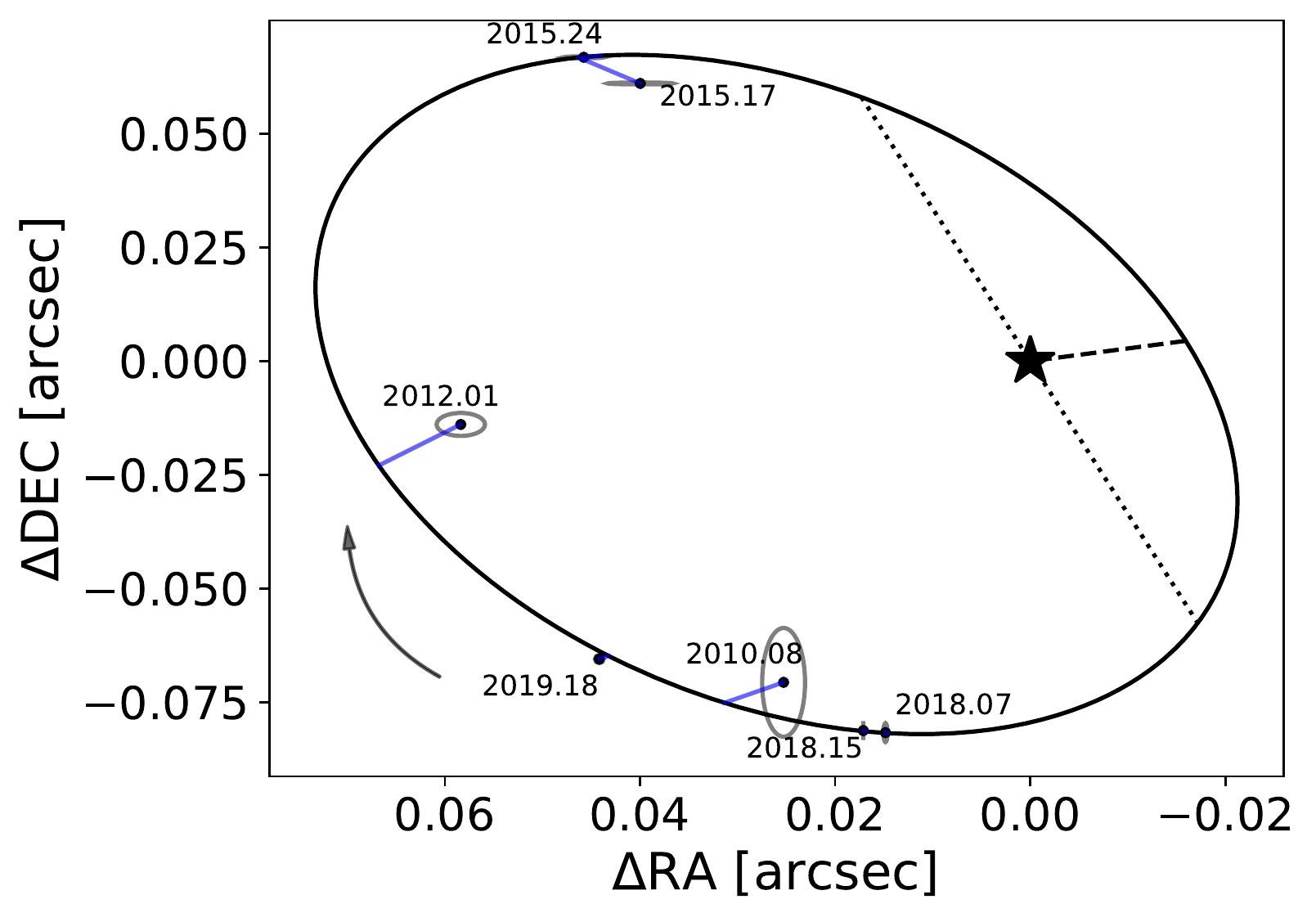}	
	\end{minipage}
\caption{Orbital parameters and fit for J0916.}
\label{fig:orb_J0916}
\end{figure*}
\begin{figure*}
	\begin{minipage}[b]{0.49\textwidth}
	    \centering  
		\renewcommand{\arraystretch}{1.3}
		\begin{tabular}{lr@{}lr@{}l}
		\noalign{\vskip1pt\hrule\vskip1pt}
		Orbital element         & \multicolumn{2}{c}{Grid}  & \multicolumn{2}{c}{MCMC}  \\
		\noalign{\vskip1pt\hrule\vskip1pt}
$t_p$ (JD)							& 2042.23& $\,^{+0.99}_{ -0.13}$
									& 20043.4.9 & $\,_{-1.2}^{+1.2}$ \\
Period $P$ (years)              	 & $     48$ & $\,^{+  24}_{  -9}$
									& $   48.5$ & $\,_{-6.5}^{+16.3}$\\
$a$ (mas)                      		 & $  196$ & $\,^{+18}_{-7}$
									& $ 197$ & $\,_{-6}^{+25}$\\
$a$ (AU)                        	 & $   13.5$ & $\,^{+  1.3}_{  -0.6}$
									& $  13.59$ & $\,_{-0.41}^{+1.72}$\\
$e$                            		& $   0.52$ & $\,^{+0.08}_{-0.14}$
									& $ 0.49$ & $\,_{-0.10}^{+0.07}$\\
$\omega$ ($^\circ$)     			 & $    169$ & $\,^{+   8}_{ -27}$
									& $   177$ & $\,_{-24}^{+33}\,^*$\\
$\Omega$ ($^\circ$)      			 & $     93$ & $\,^{+   5}_{  -6}$
									& $ 96.5$& $\,_{-6.7}^{+6.9}\,^*$\\
$i$ ($^\circ$)                 		 & $    114$ & $\,^{+   4}_{  -3}$
 									& $   112.6$ & $\,_{-2.3}^{+2.8}$\\
$M_S$ ($\rm 10^{-5} mas^3/year^2$)  & $ 0.335$ & $\,_{-0.03}^{+0.03}$
									& $   0.27$ & $\,_{-0.04}^{+0.07}$\\
$M_S$ ($M_\odot$)             	    & $  1.10$ & $\,_{-0.10}^{+0.10}$
									& $    0.87$ & $\,_{-0.13}^{+0.20}$\\

\hline
Grade								& \multicolumn{2}{c}{3} & \multicolumn{2}{c}{Tentative}\\
		\noalign{\vskip1pt\hrule\vskip1pt}
		\end{tabular}
	\end{minipage}%
	\hfill
	\begin{minipage}{0.49\textwidth}
		\centering
		\includegraphics[width=\columnwidth]{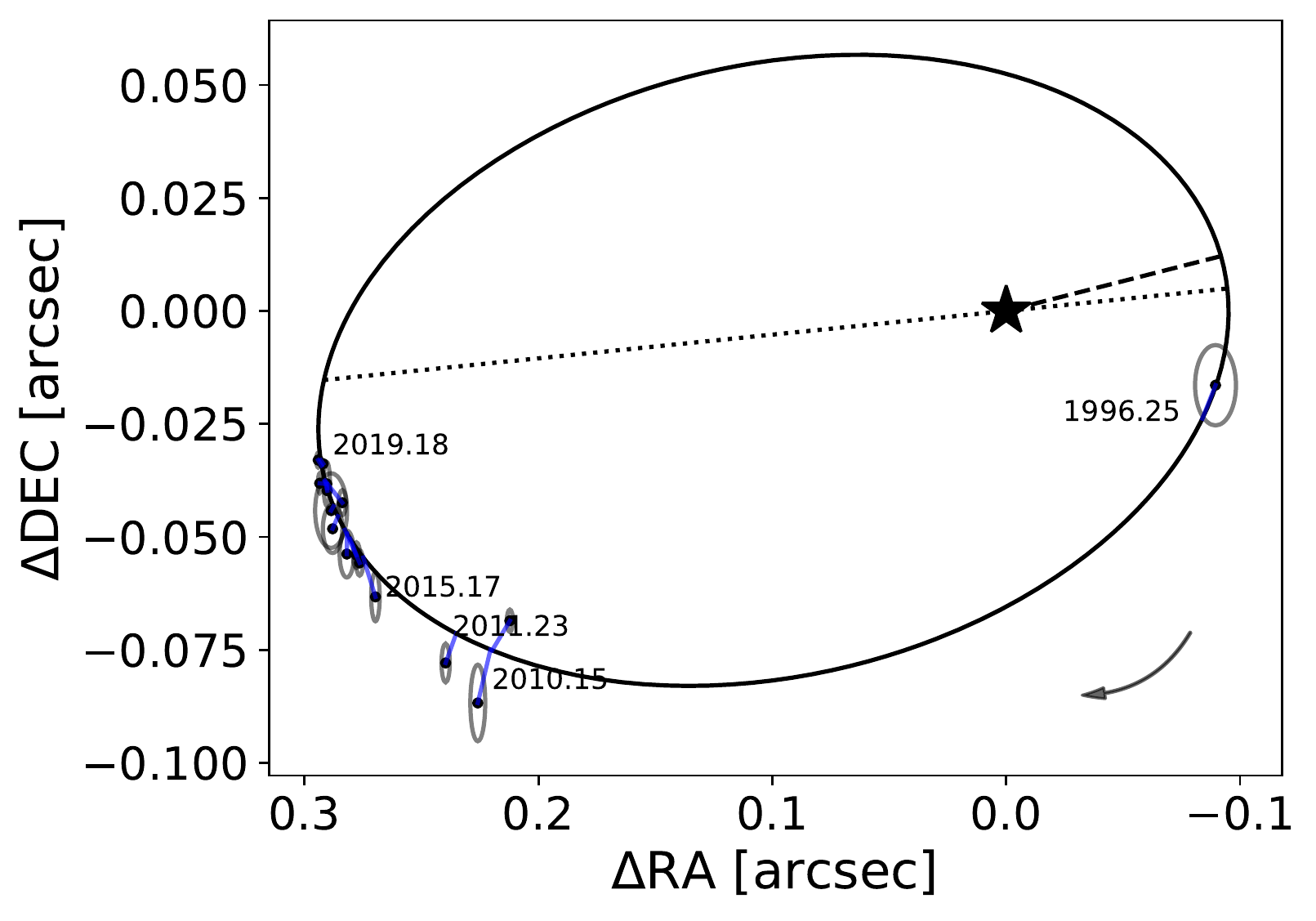}	
	\end{minipage}
\caption{Orbital parameters and fit for J1014.}
\label{fig:orb_J1014}
\end{figure*}
\begin{figure*}
	\begin{minipage}[b]{0.49\textwidth}
	    \centering  
		\renewcommand{\arraystretch}{1.3}
		\begin{tabular}{lr@{}lr@{}l}
		\noalign{\vskip1pt\hrule\vskip1pt}
		Orbital element         & \multicolumn{2}{c}{Grid}  & \multicolumn{2}{c}{MCMC}  \\
		\noalign{\vskip1pt\hrule\vskip1pt}
$t_p$ (JD)							& 2019.69 & $\,^{+ 0.01}_{ -0.01}$
									& 2019.70 & $\,_{-0.04}^{+0.04}$ \\
Period $P$ (years)              	 & $   8.56$ & $\,^{+0.02}_{-0.02}$
									& $    8.47$ & $\,_{-0.02}^{+0.02}$\\
$a$ (mas)                      		 & $  146$ & $\,^{+2}_{-2}$
									& $  148.8$ & $\,_{-1}^{+1}$\\
$a$ (AU)                        	 & $   2.92$ & $\,^{+0.03}_{-0.03}$
									& $  2.98$ & $\,_{-0.03}^{+0.03}$\\
$e$                            		 & $  0.34$ & $\,^{+0.01}_{-0.01}$
									& $ 0.36$ & $\,_{-0.01}^{+0.01}$\\
$\omega$ ($^\circ$)     			 & $   53.7$ & $\,^{+0.7}_{-0.4}$
									& $  51.6$ & $\,_{-5.6}^{+5.4}\,^*$\\
$\Omega$ ($^\circ$)      			& $   89.4$ & $\,^{+1.9}_{-0.6}$
									& $ 94.6$& $\,_{-5.6}^{+5.8}\,^*$\\
$i$ ($^\circ$)                 		& $   23.9$ & $\,^{+0.9}_{-1.6}$
 									& $  28.9$ & $\,_{-1.2}^{+1.2}$\\
$M_S$ ($\rm 10^{-5} mas^3/year^2$)  & $ 4.24$ & $\,_{-0.01}^{+0.01}$
									& $   4.59$ & $\,_{-0.09}^{+0.11}$\\
$M_S$ ($M_\odot$)             	    & $  0.34$ & $\,_{-0.01}^{+0.01}$
									& $   0.37$ & $\,_{-0.01}^{+0.01}$\\

\hline
Grade								& \multicolumn{2}{c}{1} & \multicolumn{2}{c}{Reliable} \\
		\noalign{\vskip1pt\hrule\vskip1pt}
		\end{tabular}
	\end{minipage}%
	\hfill
	\begin{minipage}{0.49\textwidth}
		\centering
		\includegraphics[width=\columnwidth]{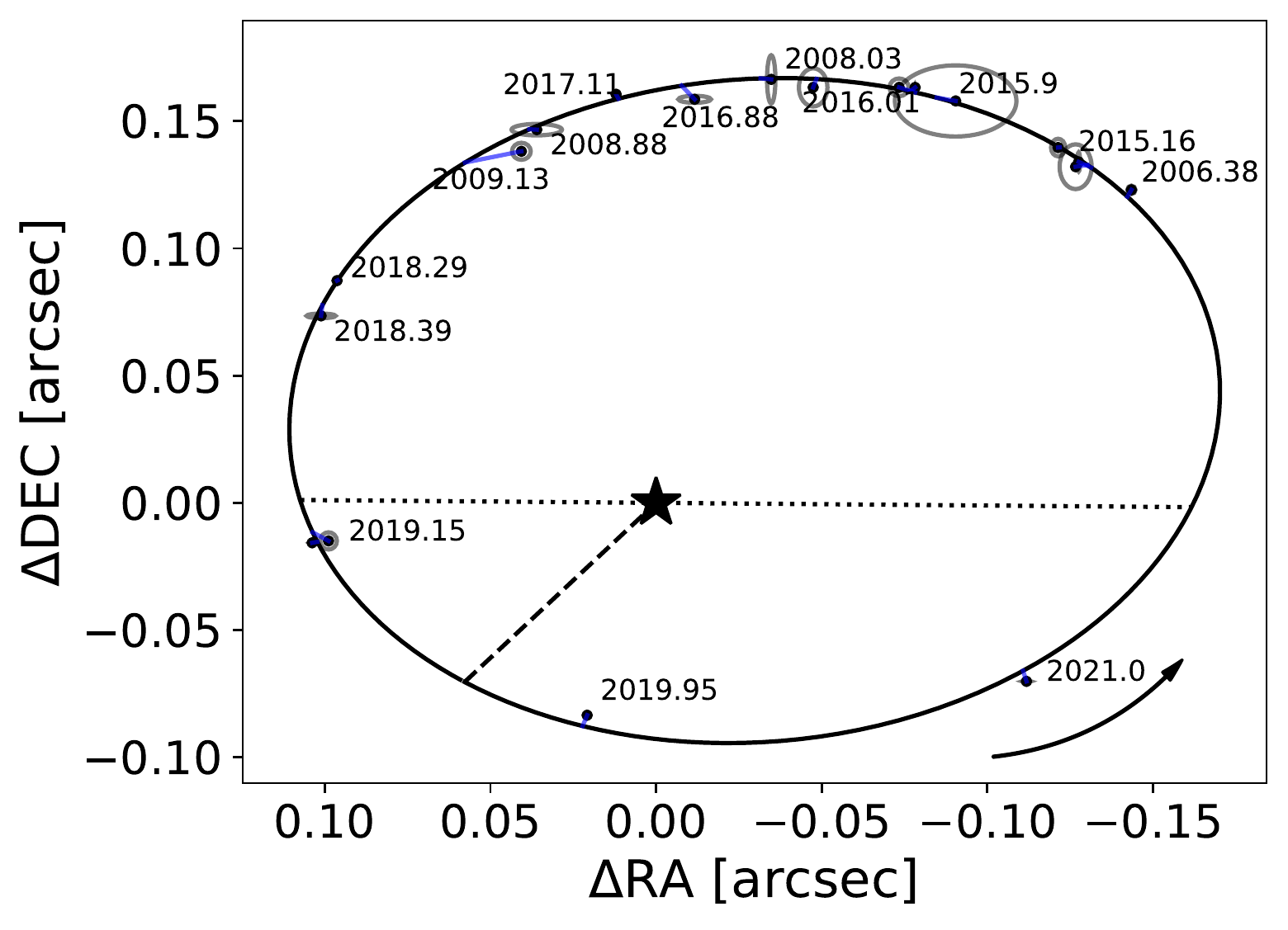}	
	\end{minipage}
\caption{Orbital parameters and fit for J1036.}
\label{fig:orb_J1036}
\end{figure*}
\begin{figure*}
	\begin{minipage}[b]{0.49\textwidth}
	    \centering  
		\renewcommand{\arraystretch}{1.3}
		\begin{tabular}{lr@{}lr@{}l}
		\noalign{\vskip1pt\hrule\vskip1pt}
		Orbital element         & \multicolumn{2}{c}{Grid}  & \multicolumn{2}{c}{MCMC}  \\
		\noalign{\vskip1pt\hrule\vskip1pt}
$t_p$ (JD)							& 2007.17 & $\,^{+ 1.45}_{ -1.58}$
									& 2007.20 & $\,_{-0.44}^{+0.63}$ \\
Period $P$ (years)              	 & $     41$ & $\,^{+4371}_{ -26}$
									& $   36$ & $\,_{-13}^{+26}$\\
$a$ (mas)                      		 & $   270$ & $\,^{+38}_{-9}$
									& $   246$ & $\,_{-33}^{+82}$\\
$a$ (AU)                        	 & $      9$ & $\,^{+  13}_{  -3}$
									& $  8.43$ & $\,_{-1.48}^{+3.15}$\\
$e$                            		 & $    0.6$ & $\,^{+0.4}_{-0.2}$
									& $ 0.59$ & $\,_{-0.07}^{+0.08}$\\
$\omega$ ($^\circ$)     			& $    108$ & $\,^{+  41}_{ -20}$
									& $  104$ & $\,_{-10}^{+25}\,^*$\\
$\Omega$ ($^\circ$)      			 & $    167$ & $\,^{+  34}_{ -17}$
									& $ 160$& $\,_{-10}^{+13}\,^*$\\
$i$ ($^\circ$)                 		& $    121$ & $\,^{+  17}_{ -24}$
 									& $  121.5$ & $\,_{-3.1}^{+3.9}$\\
$M_S$ ($\rm 10^{-5} mas^3/year^2$)  & $ 1.12$ & $\,_{-0.45}^{+2.06}$
									& $   0.80$ & $\,_{-0.18}^{+0.32}$\\
$M_S$ ($M_\odot$)             	    & $  0.45$ & $\,_{-0.09}^{+0.10}$
									& $   0.32$ & $\,_{-0.07}^{+0.09}$\\

\hline
Grade								& \multicolumn{2}{c}{4} & \multicolumn{2}{c}{Preliminary}\\
		\noalign{\vskip1pt\hrule\vskip1pt}
		\end{tabular}
	\end{minipage}%
	\hfill
	\begin{minipage}{0.49\textwidth}
		\centering
		\includegraphics[width=\columnwidth]{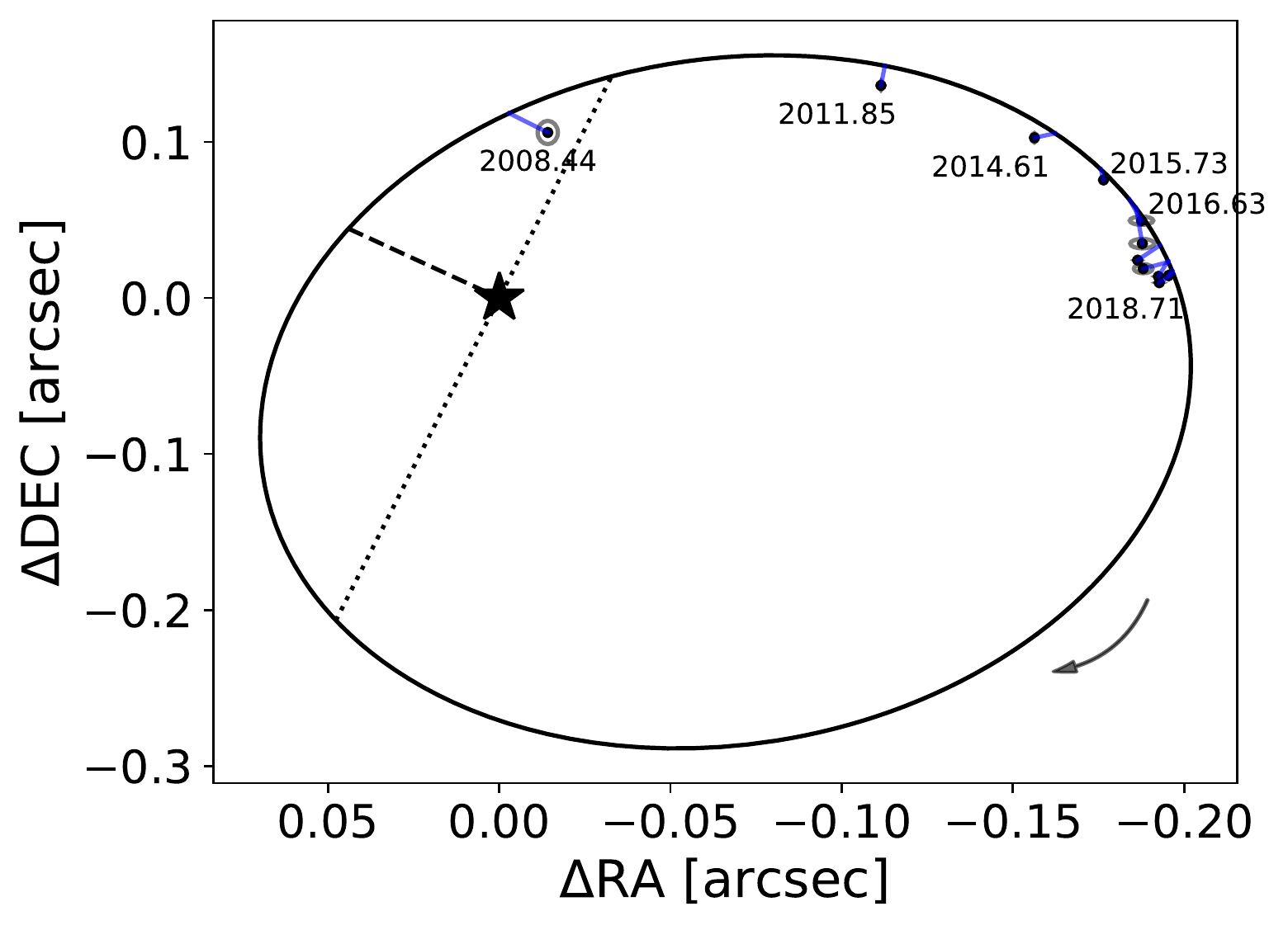}	
	\end{minipage}
\caption{Orbital parameters and fit for J2016.}
\label{fig:orb_J2016}
\end{figure*}
\begin{figure*}
	\begin{minipage}[b]{0.49\textwidth}
	    \centering  
		\renewcommand{\arraystretch}{1.3}
		\begin{tabular}{lr@{}lr@{}l}
		\noalign{\vskip1pt\hrule\vskip1pt}
		Orbital element         & \multicolumn{2}{c}{Grid}  & \multicolumn{2}{c}{MCMC}  \\
		\noalign{\vskip1pt\hrule\vskip1pt}
$t_p$ (JD)							&  2016.85 & $\,^{+ 0.76}_{ -0.21}$
									&  2016.82 & $\,_{-0.18}^{+0.32}$ \\
Period $P$ (years)              	& $     79$ & $\,^{+   7}_{  -7}$
									& $   69$ & $\,_{-17}^{+29}$\\
$a$ (mas)                      	    & $    800$ & $\,^{+800}_{-300}$
									& $   810$ & $\,_{-130}^{+190}$\\
$a$ (AU)                        	& $     14$ & $\,^{+  14}_{  -6}$
									& $  13.04$ & $\,_{-6.00}^{+6.97}$\\
$e$                            		 & $  0.64$ & $\,^{+0.01}_{-0.02}$
									& $ 0.70$ & $\,_{-0.12}^{+0.01}$\\
$\omega$ ($^\circ$)     			 & $    -174$ & $\,^{+   3}_{  -2}$
									& $ -173.3$ & $\,_{-6.8}^{+7.6}\,^*$\\
$\Omega$ ($^\circ$)      			 & $  146.6$ & $\,^{+0.5}_{-0.4}$
									& $  147.0$& $\,_{-1.5}^{+5.9}\,^*$\\
$i$ ($^\circ$)                 		 & $   81.4$ & $\,^{+0.4}_{-0.4}$
 									& $  81.5$ & $\,_{-0.3}^{+0.3}$\\
$M_S$ ($\rm 10^{-5} mas^3/year^2$)  & $ 9.44$ & $\,_{-0.48}^{+0.16}$
									& $  9.47$ & $\,_{-0.32}^{+0.46}$\\
$M_S$ ($M_\odot$)             	    & $  0.44$ & $\,_{-0.37}^{+0.36}$
									& $    0.40$ & $\,_{-0.37}^{+0.38}$\\

\hline
Grade								& \multicolumn{2}{c}{4} & \multicolumn{2}{c}{Preliminary}\\
		\noalign{\vskip1pt\hrule\vskip1pt}
		\end{tabular}
	\end{minipage}%
	\hfill
	\begin{minipage}{0.49\textwidth}
		\centering
		\includegraphics[width=\columnwidth]{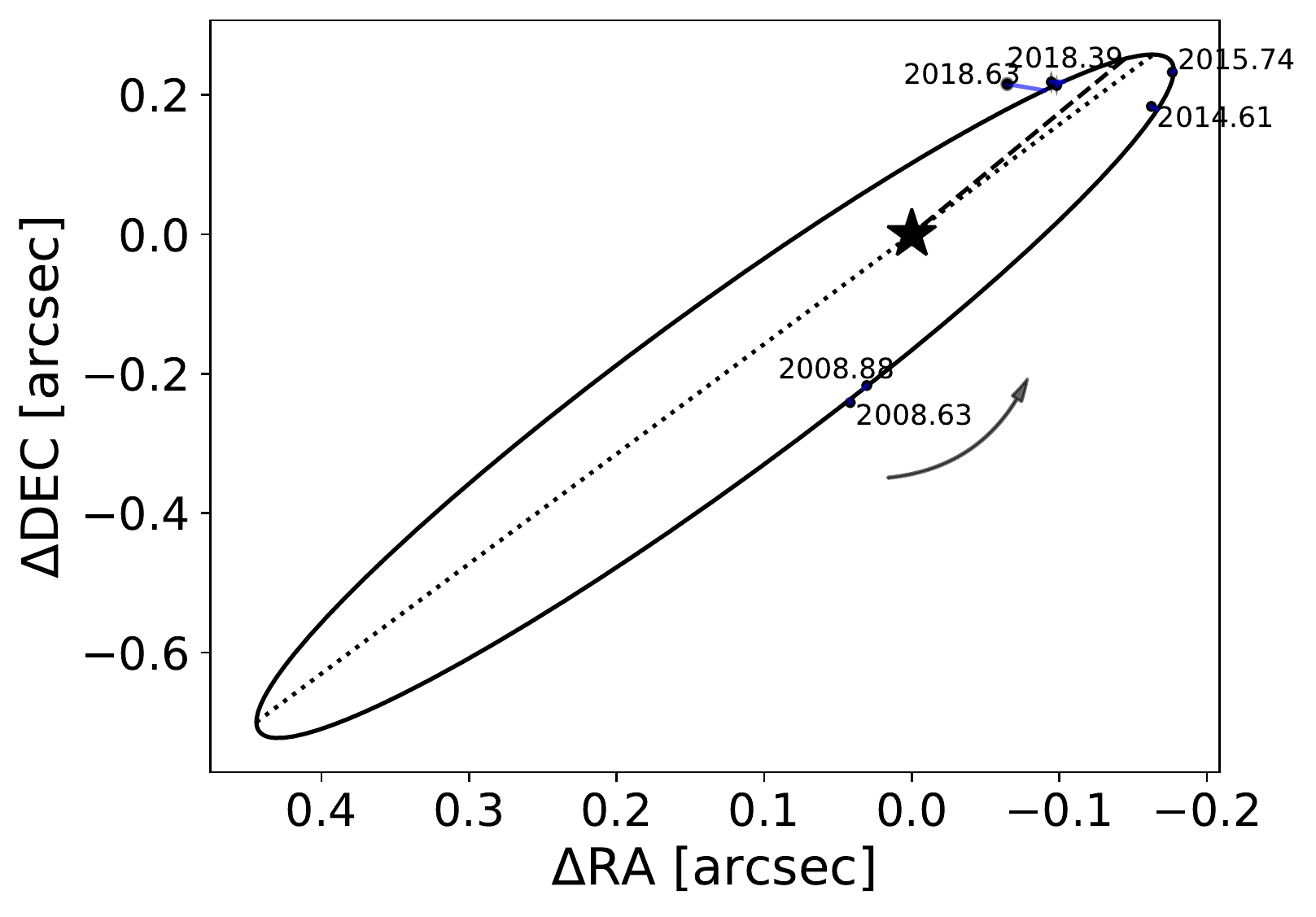}	
	\end{minipage}
\caption{Orbital parameters and fit for J2137. The parallax measurment to the system is dubious with a $\approx 30\,\%$ uncertainty.}
\label{fig:orb_J2137}
\end{figure*}
\begin{figure*}
	\begin{minipage}[b]{0.49\textwidth}
	    \centering  
		\renewcommand{\arraystretch}{1.3}
		\begin{tabular}{lr@{}lr@{}l}
		\noalign{\vskip1pt\hrule\vskip1pt}
		Orbital element         & \multicolumn{2}{c}{Grid}  & \multicolumn{2}{c}{MCMC}  \\
		\noalign{\vskip1pt\hrule\vskip1pt}
$t_p$ (JD)							& 2024.80 & $\,^{+ 0.12}_{ <-0.01}$
									& 2024.88& $\,_{-0.05}^{+0.05}$ \\
Period $P$ (years)              	& $  11.53$ & $\,^{+0.05}_{-0.03}$
									& $  11.55$ & $\,_{-0.02}^{+0.02}$\\
$a$ (mas)                      		 & $ 262.4$ & $\,^{+0.6}_{-0.9}$
									& $ 262.7$ & $\,_{-1.6}^{+1.6}$\\
$a$ (AU)                        	 & $   4.32$ & $\,^{+0.05}_{-0.06}$
									& $  4.32$ & $\,_{-0.06}^{+0.08}$\\
$e$                            		 & $ 0.46$ & $\,^{+0.01}_{<-0.01}$
									& $ 0.466$ & $\,_{-0.004}^{+0.004}$\\
$\omega$ ($^\circ$)     			& $ -155.09$ & $\,^{+0.07}_{-1.49}$
									& $    -153.5$ & $\,_{-1.6}^{+1.6}$\\
$\Omega$ ($^\circ$)      			 & $  30.69$ & $\,^{+0.08}_{-0.13}$
									& $ 30.92$& $\,_{-0.25}^{+0.23}$\\
$i$ ($^\circ$)                 		 & $  103.4$ & $\,^{+0.1}_{-0.2}$
 									& $  103.45$ & $\,_{-0.24}^{+0.29}$\\
$v_0$ (km/s)			      &  &  & $   0.114$ & $\,_{-0.089}^{+0.089}$\\	
$M_S$ ($\rm 10^{-5} mas^3/year^2$)  & $ 13.61$ & $\,_{-0.09}^{+0.16}$
									& $   13.58$ & $\,_{-0.25}^{+0.25}$\\
$M_S$ ($M_\odot$)             	    & $  0.60$ & $\,_{-0.02}^{+0.02}$
									& $    0.61$ & $\,_{-0.03}^{+0.03}$\\
$M_{\rm{A}}$ ($M_\odot$)      &  &  & $    0.34$ & $\,_{-0.02}^{+0.02}$\\
$M_{\rm{B}}$ ($M_\odot$)      &  &  & $   0.27$ & $\,_{-0.02}^{+0.02}$\\

\hline
Grade								& \multicolumn{2}{c}{1} & \multicolumn{2}{c}{Reliable}\\
		\noalign{\vskip1pt\hrule\vskip1pt}
		\end{tabular}
	\end{minipage}%
	\hfill
	\begin{minipage}{0.49\textwidth}
		\centering
		\includegraphics[width=\columnwidth]{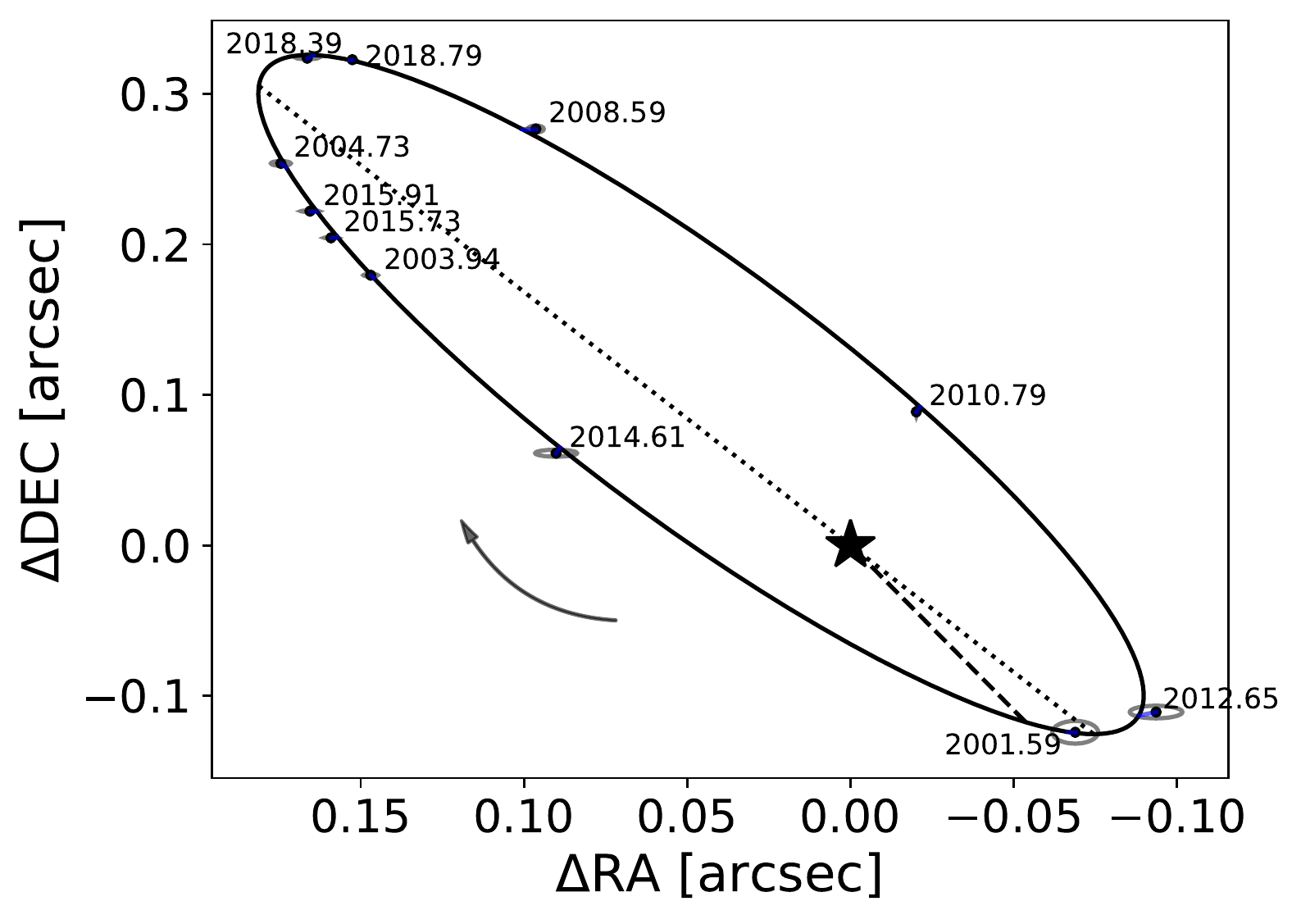}	
	\end{minipage}
\caption{Orbital parameters and fit for J2317.}
\label{fig:orb_J2317}
\end{figure*}
\begin{figure*}
	\begin{minipage}[b]{0.49\textwidth}
	    \centering  
		\renewcommand{\arraystretch}{1.3}
		\begin{tabular}{lr@{}lr@{}l}
		\noalign{\vskip1pt\hrule\vskip1pt}
		Orbital element                         & \multicolumn{2}{c}{Grid}  & \multicolumn{2}{c}{MCMC}  \\
		\noalign{\vskip1pt\hrule\vskip1pt}
$t_p$ (JD)						& 2019.13 & $\,^{+1.79}_{ -4.21}$	
								& 2020.37 & $ \,_{-2.15}^{+1.50}$ \\
Period $P$ (years)             & $     20$ & $\,^{+  34}_{ -10}$
								& $     16.4$ & $\,^{+3.1}_{-1.9}$ \\
$a$ (mas)                       & $   290$ & $\,^{+20}_{-30}$
								& $  239$ & $\,^{+30}_{-25}$\\
$a$ (AU)                       & $    6.3$ & $\,^{+0.4}_{-0.6}$
								& $   5.18$ & $\,^{+0.70}_{-0.59}$\\
$e$                             & $    0.0$ & $\,^{+0.5}_{-0}$
								& $   0.09$ & $\,^{+0.20}_{-0.09}$\\
$\omega$ ($^\circ$)     		 & $    -62$ & $\,^{+  64}_{ -74}$
								& $   -22$ & $\,^{+ 13}_{  -16}\,^*$\\
$\Omega$ ($^\circ$)      		 & $    172$ & $\,^{+   4}_{  -8}$
								& $    167.9$ & $\,^{+  7.2}_{  -7.6}\,^*$\\
$i$ ($^\circ$)                & $    126$ & $\,^{+   4}_{  -3}$
 								& $  129.5$ & $\,^{+9.3}_{-1.7}$\\
$M_S$ ($\rm 10^{-5} mas^3/year^2$)      & $6.2$ & $\,^{+0.2}_{-0.2}$
								& $4.85$ & $\,^{+0.31}_{-0.34}$\\
$M_S$ ($M_\odot$)               & $   0.62$ & $\,^{+0.12}_{-0.09}$
								& $   0.49$ & $\,^{+0.04}_{-0.04}$\\

\hline
Grade								& \multicolumn{2}{c}{3} & \multicolumn{2}{c}{Tentative}\\
		\noalign{\vskip1pt\hrule\vskip1pt}
		\end{tabular}
	\end{minipage}%
	\hfill
	\begin{minipage}{0.49\textwidth}
		\centering
		\includegraphics[width=\columnwidth]{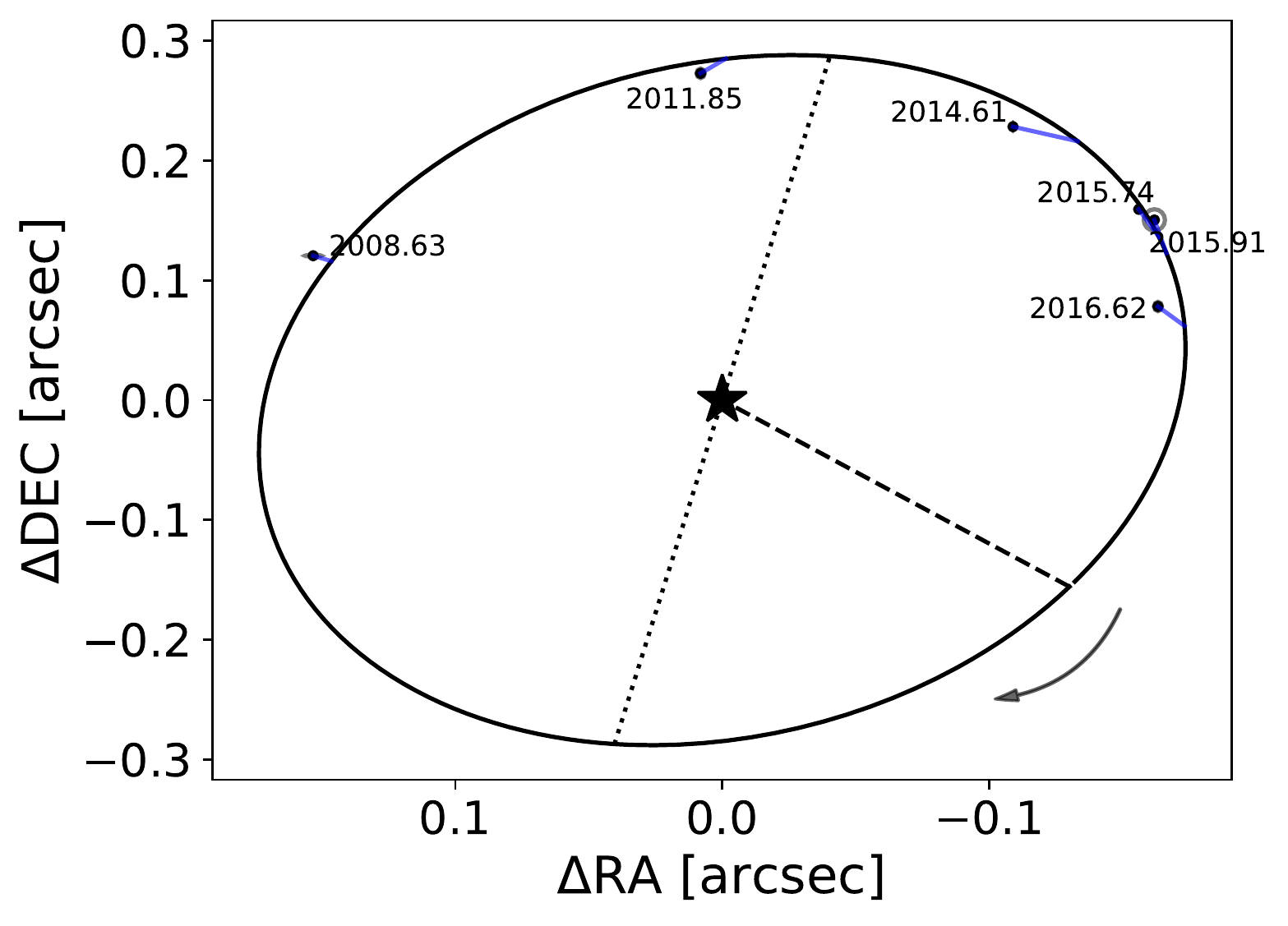}	
	\end{minipage}
\caption{Orbital parameters and fit for J232611.}
\label{fig:orb_J232611}
\end{figure*}
\begin{figure*}
	\begin{minipage}[b]{0.49\textwidth}
	    \centering  
		\renewcommand{\arraystretch}{1.3}
		\begin{tabular}{lr@{}lr@{}l}
		\noalign{\vskip1pt\hrule\vskip1pt}
		Orbital element                         & \multicolumn{2}{c}{Grid}  & \multicolumn{2}{c}{MCMC}  \\
		\noalign{\vskip1pt\hrule\vskip1pt}
$t_p$ (JD)						& 2023.73 & $\,^{+0.01}_{ -0.03}$	
								& 2023.67 & $ \,_{-0.04}^{+0.05}$ \\
Period $P$ (years)             & $   11.0$ & $\,^{+0.2}_{-0.2}$
								& $     10.95$ & $\,^{+0.16}_{-0.11}$ \\
$a$ (mas)                       & $  168$ & $\,^{+2}_{-2}$
								& $  168.4$ & $\,^{+3.3}_{-3.6}$\\
$a$ (AU)                        & $   4.33$ & $\,^{+0.06}_{-0.06}$
								& $   4.33$ & $\,^{+0.13}_{-0.14}$\\
$e$                              & $  0.546$ & $\,^{+0.007}_{-0.005}$
								& $   0.547$ & $\,^{+0.013}_{-0.012}$\\
$\omega$ ($^\circ$)     		 & $  -77.3$ & $\,^{+0.9}_{-0.4}$
								& $   -77.6$ & $\,^{+3.6}_{-4.3}\,^*$\\
$\Omega$ ($^\circ$)      	 & $    156$ & $\,^{+   1}_{  -1}$
								& $    155.6$ & $\,^{+ 3.0}_{  -4.5}\,^*$\\
$i$ ($^\circ$)                  & $  115.9$ & $\,^{+0.5}_{-0.5}$
 								& $  115.7$ & $\,^{+0.9}_{-1.1}$\\
$M_S$ ($\rm 10^{-5} mas^3/year^2$)      & $3.91$ & $\,^{+0.02}_{-0.01}$
								& $3.92$ & $\,^{+0.33}_{-0.27}$\\
$M_S$ ($M_\odot$)               & $   0.67$ & $\,^{+0.03}_{-0.03}$
								& $   0.68$ & $\,^{+0.06}_{-0.05}$\\

\hline
Grade								& \multicolumn{2}{c}{2} & \multicolumn{2}{c}{Good}\\
		\noalign{\vskip1pt\hrule\vskip1pt}
		\end{tabular}
	\end{minipage}%
	\hfill
	\begin{minipage}{0.49\textwidth}
		\centering
		\includegraphics[width=\columnwidth]{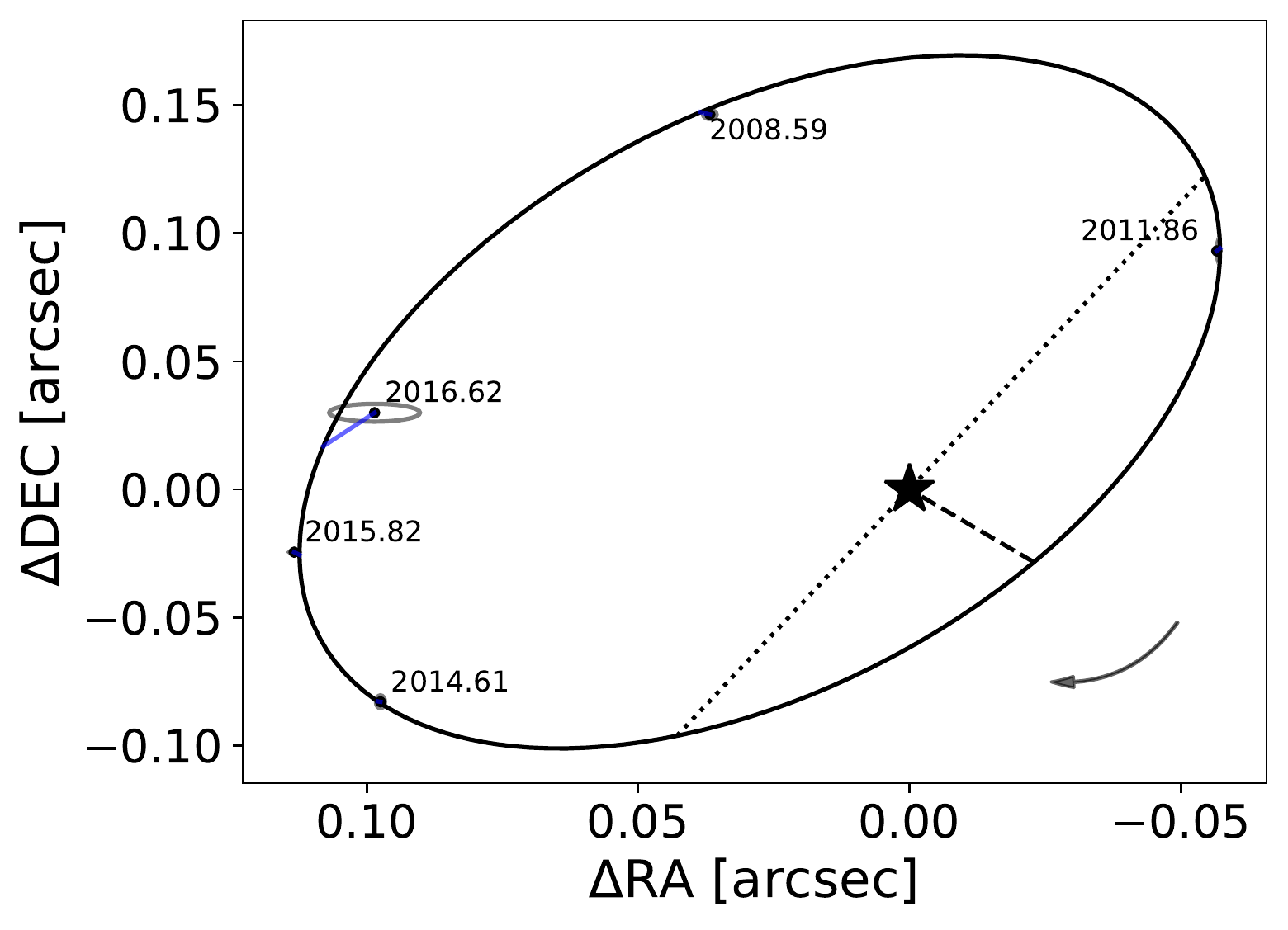}	
	\end{minipage}
\caption{Orbital parameters and fit for J232617.}
\label{fig:orb_J232617}
\end{figure*}
\begin{figure*}
	\begin{minipage}[b]{0.49\textwidth}
	    \centering  
		\renewcommand{\arraystretch}{1.3}
		\begin{tabular}{lr@{}lr@{}l}
		\noalign{\vskip1pt\hrule\vskip1pt}
		Orbital element                         & \multicolumn{2}{c}{Grid}  & \multicolumn{2}{c}{MCMC}  \\
		\noalign{\vskip1pt\hrule\vskip1pt}
$t_p$ (JD)						& 2043.99 & $\,^{+14.07}_{ -10.05}$	
								& 2009.1 & $ \,_{-7.9}^{+2.8}$ \\
Period $P$ (years)               & $     49$ & $\,^{+   8}_{ -31}$ 
								& $     40$ & $\,^{+26}_{-13}$ \\
$a$ (mas)                      & $   230$ & $\,^{+20}_{-40}$
								& $  180$ & $\,^{+43}_{-18}$\\
$a$ (AU)                      & $     11$ & $\,^{+   1}_{  -2}$
								& $   8.53$ & $\,^{+2.28}_{-1.10}$\\
$e$                            & $    0.1$ & $\,^{+0.9}_{-0}$
								& $   0.19$ & $\,^{+0.18}_{-0.19}$\\
$\omega$ ($^\circ$)     		 & $    149$ & $\,^{+ 112}_{ -93}$
								& $   -67$ & $\,^{+84}_{  -25}\,^*$\\
$\Omega$ ($^\circ$)      		 & $     73$ & $\,^{+  56}_{  -5}$
								& $    67$ & $\,^{+26}_{  -16}\,^*$\\
$i$ ($^\circ$)                 & $     58$ & $\,^{+  34}_{ -56}$
 								& $  45$ & $\,^{+7}_{-17}$\\
$M_S$ ($\rm 10^{-5} mas^3/year^2$)      & $0.49$ & $\,^{+3.02}_{-0.20}$
								& $0.22$ & $\,^{+0.20}_{-0.02}$\\
$M_S$ ($M_\odot$)               & $   0.52$ & $\,^{+0.07}_{-0.05}$
								& $   0.23$ & $\,^{+0.17}_{-0.04}$\\

\hline
Grade								& \multicolumn{2}{c}{4} & \multicolumn{2}{c}{Preliminary}\\
		\noalign{\vskip1pt\hrule\vskip1pt}
		\end{tabular}
	\end{minipage}%
	\hfill
	\begin{minipage}{0.49\textwidth}
		\centering
		\includegraphics[width=\columnwidth]{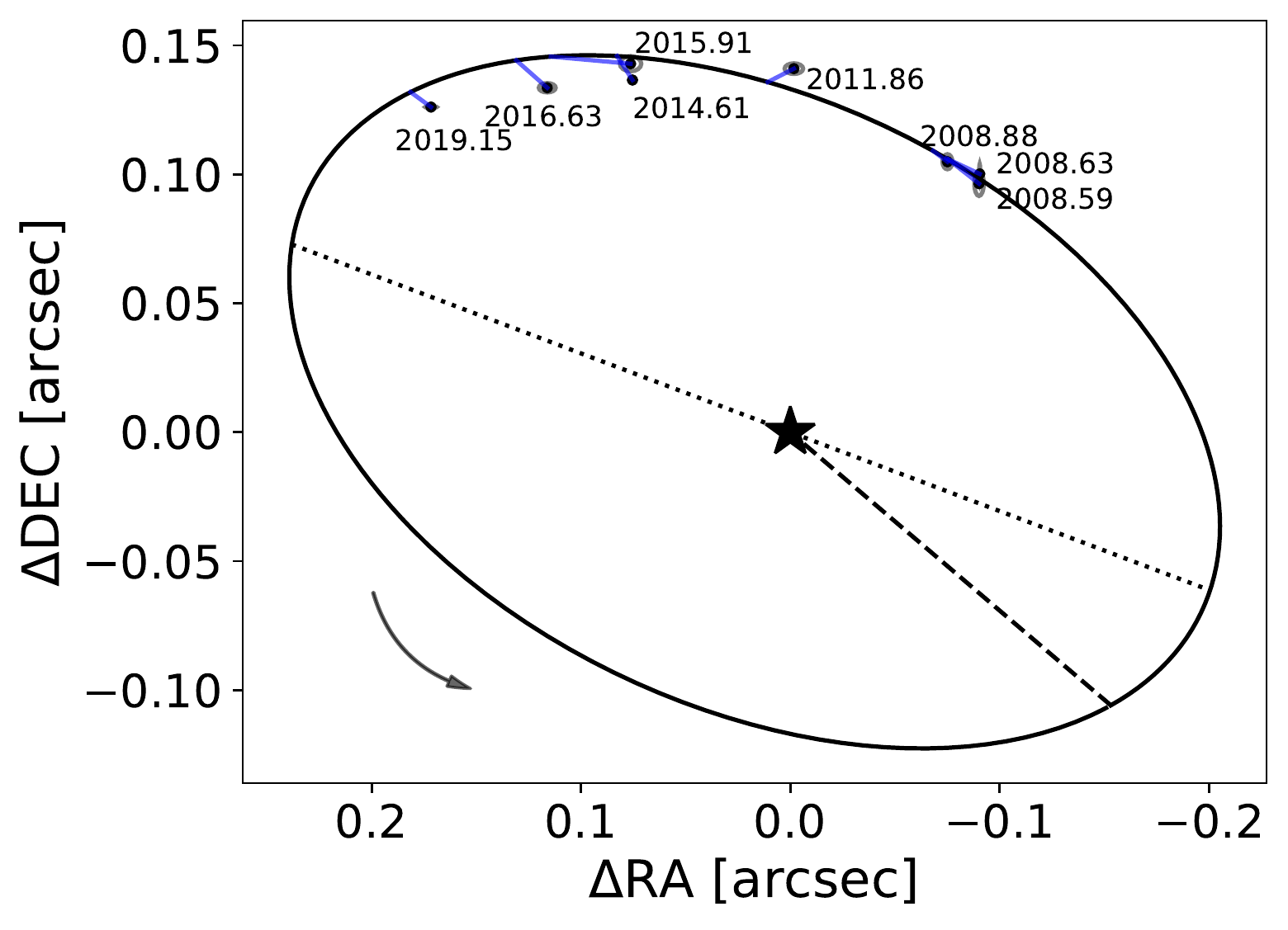}	
	\end{minipage}
\caption{Orbital parameters and fit for J2349.}
\label{fig:orb_J2349}
\end{figure*}

\clearpage
\subsection{MCMC}\label{appendixB2}
\begin{minipage}{\textwidth}
	\begin{center}
		\includegraphics[width=\linewidth]{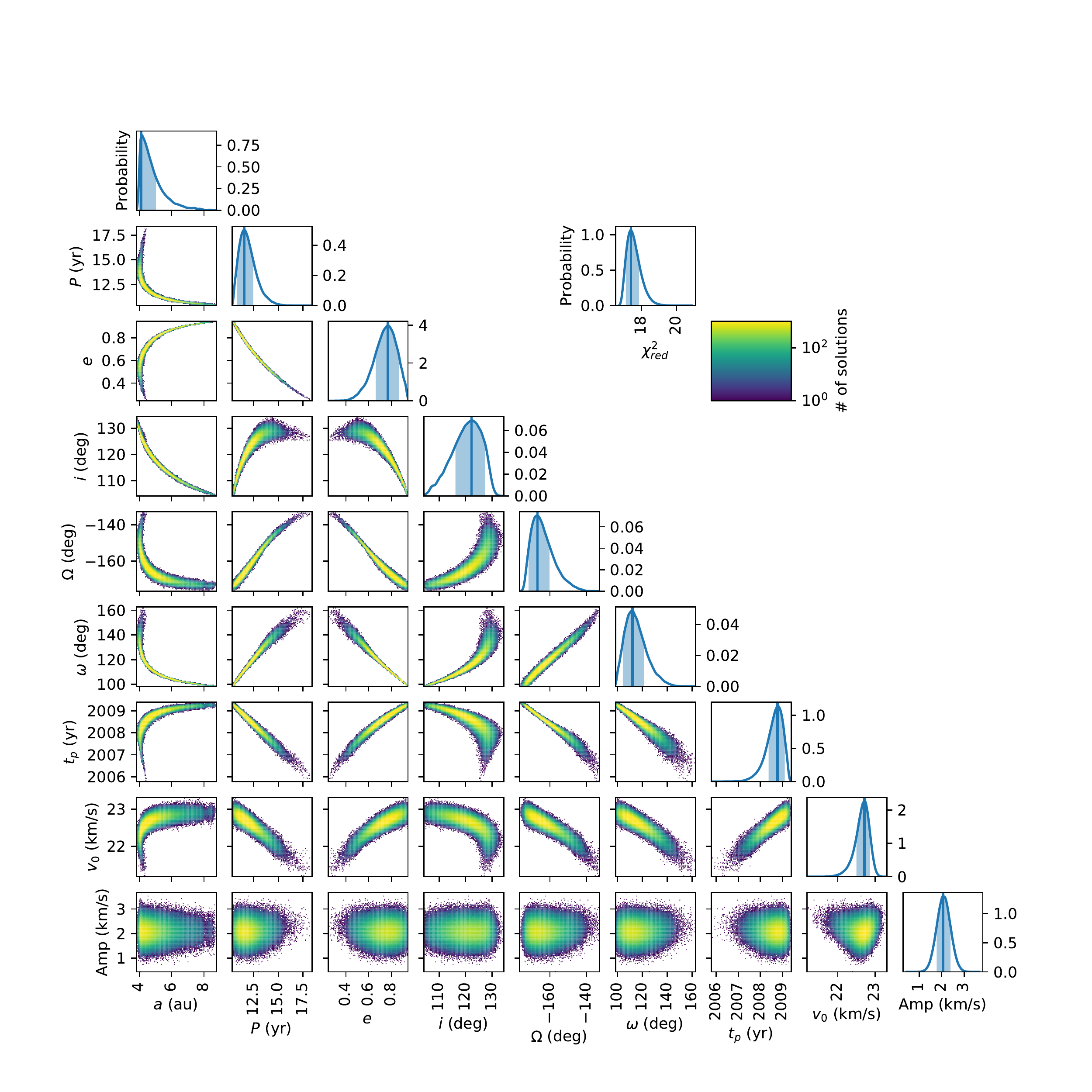}
		\captionof{figure}{Distribution and correlations of each of the orbital element fitted by the MCMC algorithm for J0613. The blue lines depict the probability peak, with the shaded light-blue area encompassing the $16-84\,\%$ confidence interval.}
		\label{fig:J0613_MCMC_pams}	
	\end{center}
\end{minipage}
\begin{figure*}
		\centering
		\includegraphics[width=.5\linewidth]{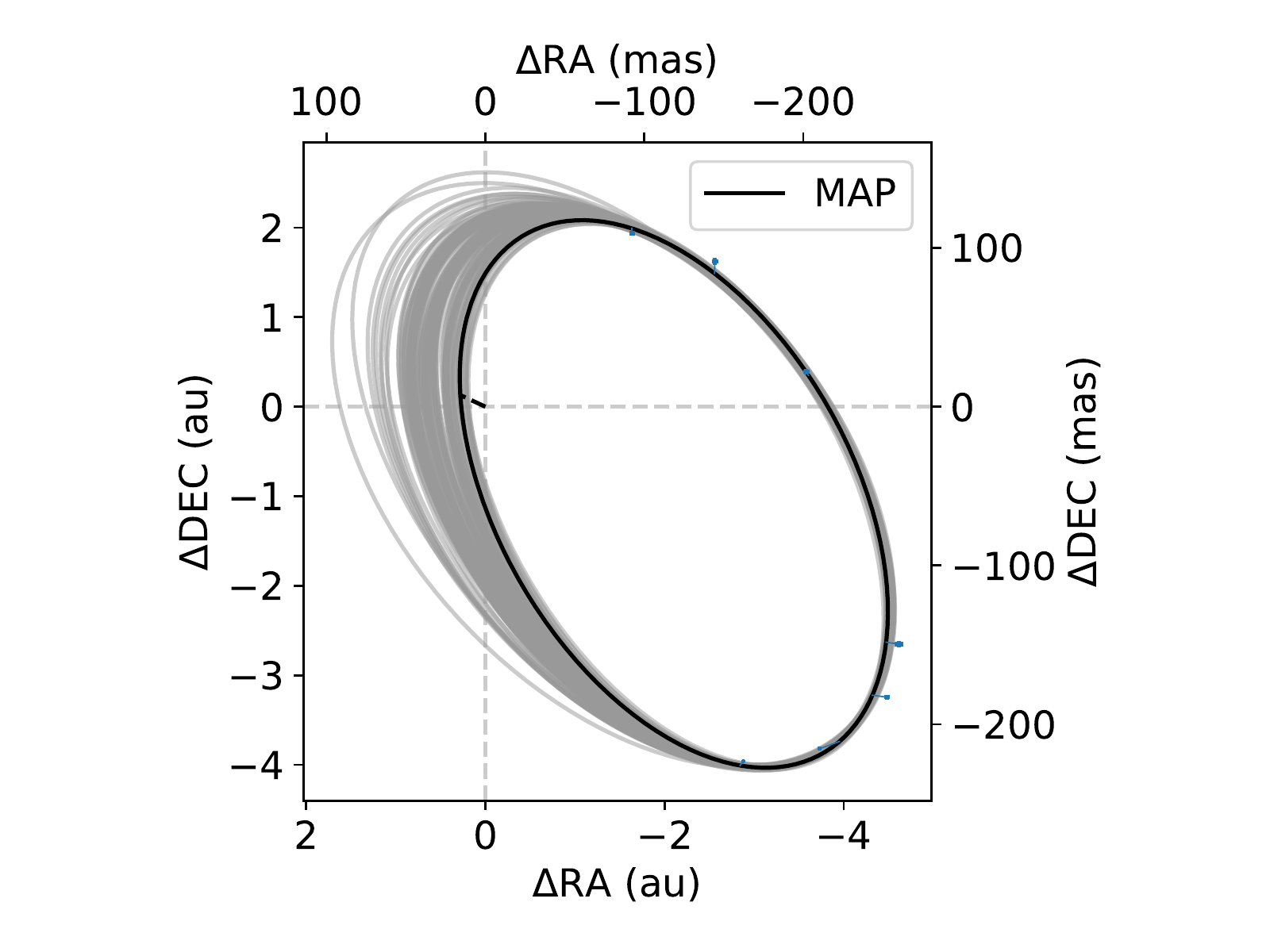}%
		\includegraphics[width=.5\linewidth]{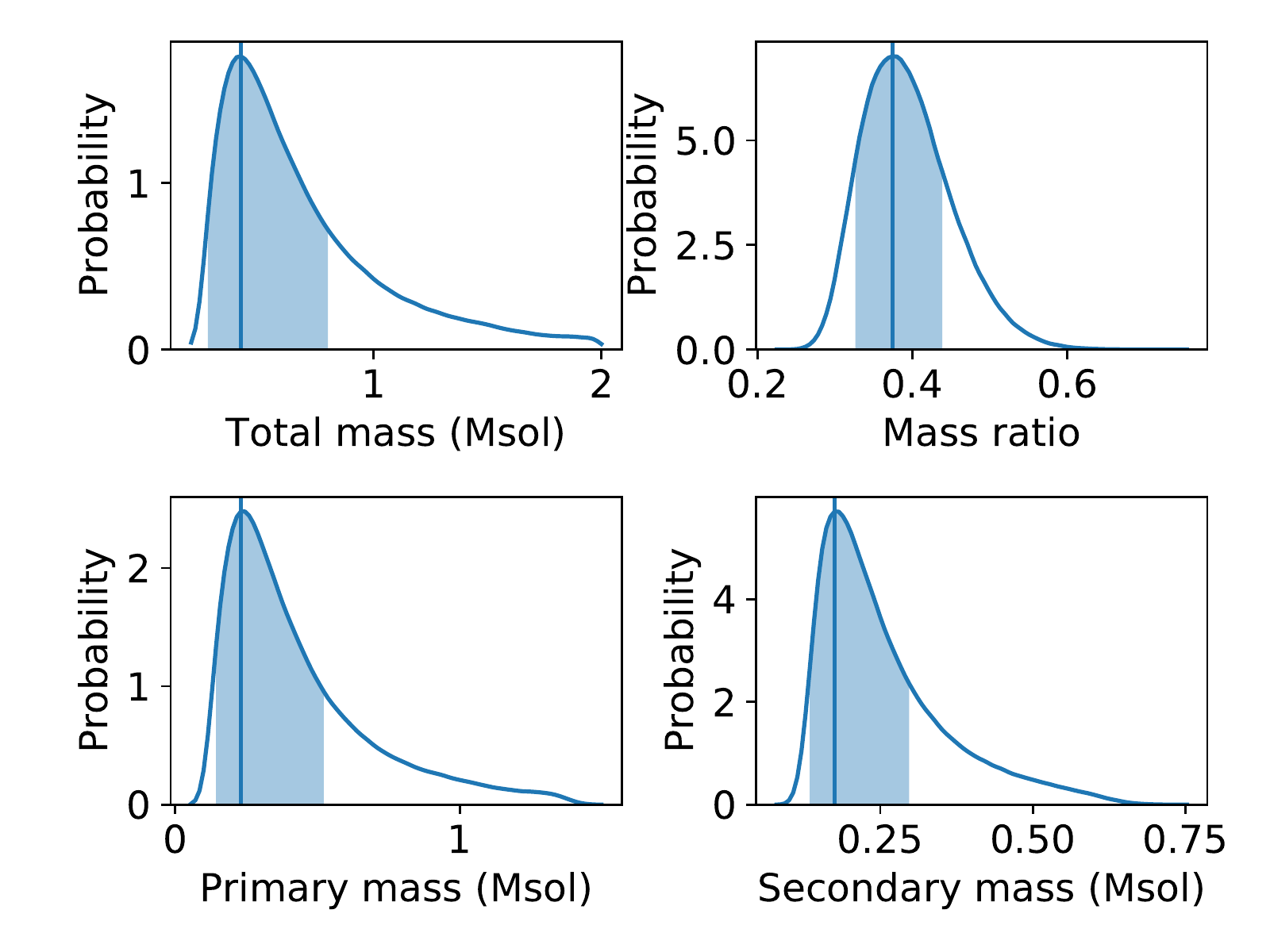}
\caption{J0613 orbits and mass-distribution from the MCMC algorithm. For the left panel, we randomly picked one hundred orbits from the MCMC solutions represented by the grey lines, and plotted the astrometry with the errorbars in blue. The black line represent the Maximum A Posteriori (MAP) estimate, i.e. the most likely solution. For the right panel the blue lines depict the peak probability and corresponding dynamical mass, with the shaded light-blue area encompassing the $16-84\,\%$ confidence interval.
}
	\label{fig:J0613_MCMC}
\end{figure*}
\begin{figure*}
	\centering
	\includegraphics[width=\linewidth]{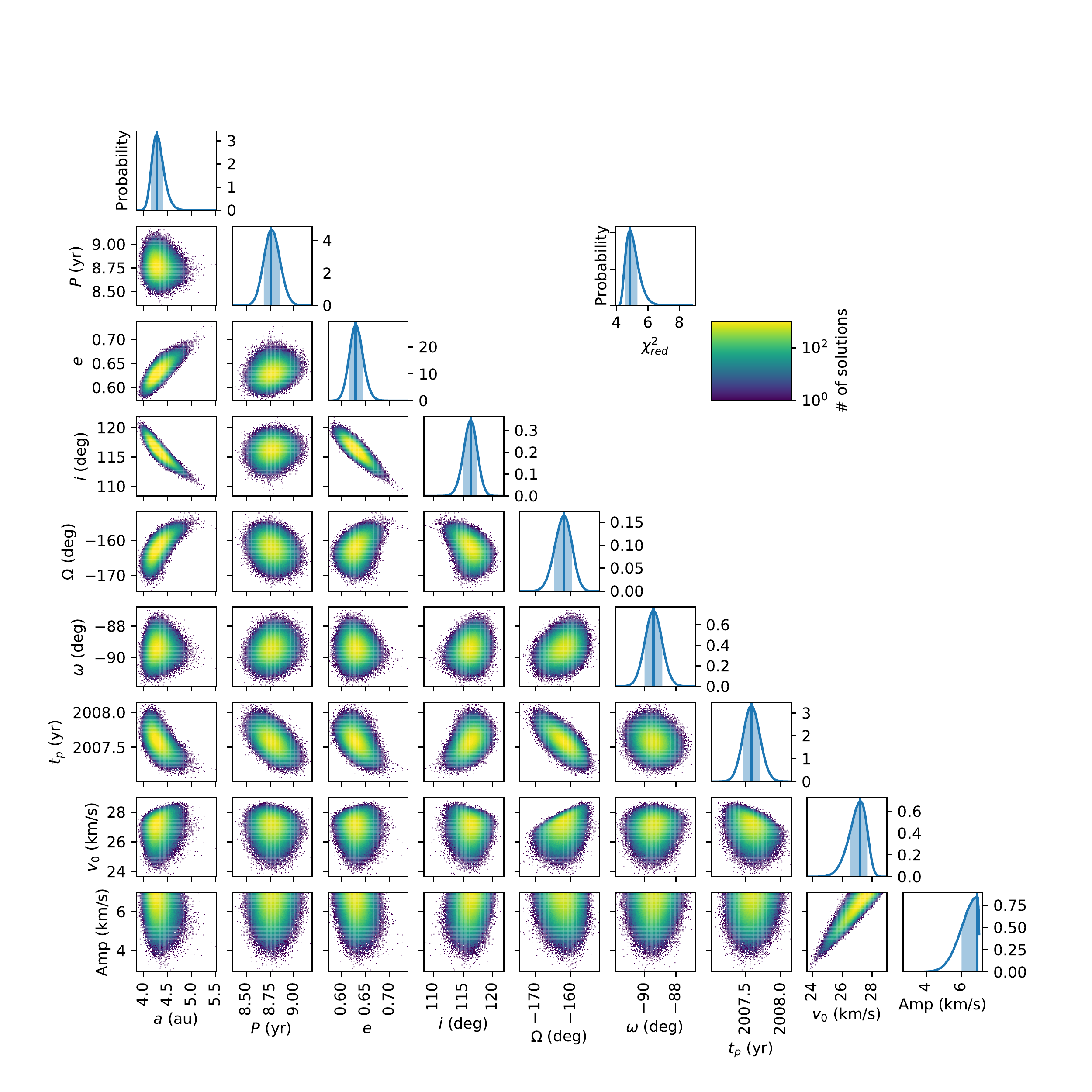}
	\caption{Distribution and correlations of each of the orbital element fitted by the MCMC algorithm for J0916. The blue lines depict the probability peak, with the shaded light-blue area encompassing the $16-84\,\%$ confidence interval.}
	\label{fig:J0916_MCMC_pams}
\end{figure*}%
\begin{figure*}
		\centering
		\includegraphics[width=.5\linewidth]{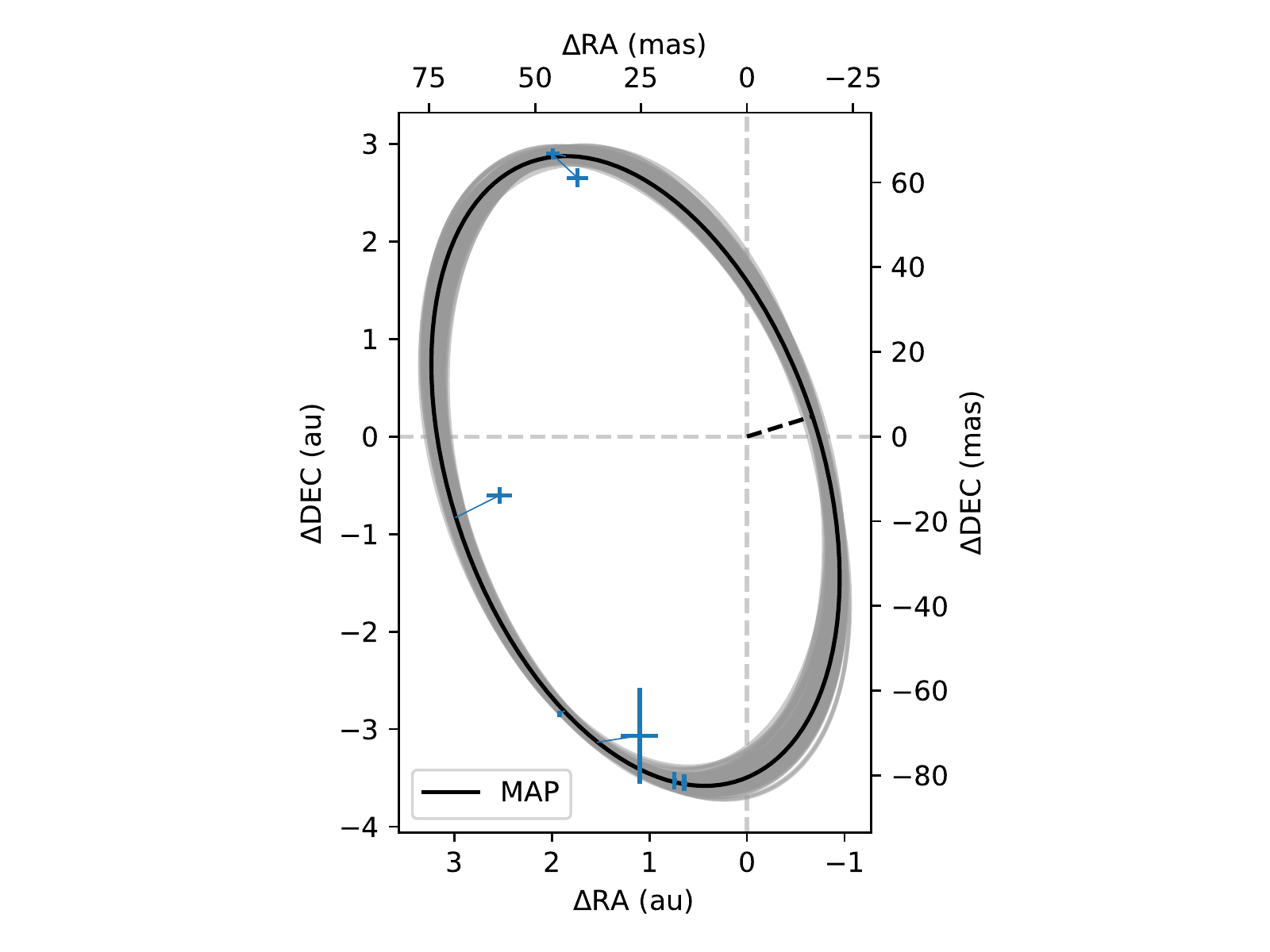}%
		\includegraphics[width=.5\linewidth]{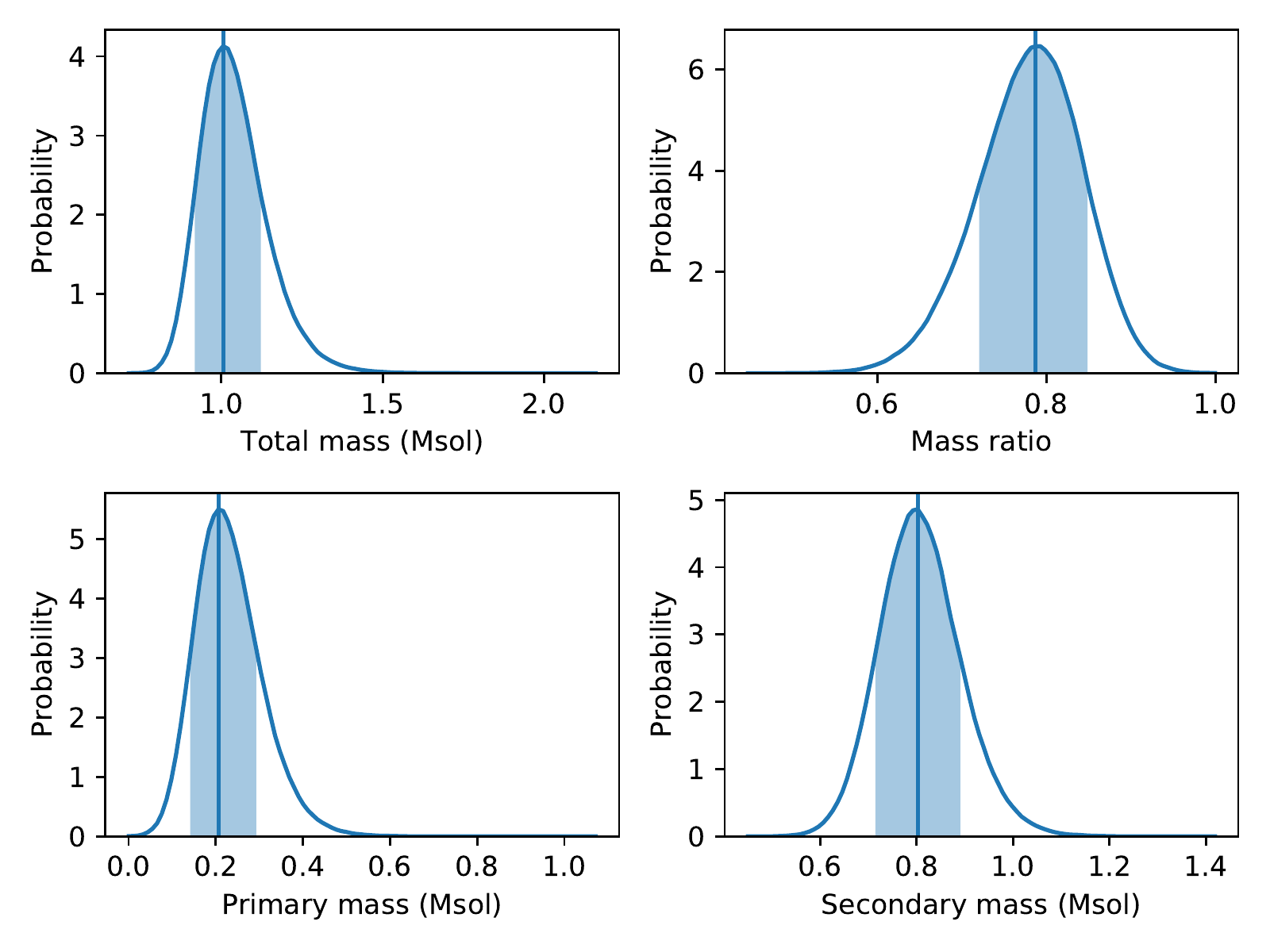}
\caption{J0916 orbits and mass-distribution from the MCMC algorithm. The blue symbols indicate the observed epoch for the astrometric data and the blue lines depict the peak probability and corresponding dynamical mass, with the shaded light-blue area encompassing the $16-84\,\%$ confidence interval.
}
	\label{fig:J0916_MCMC}
\end{figure*}
\begin{figure*}
	\centering
	\includegraphics[width=\linewidth]{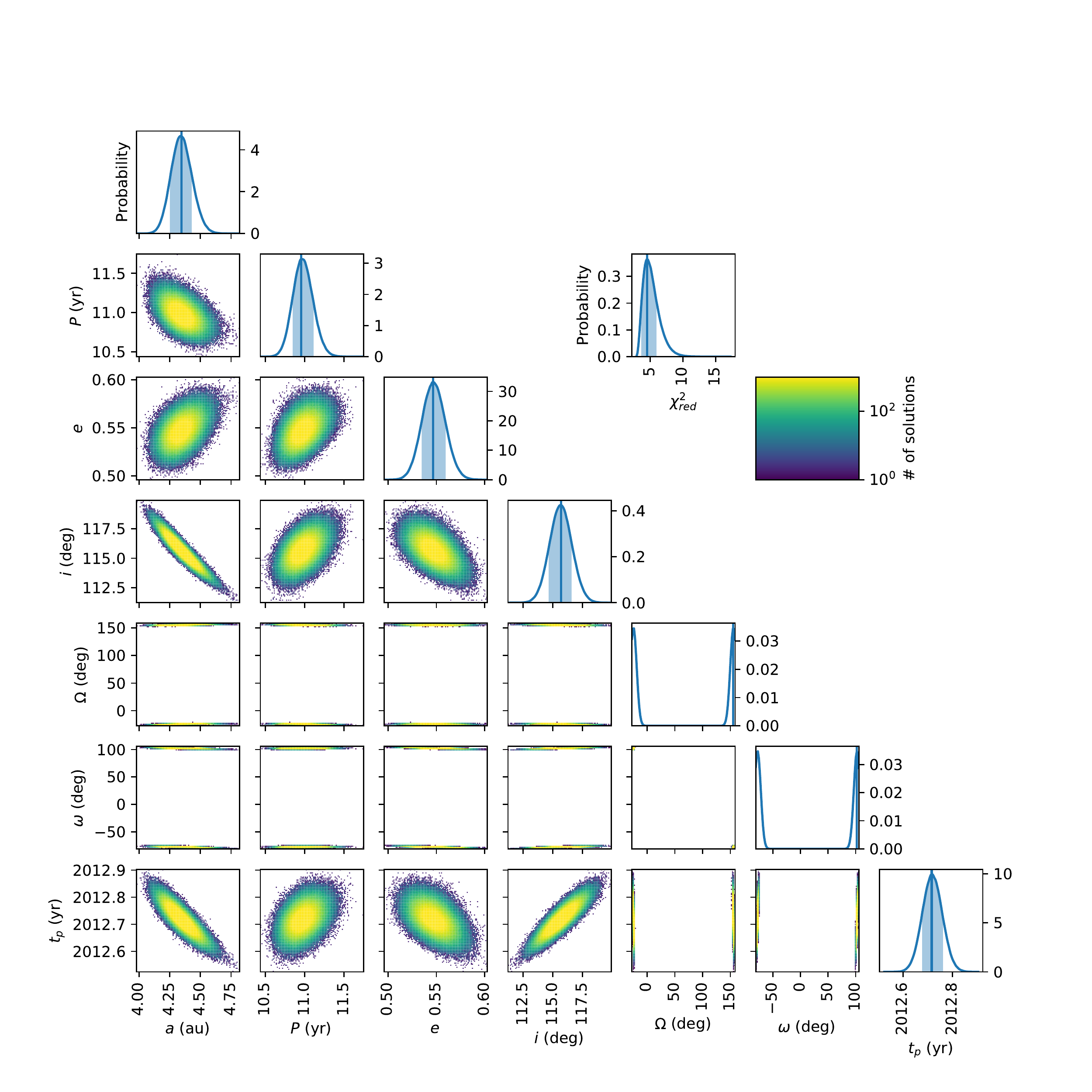}
	\caption{Distribution and correlations of each of the orbital element fitted by the MCMC algorithm for J232617. The blue lines depict the probability peak, with the shaded light-blue area encompassing the $16-84\,\%$ confidence interval.}
	\label{fig:J232617_MCMC_pams}
\end{figure*}%
\begin{figure*}
		\centering
		\includegraphics[width=.5\linewidth]{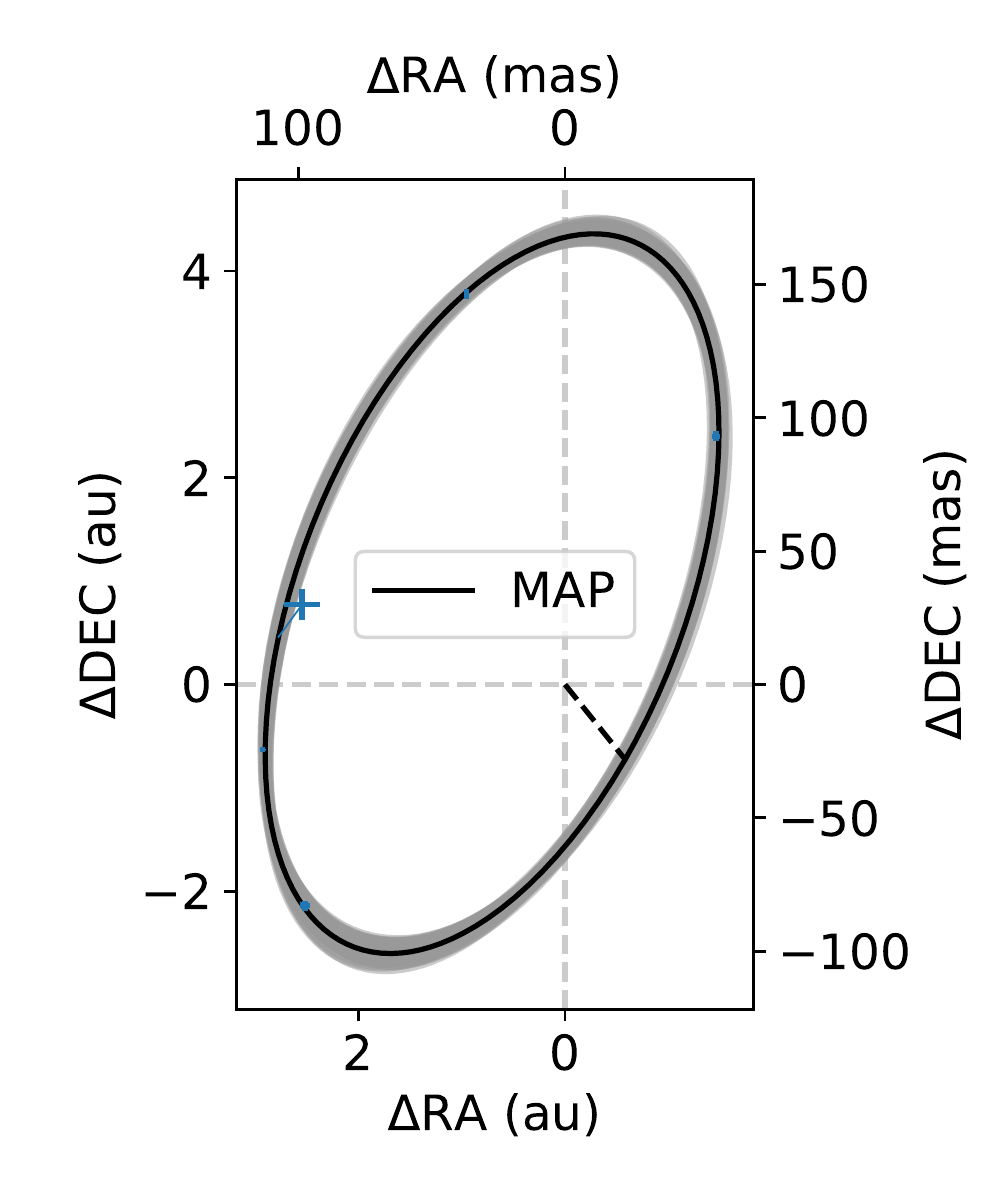}%
		\includegraphics[width=.5\linewidth]{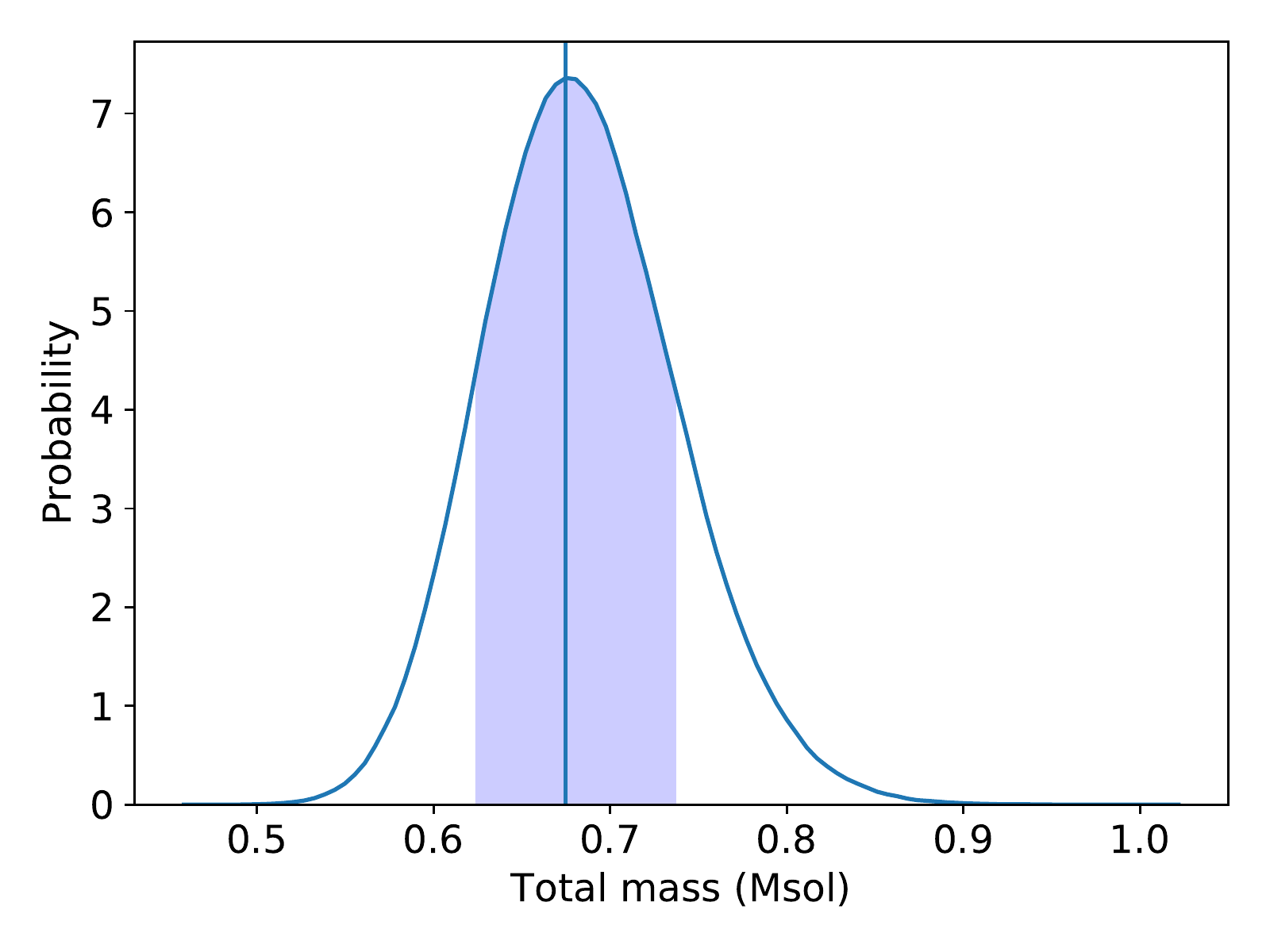}
\caption{J232617 orbits and mass-distribution from the MCMC algorithm. For the left panel, we randomly picked one hundred orbits from the MCMC solutions represented by the grey lines, and plotted the astrometry with the errorbars in blue. The black line represent the Maximum A Posteriori (MAP) estimate, i.e. the most likely solution. For the right panel the blue lines depict the peak probability and corresponding dynamical mass, with the shaded light-blue area encompassing the $16-84\,\%$ confidence interval.}
	\label{fig:J232617_MCMC}
\end{figure*}

\clearpage
\section{Evaluating orbits}\label{appendixC}
The grades for the orbits were calculated in a similar manner to that of \citet{worley_fourth_1983} and \citet{hartkopf_2001_2001}, but scaled according to our sample of 20 binaries here only. We did not include any extra weight or uncertainties for any given system depending on the quality of the observations. We determined the grade value for each orbit as
$$
{\rm Grade} = \frac{{\rm PA_{gap}} + t_{\rm gap}}{N_{\rm rev}\, N_{\rm obs}},
$$
where the positional angle coverage gap ${\rm PA_{gap}}$ is calculated as the maximum gap between observed epochs in positional angle ${\rm PA}_{i} - {\rm PA}_{i-1}$, the phase coverage gap $t_{\rm gap}$ is the maximum gap in phase coverage calculated from the period and difference between time of periastron and observed epoch $(t_0 - {\rm t_{obs}})/P$, the number of revolutions $N_{\rm rev}$ is the time between the last and first epochs divided by the orbital period $(t_{\rm last} - t_{\rm first})/P$, and the number of observed epochs $N_{\rm obs}$.


The values were then scaled between 0 and 5 logarithmically with base 10, going from lowest to highest value as best to worst for the systems J0008 and J0611 respectively, and we designated orbits with values between 0 and 1 as grade 1, values between 1 and 2 as grade 2 etc. The resulting grades for our systems are presented in Table~\ref{tab:grades}.

\begin{table}[h!]
\renewcommand{\arraystretch}{1.3}
\centering
\caption{Orbital grades}
\begin{tabular}{lccc}
\hline \hline
Name  & $\chi^2_{\rm Grid}$ & $\chi^2_{\rm MCMC}$ & Grade\\
\hline
J0008   & 3.6 & 4.11 & 1\\
J0111   & 24.1 & 9.75 & 3\\
J0225   & 6.7 & 3.53 & 5\\
J0245   & 21.5 & 6.14 & 3\\
J0437   & 23.3 & 3.78 & 2\\
J0459   & 5.5 & 4.2 & 4\\
J0532   & 1.3 & 1.62 & 4\\
J0611   & 5.9 & 11.31 & 5\\
J0613   & 1.4 & 17.39 & 2\\
J0728   & 4.8 & 1.18 & 1\\
J0907   & 2.8 & 4.24 & 3\\
J0916   & 6.5 & 4.87 & 3\\
J1014   & 3.2 & 3.23 & 3\\
J1036   & 6.7 & 3.99 & 1\\
J2016   & 13.0 & 15.26 & 4\\
J2137   & 8.2 & 8.50 & 4\\
J2317   & 1.2 & 1.52 & 1\\
J232611 & 22.6 & 17.98 & 4\\
J232617 & 2.9 & 4.52 & 3\\
J2349   & 18.0 & 15.19 & 4\\
\hline
\label{tab:grades}
\end{tabular}
{\small\\
Orbital grades: (1) Reliable, (2) Good, (3) Tentative, (4) Preliminary, (5) Indetermined.
}
\end{table}

\section{Individual systems}\label{appendixD}
{\bf 2MASS J00085391+2050252} was first reported as a binary by \citet{beuzit_new_2004} and has been monitored for almost 20 years, thus far exceeding the estimated period of $P = 5.94^{+0.02}_{-0.01}$ years, making it the binary with the shortest period in our sample. It is presumably a field binary that is unlikely to belong to any known young moving group or association. The orbital fit is the most robust in our sample according to our grading scale criteria, receiving the grade 1 (Reliable). \citet{vrijmoet_solar_2022} reported an orbital period of $P = 5.9$ years and $2.6$ semi-major axis for the system, corresponding to a total system dynamical mass of $M_{\rm tot} = 0.5$, consistent with our results. We found no resolved spectral analysis of the system.

{\bf 2MASS J01112542+1526214} does not have its parallax or proper motions measured by Gaia yet, and we obtained the distance of $d = 17.24 \pm 2.17$ pc from \citet{dittmann_trigonometric_2014} and proper motions from the fourth US Naval Observatory CCD Astrograph Catalog \citep[UCAC4;][]{zacharias_fourth_2013}. All three YMG-tools agreed that the system is a likely BPMG member. We noted that the astrometric data point for J0111 in \citet{calissendorff_characterising_2020} was reported for the wrong spatial scale for the instrument. They reported on a spatial scale of $125 \times 250$ mas/pixel, which overestimated the astrometry for the data point. We corrected for this error here and recalculated the separation at the SINFONI epoch using the $50 \times 100$ mas/pixel spatial scale which was used during the J0111 observations and obtained a projected separation of $416 \pm 5\,$mas which we used into our orbital fitting. We found a few radial velocity measurements for the system in the literature, but they were spread across several different instruments and exhibiting a large jitter, and thus not useful for aiding with constraining the orbit in this case. We found the orbit to be of grade 3 (tentative), and is likely to see some improvements in the coming years, especially with the better distance-measurements which uncertainty propagates to the dynamical mass estimate. The estimated spectral types differ largely between optical photometry and near-IR spectra for the two components, with the secondary component being of much later type than expected from the photometric estimate. This could be an effect of the secondary component being an unresolved binary itself, or an effect of the spectral type relation derived in \citet{calissendorff_characterising_2020} from the relation in \citet{rojas-ayala_m_2014} not being adequately constrained for young M dwarfs.

{\bf 2MASS J02255447+1746467} lacks Gaia parallax and proper motions, and we adopted the distance measurement of $d = 31\pm1.9$ pc from \citet{dittmann_trigonometric_2014} along with UCAC4 catalogue proper motions. The BANYAN $\Sigma-$tool suggested YMG memberships of $36.7\,\%$ for Carina-Near and $44\,\%$ for Argus, while the convergence point tool implied a $92.6\%$ probability for Carina-Near. \citet{zuckerman_nearby_2019} estimates the 40-50 Myr old Argus group to have a mean distance from Earth of 72.4 pc, which could imply the system to more likely be associated with the $\sim 200\,$Myr old Carina-Near which shares similar UVW values but is located closer to Earth at $\sim 30\,$pc \citep{zuckerman_carina-near_2006} and the adopted distance of the binary system. The orbit received a grade of 5 (indetermined) and is likely to be much better constrained with additional epochs. Both methods preferred orbits which insinuated a total dynamical mass of $0.09-0.17\,M_\odot$ for the system, which is extremely low for the observed spectral types when compared to the rest of our sample. If we restrain the orbital period to be shorter than 30 years we obtained a best-fit $\chi^2_{\nu}$ which was four times larger than for the longer periods (c.f. $\chi^2_{\nu} = 1.5$ and $6.3$), but with dynamical masses more compatible with the expected from the theoretical models. More likely is that the distance to the system is inaccurate, and a distance of $d \approx 55$pc would bring both the dynamical and theoretical masses closer together, as well as be more in line with the expected mass from the derived spectral types.

{\bf 2MASS J02451431-4344102} is a likely field binary according to all three YMG tools applied. We mainly sampled the orbit around the periastron and despite the small uncertainty in dynamical mass we obtained a grade 3 (tentative) for the orbit, which has more than half of its path yet uncharted. The MCMC method did not include the astrometric data points from 2019-2020 which explains the difference in orbital parameters determined by the two methods. The RV data in the literature \citep{durkan_radial_2018} had too short baseline to aid the orbital fitting and does not match the astrometry in this case and was therefore excluded from the fitting procedure. The spectral types are similar to that of the J0008 system which had a reliable orbital fit, also belonging to the field and having similar dynamical mass. This could suggest that the orbit we obtained for J0245 is actually better constrained than expected from our grading criteria.

{\bf 2MASS J04373746-0229282} has have its orbit constrained previously by \citet{montet_dynamical_2015}, obtaining a dynamical mass of $M = 1.11 \pm 0.04\,M_\odot$. We found a good orbital fit for the system, obtaining a grade 2 on our scale, and a slightly lower mass of $M = 0.92\,^{+0.06}_{-0.04}\,M_\odot$. This discrepancy is explained by the different distance measurements adopted for the system, where \citet{montet_dynamical_2015} resorted to using the Hipparcos distance to the comoving system 51 Eri of $29.43 \pm 0.30$ pc \citep{van_leeuwen_validation_2007} while we had access to the Gaia EDR3 parallax  corresponding to a distance of $27.77 \pm 0.37$ pc. The RV measurements for the system allow for individual dynamical masses to be derived for the system, where we scaled the flux according to the relative brightness of the components. However, the flux ratio between the two components show additional discrepancies. Indeed, we obtainad a ratio of $F_B/F_A = 0.03$ in the $i^{\prime}$-band \citep{janson_noopsorta_2014} corresponding to a mass-ratio of $M_{\rm B}/M_{\rm tot} = 0.44$. Given the magnitude difference displayed in other bands \citep[e.g. Figure 2 in ][]{montet_dynamical_2015}, the flux ratio is more likely to be $F_{\rm B}/F_{\rm A} = 0.2 \pm 0.1$, which translated to a mass-ratio of $M_{\rm B}/M_{\rm tot} = 0.62$.  As a comparison, \citet{montet_dynamical_2015} found a mass-ratio of $\approx 0.4$. However, they find that the secondary B component is missing mass when compared to evolutionary models, which could be attributed to an unseen companion. Such a companion could also be the cause for the oddity in mass-ratio being higher for the secondary than the primary component that we see. While all three of the YMG tools suggest different groups for the system, shown in Table~\ref{tab:YMGprob}, the LACEwING tool also produced a $\approx 37\,\%$ probability for BPMG membership, with most of the literature agree that the system is a member of the BPMG.

{\bf 2MASS J04595855-0333123} received a grade 4 (preliminary) from our orbital fitting. The MCMC orbital fitting did not exclude the low probability of a high-mass system, and we cut the mass-distribution at $\leq 2\,M_\odot$. The BANYAN $\Sigma$-tool suggests the system to be in the field while LACEwING proposes the system to be a HYA member, which regardless places the system as old in comparison to the younger systems in our sample. Although the system is likely belonging to the field, the isochrones in the theoretical models predict a much higher mass for the system, which further alludes to the poor orbital constraints and that more information is necessary to obtain a better mass estimate. The models could however explain the lower dynamical mass if the age of the system was $\leq 50$ Myrs. Nevertheless, the optical spectral types are similar to that of J0008 and J0245, which should indicate for similar masses given their approximate ages. The near-IR spectra on the other hand implies an earlier spectral type for the primary state, and its mass could potentially be heavily underestimated.

{\bf 2MASS J05320450-0305291} is part of a higher hierarchical sextuple system \citep{tokovinin_family_2022} and a strong candidate of being a member of the BPMG according to the BANYNA $\Sigma$-tool, but also a potential member of ABDMG according to the convergence point tool. The orbital fit is still preliminary according to our grading scale, receiving a grade of 4 (preliminary), with most of the orbital phase not yet observed. Despite the poor orbital constraints, we obtained low minimised $\chi_{\nu,\,{\rm grid}}^2 = 1.3$ and $\chi^2_{\nu,\,{\rm MCMC}} = 1.6$, suggesting that the astrometry is a good for the orbital fit. However, we find some degeneracy and the estimated period ranges from 20 to 100 years, where the shorter periods would suggest dynamical masses above $2\,M_\odot$, and thus dubious. A longer period of $\sim 80$ is more consistent with the results from \citet{tokovinin_family_2022}, where they provide two potential orbital solutions for either $P = 80$ or $P = 143$ years. We did not include the astrometric epochs from \citep{tokovinin_family_2022} in the MCMC fit which explains the discrepancy compared to the grid model.

{\bf 2MASS J06112997-7213388} received the worst grade of the orbital fits in our sample, and is assigned as undetermined. The system is however likely young, with its highest probability being a CAR member according to the BANYAN $\Sigma$-tool, with LACEwING suggesting either a COL or CAR member with the same probabilities. The convergence point tool however suggests the system to be a CARN member, which is likely just because the tool does not include CAR in its calculations. The spectral types derived from optical photometry and near-IR spectra are consistent with each other. The secondary component is estimated to move close to its apastron in its coming years and expected to show limited motion in its orbit.

{\bf 2MASS J06134539-2352077} is a likely ARG member, supported by both the BANYAN $\Sigma$ and LACEwING tools. The astrometric data from the epochs between 2019-2020 were only included in the grid-search approach, which explains the discrepancy in the orbital parameters obtained for the two methods and the different masses obtained. However, both procedures obtained masses greater than that of the evolutionary model, suggesting that perhaps there is some missing mass and an unseen companion in the system. The SPHERE observations should have been able to detect massive companions of $\sim 0.2\,M_\odot$, or at least a non-uniform PSF, down to $\sim 0.5$ AU, which is not apparent from the only epoch taken with SPHERE for the system. The orbit from the grid-search method obtained a grade 2 (good) on our scale.

{\bf 2MASS J07285137-3014490} is a well-studied system in the ABD moving group, for which we mainly contributed by adding additional astrometric data points and an updated distance parallax compared to the previous results in \citet{rodet_dynamical_2018}. The updated Gaia EDR3 parallax measurement helped to constrain the distance uncertainty to the system by a factor of $\approx 4$, resulting in a more precise mass estimate. Our results were consistent with those of \citet{rodet_dynamical_2018} for most part, with the exception of a slightly higher mass-fraction in our case of $\frac{M_{\rm B}}{M_{\rm tot}} = 0.51\,^{+0.10}_{-0.08}$ compared to $\frac{M_{\rm B}}{M_{\rm tot}} = 0.46 \pm 0.10$ in\citet{rodet_dynamical_2018}, where we applied the same flux ratio of $0.2 \pm 0.01$. The missing mass in the secondary B component that \citet{rodet_dynamical_2018} found is accounted for in our case, which potentially could be caused by the different distances adopted, but the discrepancy in mass-fractions we found instead could also be explained by the same argument of an unseen companion. The system also possesses a notably high eccentricity of $e = 0.90$ which points towards significant dynamical interactions. Nevertheless, the existence of a third unseen companion is likely uncorrelated with the eccentricity, and the close encounters required to dynamically enhance the eccentricity would make the configuration of the system unstable \citep{rodet_dynamical_2018}.

{\bf 2MASS J09075823+2154111} is a likely field binary, with no consensus of YMG membership from the different tools applied. Despite the astrometric data covering almost an entire period of 10-11 years, most of the data points are spread over a small change in positional angle of $\approx 70^\circ$, resulting in only a grade 3 (tentative) orbital fit. The MCMC method found a reasonable total mass of the system, whereas the grid-search method obtained an orbit corresponding to a dynamical mass above $14\,M_\odot$, which we rule out based on the spectral type and photometry of the system. There is nothing obvious in the Gaia EDR3 data that would elude to something being wrong with the estimated parallax to the system that could explain the high mass obtained from the grid-search orbital fit, and it would require a distance of $\approx 14$ pc to reduce the dynamical mass estimate from the grid-search to the same value as for the MCMC method. The astrometric data has rather large uncertainties compared to the rest of the sample, especially so for the epochs observed with NOT/FastCam. We are likely to see improvements for the orbit and better dynamical mass-constraints for the system in the coming years, the period is well-known and just a few or single new epoch would greatly benefit new orbital parameter estimations for the system.

{\bf 2MASS J09164398-2447428} received a grade of 3 (tentative) on our grading scale for its orbit, and we have astrometric epochs that cover more than one full revolution, allowing for a more accurate period estimation. The binary is likely to belong to the field rather than any known YMG. The RV data were consistent with a Keplerian motion and the orbit, but the adopted flux-ratio suggested a much greater mass for the secondary component compared to the primary, with a mass ratio of $\frac{M_{\rm B}}{M_{\rm tot}} = 0.79\,^{+0.06}_{-0.07}$. The total mass of the system was consistent with the prediction from the theoretical models, with a slight overestimation of photometric mass, which could imply that the system is actually younger than anticipated. However, since the RV-weighted flux-ratio was too dubious we excluded the system from the mass-magnitude diagram in Figure~\ref{fig:isochrones}.

{\bf 2MASS J10140807-7636327} is a likely CAR member according to the YMG tools, where the convergence point tool suggested CARN but does not include the CAR group in its calculations which is an approximate neighbour. The optical photometric and near-IR spectral types were consistent with each other, with a slight preference towards earlier types according to the near-IR spectra. The astrometric data spans over 20 years, but the orbital period is expected to be much larger and we obtained a grade 3 (tentative) for our orbital fits for the system, with the main uncertainties stemming from the period and the distance to the system. The dynamical mass was however consistent with the photometric mass obtained from the evolutionary models when adopting the age of the CAR moving group as suggested by the YMG tools. The system does not have a measured Gaia parallax, and instead we adopted the distance of $d = 69 \pm 2$ pc from \citet{malo_bayesian_2013} based on group member statistics, which is greater than the spectroscopic distance measured by \citet{riaz_identification_2006} of $d \approx 14$ pc. Our orbital fit favoured the greater distance which corresponded to a higher mass, as the brightness and spectral types of the binary are incompatible with the dynamical mass estimated using the smaller distance, with the mass being well below the Hydrogen burning limit in such case. We did not assume any distance to the system when assessing YMG membership probabilities. However, the BANYAN $\Sigma$-tool places the system in the field if assuming the shorter distance of $d \approx 14$ pc from \citet{riaz_identification_2006}.

{\bf 2MASS J10364483+1521394} is a well-studied triplet system which previous dynamical mass estimate depict a $\approx 30\,\%$ discrepancy between dynamical and photometric masses \citep{calissendorff_discrepancy_2017}. Only the outer BC binary pair was considered for our orbital fit here, which was well-constrained and obtained the grade 1 (reliable). The orbit of the outer binary around the main primary A star is likely over hundreds of years and thus not ready for orbital constraints yet. We found no obvious YMG membership for the system according to the YMG tools utilised, however studies have suggest the system to be a UMA candidate member, and we therefore adopted the age of $400\pm 100$ Myrs for the system. We did not procure RV data for the system, however \citet{calissendorff_discrepancy_2017} previously measured the mass-ratio between the B and C components from the relative motion around the common centre of mass for the pair on the orbit around the primary, revealing the outer binary to be of both equal brightness and mass. The previous mass estimate had most of its error budget dominated by the uncertainty in the distance to the system, which has now been remedied by Gaia EDR3 parallax. The updated distance also reduced the mass-discrepancy observed in \citet{calissendorff_discrepancy_2017}. The near-IR spectral types of the individual components derived in \citep{calissendorff_characterising_2020} were surprisingly off by more than $1-\sigma$ from each other for the BC pair, but still within the error of the optical photometric spectral types from \citet{janson_astralux_2012}. The discrepancy in the near-IR spectral types could be due to one of the components being close to the edge of the detector and the PSF not fully sampled.

{\bf 2MASS J20163382-0711456} received a grade 4 (preliminary) orbital fit and is poorly constrained. The system has an entry in Gaia EDR3, but the parallax is dubious with a distance above 1374 pc. An older entry in Gaia DR2 suggested the system to be at a distance of $d = 34.25 \pm 1.41$ pc which we adopted for our calculations. Photometric spectral type analysis in the optical indicated for early M0 and M2 types for the binary pair, which suggests that our dynamical mass estimate is underestimated, and also in agreement with the photometric mass which is about 2-3 times higher than the dynamical mass from the preliminary orbital fit. RV data exists for the system but was not helpful for constraining the orbit, exhibiting high jitter.

{\bf 2MASS J21372900-0555082} had no Gaia parallax available and we adopted the photometric parallax from \citet{lepine_all-sky_2011} for the system, which has a corresponding distance-uncertainty of $30\,\%$. The orbital fit grade was 4 (preliminary) and the dynamical mass estimate has its uncertainty budget dominated by the distance-error. It is likely that the orbital parameters are not constrained well enough, as the dynamical mass is lower than expected from the photometric mass from the models, but within the errors because of the uncertain distance. A distance of $\approx 18$ pc would bring the dynamical and photometric masses closer together. All three YMG tools suggested the system to be part of the field.

{\bf 2MASS J23172807+1936469} has previously been suggested to be part of the BPMG \citep{malo_bayesian_2013, janson_noopsortborbital_2014}, but updated space velocity parameters with the BANYAN $\Sigma$-online tool suggest it to be a field system.  The convergence point tool gave some low indication for a possibility of the system being part of the THA group. Our orbital fit of the system {was one of the more robust in our sample, receiving a grade 1 (reliable) on our scale, and the resulting orbital parameters were} consistent with the previous orbit fitted in \citet{janson_noopsortborbital_2014}. The earlier results did not have access to Gaia parallaxes, and the distance estimate by \citet{lepine_new_2005} of $d = 11.6 \pm 2.4$ pc to the system was insufficient for precise mass estimates, which uncertainty would convert the mass error to $50\,\%$ of the total mass. The two new astrometric data points included here did not change the results from the previous orbital fit, but here we also incorporated RV data into the fit and had access to the Gaia DR2 parallax which allowed for robust dynamical masses to be derived.

{\bf 2MASS J23261182+1700082} displayed a discrepancy between the two orbital fitting procedures employed, where the grid-search favoured higher masses. We estimated the grade of the orbit as 4 (preliminary), which showed surprisingly small uncertainties in mass despite the two masses from the two methods being so different. The mass from the MCMC orbit was more in line with with photometric mass from the evolutionary models compared to the grid-search that predicted a $\approx 30\,\%$ higher total dynamical mass in the system. Most of the YMG tools agreed that the system is more likely in the field, except for the convergence point tool which gave some probability for it being a THA member. If we were to assume the system to have the same age as the THA group of $45\pm4$ Myrs, the photometric mass would be reduced $0.19 - 0.27 M_\odot$, intensifying the discrepancy. The optical photometric spectral types were similar to that of J0111, and we would expect J232611 to have a mass not too dissimilar from it. New observations would greatly aid to reduce the large upper uncertainty in the orbital period that the grid-search obtained for the system.

{\bf 2MASS J23261707+2752034} is likely belonging to the field according to all three YMG tools. The system has the fewest amount of observed epochs of just five observations in our sample, and the orbital fit received the grade 3 (tentative), which span almost an entire orbital revolution. The evolutionary models predicted higher total mass than our dynamical mass estimates, and it is possible that we overestimated the field age for the system. New observations closer to periastron could help constrain the orbital parameters further.

{\bf 2MASS J23495365+2427493} receivied a grade 4 (preliminary) for its orbital fit, and it remains too uncertain to determine whether the observed epochs are close to periastron or apastron, causing a large uncertainty in the fitted orbital period. The system had previously been estimated to be part of  either the BPMG or Columba \citep{malo_bayesian_2013, janson_noopsortborbital_2014}, but updated Gaia EDR3 parameters and the BANYAN $\Sigma$-online tool suggests the system to belong to the field instead. The convergence point tool suggests the system to be a strong TWA candidate member however. The dynamical mass from the grid-search and MCMC differed for the system, where the grid-search preferred higher masses and the MCMC lower masses compared to the photometric mass from the theoretical models. If we adopted the young TWA age of $\approx 10$ Myrs however, the photometric mass is similar to the mass obtained from the MCMC method. Nevertheless, the orbit was only constrained to a preliminary level and not yet good enough to make an adequate comparison.

\section{Summary}\label{appendixE}
\clearpage
\begin{table*}[t]
\renewcommand{\arraystretch}{1.3}
\centering
\caption{Summary}
\begin{tabular}{l|cc|cc|ccc|c}
\hline \hline
Name$^{\rm grade}$  & \multicolumn{2}{|c|}{$P\,[{\rm yrs}]$} & \multicolumn{2}{|c|}{$a\,[{\rm AU}]$} & \multicolumn{3}{|c|}{$M_{\rm tot}\,[M_\odot]$} & $[M_{\rm B}/M_{\rm tot}]$ \\
 & Grid & MCMC & Grid & MCMC & Grid & MCMC & Theoretical &\\
\hline
J0008$^1$ & $5.92\pm0.01$ & $5.94^{+0.02}_{-0.01}$ & $2.60\,^{+0.05}_{-0.04}$ & $2.61 \pm0.01$ & $0.50\pm0.02$ & $0.50\,^{+0.04}_{-0.03}$ & $0.45-0.53$ &\\
J0111$^3$ & $41\,^{+  20}_{ -10}$ & $56\,^{+1}_{-15}$ & $7.8\pm1.0$ & $7.5\,^{+1.2}_{-1.1}$ & $ 0.28\,^{+0.10}_{-0.11}$ & $0.15\,^{+0.11}_{-0.05}$ & $0.11-0.20$ &\\
J0225$^5$ & $20\,^{+  20}_{ -3}$ & $49\,^{+43.6}_{-14.5}$ & $ 4.8\pm 0.3$ & $6.94\,^{+3.36}_{-1.75}$ & $0.27\,^{+0.05}_{-0.06}$ & $0.12\,^{+0.04}_{-0.03}$ & $0.22 - 0.26$ &\\
J0245$^3$ & $30\pm2$ & $69\,^{+43}_{-19}$ & $ 7.7\pm0.2$ & $14.22\,^{+0.73}_{-0.55}$ & $0.51\pm0.03$ & $0.55\,^{+0.04}_{-0.03}$ & $0.53 - 0.60$ &\\
J0437$^2$ & $26\,^{+6}_{-1}$ & $29.4\,^{+0.5}_{-0.4}$ & $ 8.6\,^{+0.1}_{-0.2}$ & $9.3\pm0.2$ & $ 0.92\,^{+0.06}_{-0.04}$ & $0.93\pm0.04$ & $0.91-1.09$ & $0.44\pm0.02$\\
J0459$^4$ & $ 28\,^{+   6}_{ -16}$ & $32\,^{+1}_{-5}$ & $6.0\,^{+0.4}_{-0.3}$ & $6.38\,^{+0.27}_{-0.72}$ & $0.28\,^{+0.05}_{-0.03}$ & $0.26\,^{+0.06}_{-0.04}$ & $0.75-0.78$ & $0.24\,^{+0.29}_{-0.04}$\\
J0532$^4$ & $87\,^{+12}_{-4}$ & $26\,^{+15}_{-7}$ & $19.7\pm0.5$ & $13.92\,^{+2.65}_{-2.02}$ & $1.01\,\pm0.07$ & $1.87\,^{+0.10}_{-0.49}$ & $0.93-1.14$ & $0.41\,^{+0.05}_{-0.04}$\\
J0611$^5$ & $121\,^{+3304}_{ -99}$ & $57\,^{+41}_{-21}$ & $54\,^{+5635}_{ -40}$ & $19\,^{+8}_{-5}$ & $11\,^{+11}_{-8}$  & $1.1\,^{+0.60}_{-0.39}$ & $0.78 - 0.93$ &\\
J0613$^2$ & $13.2\,^{+0.2}_{-0.4}$ & $11.56\,^{+0.90}_{-0.73}$ & $4.62\,^{+0.06}_{-0.04}$ & $3.91\,^{+0.83}_{-0.11}$ & $0.57\pm0.02$ & $0.42\,^{+0.38}_{-0.15}$ & $0.28-0.34$ & $0.37\,^{+0.06}_{-0.05}$\\
J0728$^1$ & $7.79\pm0.03$ & $7.76\pm0.02$ & $ 4.03\pm0.04$ & $3.99\pm0.01$ & $1.08\,\pm0.03$ & $1.06\,\pm0.03$ & $1.05-1.11$ & $0.51\,^{+0.10}_{-0.08}$\\
J0907$^3$ & $10.2\,^{+2.0}_{-0.8}$ & $10.8\,^{+0.8}_{-0.6}$ & $ 11.3\,^{+0.3}_{-0.7}$ & $4.64\,^{+1.36}_{-0.81}$ & $14\,\pm1$ & $0.78\,^{+0.60}_{-0.35}$ & $0.69 - 0.85$ &\\
J0916$^3$ & $8.6\,^{+0.1}_{-0.2}$ & $8.76\,^{+0.09}_{-0.08}$ & $4.24\,^{+0.12}_{-0.11}$ & $4.27\,^{+0.06}_{-0.17}$ & $1.02\,^{+0.08}_{-0.07}$ & $1.01\,^{+0.12}_{-0.09}$ & $1.12 - 1.13$ & $0.79\,^{+0.06}_{-0.07}$\\
J1014$^3$ & $     48\,^{+  24}_{  -9}$ & $48.5\,^{+16.3}_{-6.5}$ & $13.5\,^{+1.3}_{-0.6}$ & $13.59\,^{+1.72}_{-0.41}$ & $1.10\,\pm0.10$ & $0.87\,^{+0.20}_{-0.13}$ & $0.83 - 1.12$&\\
J1036$^1$ & $8.56\pm0.02$ & $8.47\pm0.02$ & $2.92\pm0.03$ & $2.98\pm0.03$ & $0.34\,\pm0.01$ & $0.37\,\pm0.01$ & $0.37-0.38$ & $0.50\,\pm0.02$ \\
J2016$^4$ & $     41\,^{+4371}_{ -26}$ & $36\,^{+26}_{-13}$ & $ 9\,^{+  13}_{  -3}$ & $8.43\,^{+3.15}_{-1.48}$ & $0.45\,^{+0.10}_{-0.09}$ & $0.32\,^{+0.09}_{-0.07}$ & $0.97 - 1.03$ &\\
J2137$^4$ & $     44\pm8$ & $69\,^{+29}_{-16}$ & $ 9\,^{+  11}_{  -3}$ & $13.04\,^{+6.97}_{-6.00}$ & $0.4\,\pm0.4$ & $0.40\,^{+0.38}_{-0.37}$ & $0.53 - 0.56$ &\\
J2317$^1$ & $  11.53\,^{+0.05}_{-0.03}$ & $11.55\pm0.02$ & $ 4.32\,^{+0.05}_{-0.06}$ & $4.32\,^{+0.08}_{-0.06}$ & $0.60\,\pm0.02$ & $0.61\pm0.03$ & $0.63-0.64$ & $0.45\,\pm0.03$\\
J232611$^4$ & $     20\,^{+  34}_{ -10}$ & $16.4\,^{+3.1}_{-1.9}$ & $    6.3\,^{+0.4}_{-0.6}$ & $5.18\,^{+0.70}_{-0.59}$ & $0.63\,^{+0.12}_{-0.09}$ & $0.49\,\pm0.04$ & $0.42 - 0.49$ &\\
J232617$^3$ & $   11.0\pm0.2$ & $10.95\,^{+0.16}_{-0.11}$ & $   4.33\pm0.06$ & $4.33\,^{+0.13}_{-0.14}$& $0.67\,\pm0.03$ & $0.68\,^{+0.06}_{-0.05}$ & $0.77 - 0.87$ &\\
J2349$^4$ & $     49\,^{+   8}_{ -31}$ & $40\,^{+26}_{-13}$ & $     11\,^{+   1}_{  -2}$ & $8.53\,^{+2.28}_{-1.10}$ & $0.52\,^{+0.07}_{-0.05}$ & $0.23\,^{+0.17}_{-0.04}$ & $0.42 - 0.44$ & \\ \hline \end{tabular}
\label{tab:results}
{\small\\
Orbital grades: (1) Reliable, (2) Good, (3) Tentative, (4) Preliminary, (5) Indetermined.
}
\end{table*}

\end{appendix}

\end{document}